\documentclass[a4paper,10pt]{article}
\usepackage{jheppub} 
\usepackage{lineno}
\pdfoutput=1

\usepackage[utf8]{inputenc}
\usepackage{amsmath}
\usepackage{amsfonts}
\usepackage{amsthm}
\usepackage{amssymb}
\usepackage{float}
\usepackage{braket}
\usepackage{graphicx}
\usepackage{xcolor}
\usepackage{url}
\usepackage{slashed}
\usepackage[normalem]{ulem}
\usepackage{subcaption}
\usepackage[font=small,justification=raggedright,singlelinecheck=false]{caption}

\graphicspath{{./Figures/}}

\newcommand{\tr}{\mathrm{tr}}

\newcommand{\coupling}{g_f}
\newcommand{\mf}{m_f}

\newcommand{\zweibein}[2]{\mathfrak{e}^{#1}_{#2}}
\newcommand{\invweibein}[2]{\mathcal{E}^{#1}_{#2}}

\title{Quantum dynamics of cosmological particle production: interacting quantum field theories with matrix product states}

\author[a]{Evan Budd}
\author[b]{Adrien Florio}
\author[c]{David Frenklakh}
\author[c]{Swagato Mukherjee}

\affiliation[a]{Department of Physics, North Carolina State University, Raleigh, NC 27695, USA}
\affiliation[b]{Fakultät für Physik, Universität Bielefeld, D-33615 Bielefeld, Germany}
\affiliation[c]{Department of Physics, Brookhaven National Laboratory, Upton, New York 11973, USA}

\emailAdd{dfrenklak@bnl.gov}

\abstract{Understanding real-time dynamics of interacting quantum fields in curved spacetime remains a major theoretical challenge. We employ tensor network methods to study such dynamics using interacting scalar and gauge theories in 1+1 spacetime dimensions, subject to 
a quench modeling 
a homogeneously expanding gravitational background. The models considered are the scalar $\lambda\phi^4$ theory and the Schwinger model, i.e. a Dirac fermion coupled to a $U(1)$ gauge field which is equivalent via bosonization to a scalar field with a cosine self-interaction. In the free scalar limit, both theories reproduce known analytical results, providing a nontrivial numerical validation of bosonization in curved spacetime for the Schwinger model. Our central finding is that self-interactions lead to a suppression of gravitational particle production compared to the free-field case, as evidenced by two-point functions and the spectra of produced particles.  We further examine the behavior of entanglement generation and find that interactions suppress entanglement growth in the $\lambda\phi^4$ theory, while in the Schwinger model, the interplay between suppressed particle production and enhanced inter-particle correlations leads to more complex entanglement behavior. Our results pave the way for further explorations of nonperturbative quantum real-time dynamics of interacting scalar and gauge theories in arbitrary gravitational backgrounds.}

\begin{document}

\maketitle


\section{Introduction}

A compelling way to generate  a Universe that mostly evolves around thermal equilibrium is to assume an initial period of exponential inflation, where the volume of the Universe grew by at least $60$ e-folds.
Inflation is postulated to have been dynamically sourced by an ``inflaton" field. The inflaton decays into Standard Model particles, creating a thermal medium and thereby ``gracefully" exiting inflation. This phase of ``(p)re-heating" often proceeds through nonperturbative dynamical instabilities and is associated with exponential particle production, whose semi-classical treatment using numerical simulations is an active area of research; see \cite{Figueroa:2020rrl, Baeza-Ballesteros:2025tme} for recent reviews. 

A key feature of this process is the interaction between the matter fields and the gravitational field. Indeed, even when backreaction of matter on the spacetime geometry is neglected, quantum field theory (QFT) in a background gravitational field gives rise to some of the most intriguing phenomena in modern physics. In addition to the early Universe phenomena, a famous  example is that of Hawking radiation~\cite{Bekenstein:1973ur, Hawking:1975vcx}. Beyond genuinely gravitational systems, QFT in curved spacetime is also relevant for a variety of systems across high-energy, condensed-matter and atomic physics. Heavy ion collisions at the Relativistic Heavy Ion Collider (RHIC) and the Large Hadron Collider (LHC), which recreate short-lived quark-gluon plasma undergoing rapid expansion, provide an experimentally accessible analogue to certain aspects of early-Universe dynamics~\cite{Heinz:2000bk}. Another important class of analogues arises in fluid flows~\cite{Unruh:1980cg}, most notably using Bose-Einstein condensates, which have enabled a wide range of experiments and simulations in analogue gravity~\cite{Garay:1999sk,Visser:2001fe, Novello:2002qg, Barcelo:2003wu, Fedichev:2003bv, 
Fedichev:2003dj,
Fedichev:2003id,
Fischer:2004bf,
Jain:2007gg, Viermann:2022wgw}. Additionally, certain non-Hermitian Hamiltonians can be mapped to Hermitian systems placed in curved space \cite{Lv:2021dgf}, offering the possibility to explore the interplay of gravitational and QFT phenomena in controlled laboratory environments.

By all accounts, a major theoretical challenge is that perturbative calculations, highly successful in the Standard Model of particle physics, are significantly more challenging in curved spacetime. The difficulties result from the non-uniqueness of vacua and of positive- and negative-frequency mode decompositions, the lack of Poincaré invariance, and other related effects; see, e.g.~\cite{Birrell:1982ix}. An effective field theory description remains viable at sufficiently low energies~\cite{Cabass:2022avo}, but it breaks down in regimes of high curvature or strong gravitational fields. This motivates the search for a fully nonperturbative, first-principles framework capable of bridging regimes where existing methods are applicable and extending into those where they fail. Although Euclidean lattice field theory has been remarkably successful in elucidating nonperturbative aspects of strongly-coupled QFTs, most notably Quantum Chromodynamics (QCD)~\cite{Kogut:1982ds, Brambilla:2014jmp}, it is not well suited to settings where real-time evolution is essential because of the famous sign problem. Gravitational, and particularly cosmological, phenomena are notable examples of such situations, and call for new approaches that provide direct access to their real-time nonperturbative dynamics.

A promising route towards accessing real-time dynamics is offered by the rapidly advancing field of Hamiltonian lattice simulation techniques~\cite{Banuls:2019bmf, Bauer:2022hpo,DiMeglio:2023nsa}. These include both classical approaches, most notably tensor network techniques, and quantum simulation, implemented either on quantum hardware or through classical emulation. While simulating realistic high-energy theories such as QCD in 3+1 dimensions remains far beyond current hardware capabilities, recent years have seen substantial progress in developing quantum algorithms and high-precision classical benchmarks in lower-dimensional lattice models, particularly in 1+1 dimensions~\cite{Klco:2018kyo, Kharzeev:2020kgc, Ciavarella:2021nmj, Rigobello:2021fxw, deJong:2021wsd, Zhou:2021kdl, Czajka:2021yll, Farrell:2022wyt, Farrell:2022vyh, Davoudi:2022xmb, Florio:2023dke, Farrell:2023fgd, Belyansky:2023rgh, Barata:2023jgd, Farrell:2024fit, Davoudi:2024wyv, Papaefstathiou:2024zsu, Florio:2024aix, Grieninger:2024axp, Grieninger:2024cdl, Araz:2024dcy, Barata:2024apg, Farrell:2025nkx, Florio:2025hoc, Barata:2025jhd, Barata:2025rjb, Grieninger:2025mbm, Kang:2025xpz}.

In this paper, we propose a pathway toward leveraging the progress in quantum techniques to study cosmological particle production and other dynamical effects of an expanding Universe using simple scalar-field toy models. Recent investigations of gravitational particle production~\cite{Steinhauer:2021fhb, Viermann:2022wgw, Fulgado-Claudio:2024xvk, Schmidt:2024zpg,Kinoshita:2024ahu,Gong:2025lmj, Kharel:2025lek, Chandran:2023ogt} and related dynamical gravitational phenomena~\cite{Fedichev:2003dj,Fedichev:2003id,Steinhauer:2015saa, Rodriguez-Laguna:2016kri, Hu:2018psq, Lewis:2019oyx, Ikeda:2025yqq, Ikeda:2025lig, Catterall:2025ofz} have employed both classical Hamiltonian methods and quantum analogue as well as digital simulation platforms, but have primarily focused on free and, to a lesser extent, interacting fermionic theories. The behavior of interacting scalar and gauge fields in dynamically expanding backgrounds remains largely unexplored. In particular, the challenge of formulating Hamiltonian lattice gauge theories in a curved spacetime background has, to the best of our knowledge, not been tackled.

We address this gap by studying two representative models in 1+1 spacetime dimensions: the $\lambda\phi^4$ scalar theory and the massive Schwinger model. The former is arguably the simplest interacting scalar field theory and prepares the stage for comparison to cosmological models, since it has long been used as a toy inflaton potential in studies of inflation and preheating. The latter is especially interesting because it admits a dual interpretation: it can be viewed either as a fermionic theory coupled to a $U(1)$ gauge field or, via bosonization, as a massive scalar field with a cosine self-interaction potential. We show that the fermionic lattice  discretization of the Schwinger model retains a consistent bosonic interpretation even in a dynamically expanding background~\cite{Barcelos-Neto:1985yfn, Eboli:1987mu}, and we compare it to $\lambda\phi^4$ to demonstrate that they exhibit qualitatively similar dynamical phenomena. Importantly, in both cases our approach enables a fully nonperturbative treatment of the interactions.

The main findings of our work can be summarized as follows. First, we construct lattice Hamiltonian formulations of $\lambda \phi^4$ theory and the Schwinger model that are valid for an arbitrary spacetime geometry in 1+1 dimensions. The latter provides, to the best of our knowledge, the first example of a Hamiltonian lattice gauge theory formulated in general gravitational background. We work in conformal coordinates, which can always be adopted in 1+1D, simplifying the form of the metric. For a homogeneous Friedmann-Lemaitre-Robertson-Walker (FLRW) expanding spacetime, this construction leads to a particularly simple and intuitive result: the dimensionful parameters of either theory are rescaled with the powers of the scale factor, matching their mass dimension. 

By numerically simulating time-dependent global parameter quenches in these interacting lattice models, we investigate the dynamics of quantum fields in a dynamically expanding background. In the limit of a free scalar field both models accurately reproduce known analytical results, as verified through accessible two-point correlation functions and through the particle content of the final state. We further find strong numerical evidence that self-interaction of the scalar field suppresses gravitational particle production. Finally, we analyze the generation of real-space entanglement due to particle production. Our findings suggest that while interaction suppresses entanglement growth in the $\lambda\phi^4$ theory, in the Schwinger model the entanglement behavior is more complex. This indicates a nontrivial interplay between suppressed particle production and enhanced entanglement between the produced excitations, enabled by stronger coupling.

For the numerical implementation, we employ classical tensor network methods. Although state-of-the-art digital quantum computers have begun to produce reliable results for selected physical problems, substantial error mitigation remains necessary, and classical algorithms remain superior for near-term studies in 1+1 dimensions. Tensor network algorithms efficiently exploit the entanglement structure of low-energy states in gapped theories, enabling accurate computation of ground-state properties and, for moderate evolution times, real-time dynamics, see e.g.~\cite{Orus:2013kga} for a review. Such methods have been developed for both fermionic and bosonic theories, and can incorporate gauge fields. By enabling simulations in systems significantly larger than those tractable by exact diagonalization, tensor network methods allow us to explore trends towards the continuum limit, even if we do not attempt a controlled continuum extrapolation. For tasks where tensor networks become inefficient, such as accurately resolving excited states, we supplement our approach with exact diagonalization on smaller systems. These smaller systems are nevertheless sufficiently large for our purposes, enabled in the fermionic Schwinger model by the Pauli exclusion principle, constraining the dimension of the local Hilbert space.

This paper is organized as follows. In Section~\ref{sec:prelim} we formulate the two models in a curved spacetime background in Hamiltonian terms, first in the continuum, and then on the lattice, and conclude with a description of our numerical methods. Section~\ref{sec:2-point} discusses the computation of two-point correlation functions, their comparison with known analytical results in the free-field limit, and the effects of self-interaction on these observables. In Section~\ref{sec:prob} we analyze the dynamically generated particle content, using excited-state structure in the Schwinger model and the  momentum-space occupation number in the $\lambda\phi^4$ theory. Section~\ref{sec:entanglement} examines the dynamical generation of position-space entanglement due to particle production and its dependence on the interactions in both theories. We present our conclusions and outlook in Section~\ref{sec:conclusion}. 

The appendices contain additional technical details. Appendix~\ref{app:Hamiltonian} provides a complete first-principles derivation of the lattice Hamiltonian for the Schwinger model in a curved background. In Appendix~\ref{app:free:analytic} we review analytical results for the dynamics of a continuum free scalar field theory in the FLRW-type background considered in the main text. Appendix~\ref{app:free_2pt} details the calculation of the two-point functions in this analytically solvable setting. In Appendix~\ref{app:Schroedinger_Heisenberg} we demonstrate the equivalence between the Schr\"odinger and Heisenberg pictures for a harmonic oscillator with time-dependent frequency, offering an alternative derivation of several results from Appendix~\ref{app:free:analytic}. Finally, Appendix~\ref{app:survival} analyzes the behavior of the ``survival probability" --- the probability for the system to remain in the instantaneous ground state of the time-dependent Hamiltonian --- analytically both in the small- and large-volume limits. 
\section{Preliminaries} \label{sec:prelim}

To set the stage for our study of real-time scalar field dynamics in an expanding (1+1)-dimensional spacetime, we will consider both a scalar field with a $\lambda\phi^4$ self-interaction, and the massive Schwinger model, which in the continuum is bosonized to a scalar field with a cosine interaction. We start by recalling key features of these theories in the continuum. Then, in Sec.~\ref{subsec:prelim_lattice} we address their discretization in an expanding background and in Sec.~\ref{subsec:prelim_setup} we provide a description of our numerical methods.

\subsection{Continuum theories: expanding spacetime and bosonization} \label{subsec:prelim_cont}

\subsubsection{$\lambda\phi^4$ theory}

We consider a real scalar field $\phi$ with quartic self-interaction in 1+1 dimensions. In flat Minkowski spacetime, the Lagrangian density is
\begin{equation}
\mathcal{L}_{\lambda\phi^4}^{\rm flat} = \frac{1}{2} \partial_\mu \phi\partial^\mu\phi -\frac{1}{2}m_0^2 \phi^2 -\frac{\lambda}{4!}\phi^4 \,,
\end{equation}
where $m_0$ is the bare mass and $\lambda$ is the quartic coupling constant. This theory possesses a mass gap and excited bound states. 

To study the dynamics of this theory in a homogeneous expanding background, we work with a conformal metric of the type $g_{\mu\nu} = \Omega(t)^2 \eta_{\mu\nu}$, where $t$ denotes the conformal time, $\Omega(t)$ is the scale factor, and $\eta_{\mu\nu} = {\rm diag}(1,-1)$ is the flat Minkowski metric. Coupling the scalar field to the gravitational background in a minimal way, the theory becomes

\begin{align}
\mathcal{L}_{\lambda\phi^4} &= \sqrt{-g}\left(\frac{1}{2} \partial_\mu \phi\partial^\mu\phi -\frac{1}{2}m_0^2 \phi^2 -\frac{\lambda}{4!}\phi^4\right) \\
&= \frac{1}{2}\left(\left(\partial_t\phi\right)^2 - \left(\partial_x \phi\right)^2\right)  - \Omega(t)^2\left(\frac{1}{2}m_0^2 \phi^2 +\frac{\lambda}{4!}\phi^4\right) \,,
\end{align}
where $x$ denotes the spatial coordinate; in the second line we used $\sqrt{-g}  = \Omega^2(t)$, so in conformal coordinates the kinetic term retains its Minkowski form. The only effect of an expanding metric is then to uniformly rescale the potential.

Passing to the Hamiltonian formulation, the conjugate momentum is $\Pi = \partial_t\phi$, and the Hamiltonian density reads
\begin{equation}
\mathcal{H}_{\lambda\phi^4} = \frac{1}{2}\Pi^2 + \frac{1}{2}\left(\partial_x\phi\right)^2 + \Omega(t)^2\left(\frac{1}{2}m_0^2 \phi^2 +\frac{\lambda}{4!}\phi^4\right) \,. \label{eq:phi4_cont_Ham}
\end{equation}
In a particular expanding background, specified below in Eq.~(\ref{eq:omega_profile}), the free theory (at $\lambda=0$) admits an exact solution, which we  summarize in Appendix B. While perturbative expansions around the free theory are formally possible, they are significantly more involved in a time-dependent background compared to Minkowski spacetime, see for instance \cite{Akhmedov:2022whm}. To overcome this challenge and to study the real-time dynamics of this model, in this work we employ fully nonperturbative Hamiltonian lattice simulations.
The interacting 1+1D $\lambda\phi^4$ theory serves as a useful testbed for understanding nonperturbative scalar field dynamics in curved spacetime, as its $(3+1)$-dimensional generalization is of direct relevance to inflationary cosmology and preheating scenarios.

\subsubsection{Schwinger model}

We now turn to the Schwinger model, which is quantum electrodynamics (QED) in 1+1 spacetime dimensions. This theory describes a Dirac fermion field $\psi$ interacting with a $U(1)$ gauge field $A_\mu$. The Lagrangian density in flat Minkowski spacetime is
\begin{equation}
\mathcal{L}_{\rm Schwinger}^{\rm flat} = \bar\psi\,[i\,\gamma^\mu (\partial_\mu + i  A_\mu)  - \mf]\psi -\frac{1}{4\coupling^2}F_{\mu\nu}^2 \,,
\end{equation}
where $\gamma^\mu$ are the Dirac gamma matrices (in the following, we use the representation $\gamma^0 = \sigma^z$, $\gamma^1 = i\sigma^y$), $A_{\mu}$ is a $U(1)$ gauge field, $F_{\mu\nu} = \partial_\mu A_\nu - \partial_\nu A_\mu$ is the field strength tensor, $\mf$ is the fermion mass, and $\coupling$ is the gauge coupling which has dimensions of mass in 1+1D.

The Schwinger model is particularly interesting for several reasons. First, in flat spacetime it exhibits QCD-like phenomena including confinement and chiral symmetry breaking, making it a valuable toy model for understanding strongly-coupled gauge theories \cite{Schwinger:1962tp, Casher:1974vf, Coleman:1975pw, Coleman:1976uz}. The linearly rising Coulomb potential in one spatial dimension implies that isolated charges have infinite energy and must be confined. Second, the charge-neutral states are bosonic and the theory admits an exact solution via bosonization: it can be mapped to a massive sine-Gordon model for a scalar field $\varphi$ \cite{Coleman:1975pw, Coleman:1976uz}:
\begin{equation}
\mathcal{L}^{\rm flat}_{\rm boson} = \frac{1}{2}\partial_\mu\varphi\partial^\mu\varphi - \frac{m_B^2}{2}\varphi^2 - c \cos (2\sqrt{\pi}\varphi) \,,
\end{equation}
where $m_B = \frac{\coupling}{\sqrt{\pi}}$ and $c = \frac{e^\gamma}{2\pi}m_B \mf$ with the Euler constant $\gamma\approx0.577$; normal ordering with respect to $m_B$ is implied. In the massless fermion case, the corresponding bosonic field has mass equal to $m_B$ and is free: the theory is exactly solvable. For $\mf\neq0$, the lightest meson mass, which we will denote as $M_1$, receives corrections that can be computed perturbatively at small $\mf/\coupling$ or numerically in the nonperturbative regime, using lattice techniques \cite{Byrnes:2002nv,Banuls:2013jaa, Dempsey:2025wia}. Furthermore, the bosonization dictionary establishes connections between fermionic and bosonic operators. We recall such a relation for the $U(1)$ current density
\begin{equation}
j^\mu = \bar\psi\gamma^\mu\psi =  \frac{1}{\sqrt{\pi}}\epsilon^{\mu\nu}\partial_\nu \varphi \,,
\end{equation}
that will be useful to us later.

To study the Schwinger model in an expanding universe, we consider its Lagrangian density in curved spacetime
\begin{equation}
    \mathcal{L}_{\rm Schwinger} = \sqrt{-g} \left[\bar\psi(i \gamma^a \invweibein{\mu}{a} \overleftrightarrow\nabla_\mu - \mf)\psi - \frac{1}{4\coupling^2}F_{\mu\nu}F^{\mu\nu}\right],
\end{equation}
where $\invweibein{\mu}{a}$ is the inverse zweibein and the covariant derivative $\overleftrightarrow\nabla_\mu$ includes both gravitational and gauge connection terms. As detailed in Appendix~\ref{app:Hamiltonian}, for a conformally flat metric in 1+1 dimensions, the effect on the fermionic part is a simple overall factor of $1/\Omega$, while the gauge field dynamics remains standard. Canonical quantization has to be performed with care, see also Appendix~\ref{app:Hamiltonian}. The resulting Hamiltonian density in temporal gauge $A_0 = 0$ takes the form
\begin{equation}
    {\cal H}_{\rm Schwinger} = \frac{1}{2} \coupling^2\Omega(t)^2 \Pi_E^2  + \Omega(t)\bar\psi[-i\gamma^1\partial_x + \gamma^1A_1]\psi + \Omega(t)^2 \mf \bar\psi\psi \,.\label{eq:cont_Ham_curved}
\end{equation}
where the canonical momentum of the gauge field is $\Pi_E = \dfrac{1}{\coupling^2\Omega^2}\partial_0A_1$.
It is apparent that the net effect of the time-dependent metric in this frame is to rescale $\coupling$ and $\mf$ by $\Omega(t)$, together with rescaling the fermionic fields by $\sqrt{\Omega(t)}$. This Hamiltonian is supplemented by the covariant Gauss's law, which reads
\begin{equation}
    \partial_x \Pi_E - \Omega \bar\psi\gamma^0\psi = 0 \,, \label{eq:cont_Gauss}
\end{equation}
and implies in 1+1D that the electric field can be fully represented in terms of fermionic variables.

One expects the bosonization procedures to carry through in the expanding background~\cite{Barcelos-Neto:1985yfn, Eboli:1987mu}, with the theory given by Eq.~(\ref{eq:cont_Ham_curved}) corresponding to 
\begin{equation}
\mathcal{L}_{\rm boson} =  \frac{1}{2}\left(\left(\partial_t\varphi\right)^2 - \left(\partial_x \varphi\right)^2\right)   - \Omega^2(t)\left(\frac{m_B^2}{2}\varphi^2 + c \cos (2\sqrt{\pi}\varphi)\right ) \,,
\end{equation}
where again only the interaction potential is rescaled by $\Omega(t)^2$. We find compelling numerical evidence that evolution of the massless Schwinger model under the expansion is consistent with the one expected of a free scalar field. Therefore, it is reasonable to expect that the massive Schwinger model can be used as a model of the scalar field with a cosine self-interaction.

\subsection{Lattice formulations} \label{subsec:prelim_lattice}
\subsubsection{$\lambda\phi^4$ theory} \label{sec:prelim_lattice_lambda}

We discretize the spatial coordinate on a lattice with $N$ sites and lattice spacing $a$, the resulting one-dimensional volume of the system is $L=Na$. The field $\phi(x)$ is replaced by discrete variables $\Phi_n$ at each lattice site $n$, and the canonical momentum $\Pi(x)$ by discrete momenta $\Pi_n= a\Pi(x_n)$, rescaled  so that the lattice operator $\Pi_n$ is dimensionless and lattice operators obey the canonical commutation relations $[\Phi_n, \Pi_m] = i\delta_{nm}$. The lattice Hamiltonian then reads 
\begin{equation}
  H_{\lambda\phi^4}(t) = \sum_{n=1}^N \frac{1}{2a} \Pi_n^2 +\Omega(t)^2\left(\frac{am_0^2}{2} \Phi_n^2 +\frac{a\lambda}{4!}\Phi_n^4 \right)+ \sum_{n=1}^{N-1}  \frac{1}{2a} \left(\Phi_n - \Phi_{n+1}\right)^2 \,. \label{eq:lphi4disc}
\end{equation}
We express the field operators in terms of position-space annihilation and creation operators $a_n$ and $a_n^\dagger$ via
\begin{align}
\Phi_n &= \frac{1}{\sqrt{2}}\left(a_n^\dagger +a_n \right) \,,\\
\Pi_n &= \frac{i}{\sqrt{2}}\left(a_n^\dagger -a_n \right) \,.
\end{align}
Truncating the local Hilbert space to dimension $K$, an explicit matrix representation that satisfies the canonical commutation relations is
\begin{equation}
 a_n = \begin{pmatrix}
0  & \sqrt{1} & 0 & 0& \cdots \\
0 & 0 & \sqrt{2} & 0 & \cdots \\
\vdots &\vdots & \ddots &\ddots \\
0  && \cdots & 0 & \sqrt{K-1} \\
0 && \cdots & 0 & 0
 \end{pmatrix} \,.
\end{equation}
Matrix representations of $\Phi_n$, $\Pi_n$ and powers thereof follow directly from the definition. 

To explore particle production, we will later compute the ``occupation number per mode'', often considered in the cosmology literature \cite{Felder:2000hr}
\begin{align}
    n_k = \frac{1}{2}\left(\omega_k \langle\Phi_k^\dagger\Phi_{k}\rangle + \frac{1}{a^2\omega_k}\langle\Pi^\dagger_k\Pi_{k}\rangle\right) - \frac{1}{2}  \, ,
\end{align}
where we subtracted the zero-energy contribution. We define the discrete Fourier transforms as
\begin{align}
    \langle\Phi^\dagger_k\Phi_{k}\rangle = \frac{a^2}{L}\sum_{n,m} \langle\Phi_{n}\Phi_{m}\rangle \exp\left[-ika(n -m) \right] \,, \\  \langle\Pi^\dagger_k\Pi_{k}\rangle = \frac{a^2}{L}\sum_{n,m} \langle\Pi_{n}\Pi_{m}\rangle \exp\left[-ika(n -m) \right]\,,
\end{align}
and the time-dependent dispersion relation in the mean-field approximation reads
\begin{align}
    \omega_k^2 = k^2 + m_{\rm eff}^2(t) \equiv k^2 + \Omega^2(t) \left( m^2 + \frac{\lambda}{2}\langle\Phi^2\rangle\right)\ .
\end{align}

\subsubsection{Schwinger model}
\label{sec:prelim_lattice}

In this work, we introduce staggered fermions and integrate out the gauge field with the help of Gauss's law to place the Schwinger model in Hamiltonian formulation on the lattice. For more details on this procedure, see e.g.~\cite{Kharzeev:2020kgc, Florio:2023dke}.  In Appendix~\ref{app:Hamiltonian} we show the detailed derivation of the following Hamiltonian, presenting it in a generalized case of an inhomogeneous conformal metric, with $\Omega = \Omega(x,t)$. The Hamiltonian, representing the discrete version of Eq.~\eqref{eq:cont_Ham_curved}, reads
\begin{align}
H_{\rm Schwinger}(t) &= \frac{a \coupling^2  \Omega^2(t)}{2}H_E +  \frac{1}{2a} H_k
    + \mf \Omega(t) H_m, \label{eq:coord_Ham}
\end{align}
with
\begin{align}
    H_E &= \sum_{n=1}^{N-1} \bigg[\sum_{i=1}^n \chi_i^\dagger\chi_i + {\rm floor}\bigg(\frac{n + 1}{2}\bigg)\bigg]^2 \,, \label{eq:H_E} \\
    H_k &= - i \sum_{n=1}^{N-1} (\chi^\dagger_n\chi_{n+1} - \chi^\dagger_{n+1}\chi_n)   \,, \\
    H_m &= \sum_{n=1}^N(-1)^n\chi_n^\dagger\chi_n \,,
\end{align}
where $\chi^\dagger_n, \chi_n$ are discrete fermionic variables with the canonical anticommutation relation, $\{\chi_n, \chi_m^\dagger\} = \delta_{nm}$, used to represent the continuum Dirac fermion in a staggered way~\cite{Lewis:2019oyx}
\begin{align}
    \psi^1(x_n) &= \frac{1}{\sqrt{a\Omega(t,x_n)}}\chi_n~,~ \text{even } n \,, \\
    \psi^2(x_n) &= \frac{1}{\sqrt{a\Omega(t,x_n)}}\chi_n ~,~ \text{odd } n\,. \label{eq:staggering} 
\end{align}
Furthermore, in Eq.~(\ref{eq:H_E}) we used the discretized version of Gauss's law
\begin{equation}
    E_{n}-E_{n-1} = \chi^\dagger_n\chi_n - \frac{1-(-1)^n}{2}\,, \label{eq:lat_Gauss}
\end{equation}
to eliminate the electric field with open-boundary conditions (OBC) in the absence of background electric field. Note that, as explained in Appendix~\ref{app:Hamiltonian}, the discrete electric field operator $E_n$ directly corresponds to the canonical momentum of the gauge field $\Pi_E$. Same as with the scalar field, $N$ denotes the total number of (staggered) lattice sites and $a$ is the lattice spacing, with the total volume $L=Na$.

There is a simple intuition behind this discretization. Using a spatial discretization of the $x^1$ coordinate with a constant lattice spacing $a$, the physical distance between the lattice points changes in accord with the metric: 
\begin{equation}
    \Delta(t) = a\,\Omega(t).
\end{equation}
The static system is characterized by two dimensionless parameters, $\coupling a$ and $\mf a$. The time-dependent expansion then implies that their physical values must be time-dependent, for instance $\coupling a\,\Omega(t)$  for the former. Keeping $a$ fixed, we conclude that effectively
\begin{align}
    \coupling&\rightarrow \coupling(t)\equiv \coupling\, \Omega(t) \,, \\
    \mf&\rightarrow \mf(t)\equiv \mf \,\Omega(t) \,.
\end{align}
Equation \eqref{eq:coord_Ham} is indeed the usual  Schwinger model Hamiltonian with staggered fermions and integrated out electric field, subject to such rescaling of the fermion mass and coupling.

We recall here also the definition of the quasimomentum in the Schwinger model with OBC, given by \cite{Banuls:2013jaa}
\begin{align}
    \hat P = \int dx\, T^{01}(x) = \frac{i}{4a^2}\sum_{n=1}^{N-2} (\chi_n  \chi^\dagger_{n+2} - \chi^\dagger_n  \chi_{n+2} ) \,, \label{eq:P_def}
\end{align}
where $T^{\mu\nu}$ stands for the energy-momentum tensor. We refer to this operator as a \textit{quasi}momentum because open-boundary conditions break translational invariance and momentum is not a conserved quantity.
In the infinite-volume limit, $L\rightarrow\infty$, the distinction between open and periodic boundary conditions (PBC) disappears, and $\hat P$ becomes the momentum operator in the usual sense. In finite volume, we use it as a well-motivated approximation of momentum, allowing us to examine the dispersion relation. 

Another quantum number, precise with periodic boundary conditions, is the charge conjugation, given by the following operator (\cite{Banuls:2013jaa, Itou:2024psm}): 
\begin{equation}
    S_R = \bigotimes_{k=1}^{N} (i \chi_k^\dagger e^{i\pi\sum_{l<k}n_l} + h.c.) T^{(1)} \,, \label{eq:S_R_def}
\end{equation}
where $n_l = \chi_l^\dagger\chi_l$. The string accompanying $\chi_k^\dagger$ counts the fermion number to the left of site $k$ and results from the Jordan-Wigner transformation \cite{Jordan:1928wi}, ensuring Fermi statistics. $T^{(1)}$ denotes a cyclic translation by one staggered lattice site to the right.

The two lowest-mass excitations of the massive Schwinger model are a pseudoscalar and then a scalar meson. In the strong coupling limit, the latter represents a bound state of the former. In the case of PBC, the eigenstates of the Hamiltonian are classified according to the expectation value ($\pm 1$) of the operator (\ref{eq:S_R_def}). The $(+1)$ sector contains the vacuum and the scalar meson, while the $(-1)$ sector contains the pseudoscalar meson, which is the lowest-lying excitation both in the continuum and on the lattice in charge-0 sector. With OBC, charge conjugation symmetry is not exact and this operator can assume any expectation value in the interval $[-1,1]$. We will use the sign of this value as a proxy of charge parity of the state, which is known to work well at least for the low-lying part of the spectrum \cite{Banuls:2013jaa}.

Finally, we define the conserved total electric charge operator
\begin{equation}
    Q = \sum_{n=1}^N \bigg[\chi_n^\dagger\chi_n + \frac{(-1)^n-1}{2}\bigg] \,,
\end{equation}
commuting with the Hamiltonian. In the infinite volume limit confinement implies that any states with $\langle Q\rangle \neq 0$ have infinite energy and are not a part of the physical spectrum. For this reason, even working on a finite lattice we focus on the charge-zero sector only.

\subsection{Simulation details} \label{subsec:prelim_setup}

\begin{figure}
    \centering
    \includegraphics[width=0.5\linewidth]{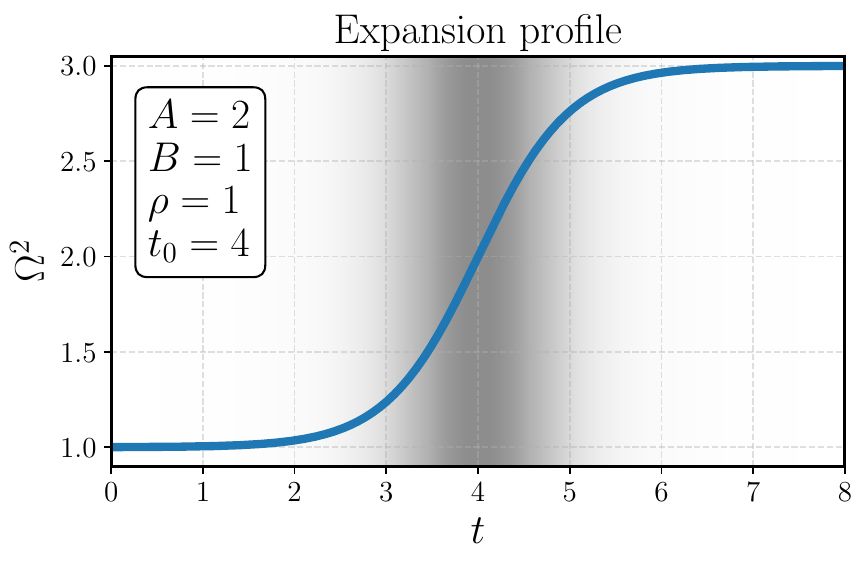}
    \caption{Scale factor as a function of time, given by Eq.~(\ref{eq:omega_profile}). In this work, we fix $A=2$, $B=1$, and vary $\rho$ and $t_0$ depending on the model analyzed and its parameters. The parameters presented here correspond to our studies in the Schwinger model (except when exploring the effects of $\rho$ variation). The shaded region helps visualize the expansion rate, with the shading intensity proportional to $\partial_t \Omega^2$.}
    \label{fig:scale_factor}
\end{figure}

In this work, we consider the asymptotically static expansion profile. This allows us to introduce well-defined notions of particle states both in the distant past and in the distant future. As a result, Fock-space concepts from  the conventional QFT in Minkowski spacetime can be employed in this setup, allowing us to study particle production directly. A particularly useful choice of the expansion profile is
\begin{align}
    \Omega^2(t) = A + B\tanh[\rho(t-t_0)] \,, \label{eq:omega_profile}
\end{align}
where $\rho$ sets the expansion rate while $A$ and $B$ affect the asymptotic scale factor values. In Fig.~\ref{fig:scale_factor} such an expansion profile is displayed for one of the parameter sets that we use in this study. For this functional form of $\Omega^2$, exact results are available in the free massive scalar field theory~\cite{Bernard:1977pq, Birrell:1982ix}, as reviewed in Appendix \ref{app:free:analytic}.

Let us now describe the protocol for numerical simulations, referencing the definitions in the Schwinger model; it proceeds in the same way for the $\lambda\phi^4$ model with the appropriate definition replacements. First, we prepare the system in the ground state of the Hamiltonian (\ref{eq:coord_Ham}) at $t\rightarrow-\infty$, $|\Psi_0\rangle = |0_{t\rightarrow-\infty}\rangle$. In practice, we choose $t_0$ such that $\tanh(\rho \,t_0)\approx1$; then at $t=0$ the Hamiltonian is sufficiently close to the asymptotic past one. Afterwards, we numerically perform time evolution of the state using the time-dependent Hamiltonian. We begin at $t=0$ and finish the evolution at $t_f$ such that $H(t\rightarrow\infty)\approx H(t=t_f)$. This way we obtain the time-dependent quantum state:
\begin{align}
    |\Psi_t\rangle = {\mathcal T} \exp \left[-i\int_0^t H(t') dt'\right]|\Psi_0\rangle . \label{eq:psi_time_evol}
\end{align}
From the time-dependent state we can compute expectation values of various observables as functions of time with a generic form
\begin{align}
    \langle {\cal O}\rangle(t) = \langle\Psi_t|{\cal O}|\Psi_t\rangle .
\end{align}
Sometimes we find it useful to compare these observables to the \textit{instantaneous vacuum} ones, as at every moment in time the system is described by a different Hamiltonian, with correspondingly different ground states. We denote the ground state of the Hamiltonian in Eq.~(\ref{eq:coord_Ham}) at time $t$ as $|0_t\rangle$ and more generally the $n$-th excited state of the same Hamiltonian as $|n_t\rangle$. 

To implement this procedure numerically, we mainly use tensor network methods. We use a matrix product state (MPS) representation of quantum states and matrix product operator (MPO) representation of the operators. We utilize tensor network algorithms, such as density matrix renormalization group (DMRG)~\cite{White:1993zza, Schollwock:2005zz} to find the ground states, and the time-dependent variational principle (TDVP)~\cite{Haegeman:2011zz, Haegeman:2016gfj} for real-time evolution. 
These methods are implemented using the iTensor package for Julia programming language \cite{itensor, itensor-r0.3}. In the case of the Schwinger model, we complement these studies with exact diagonalization explorations in small systems, with $N\leq 20$, to extract the low-lying spectrum of the theory after the quench. This helps us characterize in more detail the type of excitations created by the expansion. 

\begin{table}[h]
\centering
\begin{tabular}{|c| c|}
\hline
$\lambda/m^2$ & gap $(E_1 - E_0)/m$ \\
\hline
1  & 0.9999999652827682 \\ \hline
6  & 0.9999955927387987 \\ \hline
12 & 0.9118149944982861 \\ \hline
24 & 0.746486786151138\\
\hline
\end{tabular}
\caption{Energy gap $(E_1 - E_0)$ computed in the lattice $\lambda\phi^4$ model as a function of $\lambda$ for fixed $a=0.25, N=100, m=1$.}
\label{tab:gap_lambda}
\end{table}

To fix the model parameters in the $\lambda\phi^4$ theory, we follow the method described in~\cite{Sugihara:2004qr}. The model contains a single one-loop diagram with an ultraviolet (UV) divergence, describing the loop correction for the scalar propagator. It can be renormalized by redefining the bare mass parameter as 
$m_0^2 = m^2 - \delta m^2$.
The counterterm $\delta m^2$ in the lattice setup reads ~\cite{Sugihara:2004qr} 
\begin{equation}
  \delta m^2 = \frac{\lambda}{2} \frac{1}{2N}\sum_{n=1}^N \frac{1}{\sqrt{m^2 a^2+4\sin^2\left(\frac{\pi n}{N}\right)}}  \ .
\end{equation}
For all our results, we work at $m^2 = 1$ and vary $\lambda$ in the interval $\lambda\in[0,12]$. All the simulations we perform are in the symmetric phase of $\lambda \phi^4$. Numerically, we find that after the above renormalization the mass of the scalar field, measured through the energy of the first excited state, is sufficiently close to $m=1$ for all values of $\lambda$ under consideration, see Table~\ref{tab:gap_lambda}. Most of the time we work with lattice parameters $a=0.25, N=100$, and change them occasionally to study convergence to the continuum limit; we specify this explicitly. The majority of the $\lambda \phi^4$ simulations are performed with the expanding metric parameters $A=2, B=1$, $\rho = 2$ and $t_0=2.5$; we note explicitly if different values are used. We set the cutoff in the on-site Hilbert space to $K=10$.

For the Schwinger model, we use a similar strategy to fix the mass of the lightest meson, which we denote $M_1$, while varying the ratio of the fermion mass and coupling. In contrast with $\lambda\phi^4$, renormalization is performed based on existing numerical input. The meson mass is decomposed as 
\begin{align}
    M_1 = 2\mf+\kappa\coupling \,,
\end{align}
where $\kappa$ denotes a numerical coefficient, that depends on $\mf/\coupling$. 
We consider the set $\mf/\coupling = \{0,1/8, 1/4, 1/2\}$ and list the corresponding (continuum- and infinite-volume-extrapolated \cite{Byrnes:2002nv,Banuls:2013jaa, Dempsey:2025wia}) value of $\kappa$ in Table \ref{tab:m_to_g_set}. Using these values, we also present in Table \ref{tab:m_to_g_set} a set of couplings $\coupling$, such that the lightest meson mass is the same, namely $M_1 = 0.25$. In rare exceptions to this prescription, we specify the values of $\mf$ and $\coupling$ explicitly. The values of $N$ and $a$ we use in Schwinger model simulations vary; these values are specified for every result we report. Most simulations are performed with $A=2, B=1$, $\rho = 1$ and $t_0=4$; we again make an explicit note wherever different metric parameters are used. 

\begin{table}[]
    \centering
    \begin{tabular}{|c|c|c|c|c|}
    \hline
        $\mf/\coupling$ & 0 & 1/8 & 1/4 & 1/2 \\ \hline
        $\kappa$ & $1/\sqrt\pi$ & 0.5395 & 0.519 & 0.4875 \\ \hline
        $\coupling|_{M_1=0.25}$ & 0.443 & 0.316 & 0.245 & 0.168\\
    \hline
    \end{tabular}
    \caption{A compilation of results from \cite{Byrnes:2002nv,Banuls:2013jaa, Dempsey:2025wia} on the mass of the lightest meson in the massive Schwinger model, $M_1$. Here $\kappa \equiv \frac{M_1 - 2\mf}{\coupling}$. In the third row, we list values of $\coupling$ at different $\mf/\coupling$ such that the meson mass is the same at all values of fermion mass, namely $M_1 = 0.25$. In the context of time-dependent $\mf$ and $\coupling$ studied in this work the values in this table refer to the $t=0$ values.}
    \label{tab:m_to_g_set}
\end{table}
\section{Two-point correlators} \label{sec:2-point}

We start the numerical exploration by studying two-point correlation functions in our models, which are important for several reasons. First, they give an indirect indication of the particle content of the state through their spectral representation, which we will study later, in Section \ref{sec:prob}. Moreover, the time-dependent two-point function we will study can be computed analytically in the free-field limit, providing a crucial benchmark for our analysis. Finally, these observables are sensitive to the field self-interactions and can be used to quantify their effects even in scenarios when the definition of a particle becomes ambiguous.

As described in the Introduction, our strategy in this exploratory work is to study simultaneously a $\lambda\phi^4$ theory and the Schwinger model, both in the same expanding background. As we will see in the next section, the main benefit of this approach is to have access to different observables, in particular in relation to particle production. The correlation functions, naturally accessible in these models, also differ slightly. 

\subsection{$\lambda \phi^4$ theory}

The simplest correlation function we have at our disposal in the continuum $\lambda \phi^4$ theory is 
\begin{equation}
{\cal C}_2^{\rm cont.}(x,t)\equiv\langle\phi(0,t) \phi(x,t)\rangle \ . \label{eq:C2_cont}
\end{equation}
At $\lambda=0$, this correlator admits an analytical expression in the metric  \eqref{eq:omega_profile}; we recall it in App.~\ref{app:free_2pt}. On the lattice, the corresponding two-point function of the operator $\Phi_n$
\begin{align}
   {\cal C}_2^{\rm lat.}(n, t) = \langle \Psi_t|\Phi_{N/2}\Phi_{N/2+n}|\Psi_t\rangle \, \label{eq:C2_lat}
\end{align}
is already finite after the mass renormalization described in the previous section. We defined this two-point function keeping one of the positions of field operators at the center of the lattice, corresponding to $x=0$ in the continuum. This prescription helps maintain control over finite-volume effects. 

Starting with the free case, $\lambda=0$, we compare the numerical results to the analytical prediction in Fig.~\ref{fig:2pt_vs_x}, in the right panel. Namely, we display the spatial dependence of the correlators ${\cal C}_2^{\rm cont.}(x,t)$ and ${\cal C}_2^{\rm lat.}(n,t)$, with $n$ in the latter translated to $x = an$, for several fixed moments of time $t$.  Even at late times we find that lattice discretization effects are minimal and the agreement is remarkable. The only noticeable effect is a progressive dephasing occurring at later times and larger distances; it provides an estimate of the remaining systematic uncertainties. 

We then move on to study the time dependence of the correlator ${\cal C}_2^{\rm lat.}(n,t)$ for a distance $x$ where the systematics are small. In the right panel of Fig.~\ref{fig:2pt_vs_t}, we show its time evolution for different values of $\lambda$. First, in accordance with the previous analysis, the $\lambda=0$ data shows remarkable agreement with the analytical prediction, even at late times. We point out that this agreement is not trivial, as tensor network algorithms often fail at late times because of a significant amount of entanglement generated during a quench. The observed agreement suggests that the regime under consideration generates a limited amount of entanglement, making the studies of real-time evolution in this setup particularly well-suited for MPS techniques. 

Next, we look at the effect of interactions on the correlator. We see that already at $t=0$, the effect of the interaction is significant. One could be tempted to attribute this effect to a varying mass generated by interactions. However, as shown in Table~\ref{tab:gap_lambda}, with our renormalization prescription, the physical mass $m$ is still 1 with subpercent correction at $\lambda=6.0$, where the $t=0$ value of the correlator is significantly modified. In other words, the difference in the correlator is not just a mass renormalization effect, but a nontrivial result of the self-interaction. The effect of the interaction is more dramatic during the evolution. The inclusion of interaction suppresses the oscillations; in the theory with stronger interaction the ground state is disturbed less by the expansion. In other words, the effects of self-interaction dominate over those of the expansion at strong coupling.   

\subsection{Schwinger model}

\begin{figure}
  \centering
  \begin{subfigure}[t]{0.49\textwidth} 
    \centering
    \includegraphics[width=\textwidth]{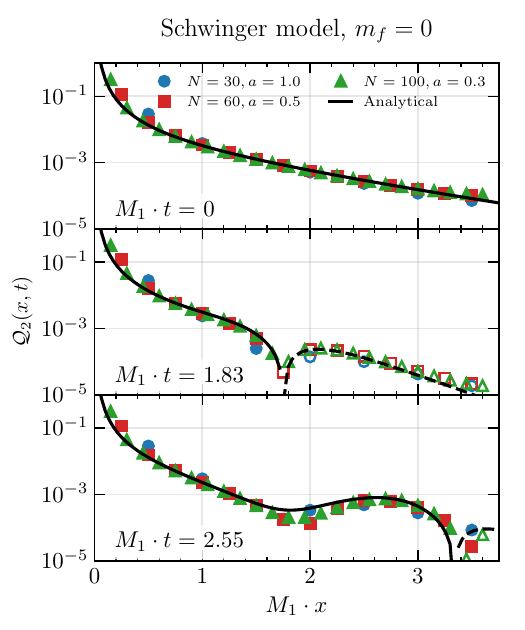}
  \end{subfigure}
  \hfill
  \begin{subfigure}[t]{0.49\textwidth}
    \centering
    \includegraphics[width=\textwidth]{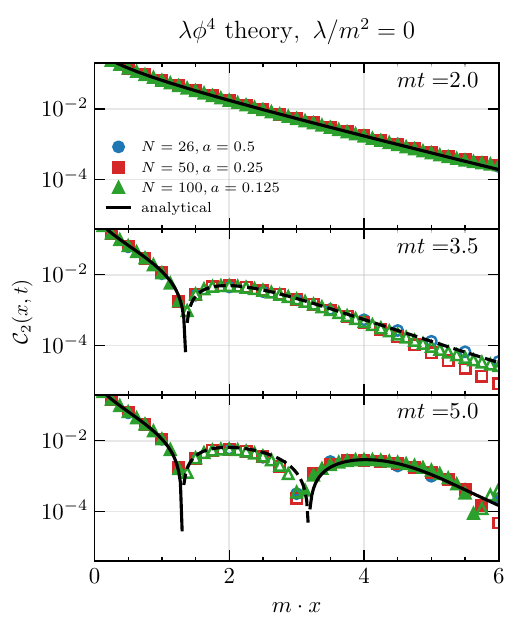}
  \end{subfigure} 
\caption{Left: two-point charge density correlator ${\cal Q}_2^{\rm lat.}(n,t)$, given by Eq.~(\ref{eq:Q2_lat}), in the massless Schwinger model with $M_1 = 0.25$, as a function of distance $x=2an$ for several fixed values of time $t$.  Different markers correspond to different choices of lattice spacing; physical volume is fixed. Right: same for the two-point field correlator ${\cal C}_2^{\rm lat.}(n,t)$, given by Eq.~(\ref{eq:C2_lat}),  calculated in the $\lambda\phi^4$ theory with $\lambda=0$ and $m=1$; $x=an$ in this case. In both panels, black curves show the corresponding analytical continuum results for a free scalar theory, computed as described in App. \ref{app:free_2pt}. Solid curve and filled markers correspond to the positive value of the correlator; dashed curve and empty markers - to the negative one.}
  \label{fig:2pt_vs_x}
\end{figure}

In the Schwinger model, two-point correlators exhibit similar behavior, as we show below. First, we note that the correlator whose bosonized version in the continuum corresponds to (\ref{eq:C2_cont}) is not directly accessible in the lattice Schwinger model. On the other hand, we can measure the charge density correlators
\begin{align}
{\cal Q}_2^{\rm lat.}(n,t)\equiv\langle \Psi_t| q_{N/4}q_{N/4+n}|\Psi_t\rangle -   \langle \Psi_t| q_{N/4}|\Psi_t\rangle\langle \Psi_t| q_{N/4+n}|\Psi_t\rangle \,, \label{eq:Q2_lat}
\end{align}
where we introduced a connected correlator, as the disconnected part generally does not vanish in the Schwinger model due to inhomogeneity introduced by the boundary. In order to mitigate staggering effects, we consider the charge density operator per physical site
\begin{equation}
    q_n = \frac{Q_{2n} + Q_{2n+1}}{2a} \,. \label{eq:q_phys}
\end{equation}
 with the local charge operator given by $Q_n = \chi_n^\dagger\chi_n + \frac{(-1)^n-1}{2}$; see Appendix \ref{app:Hamiltonian} where this conventional flat spacetime expression is shown to hold also in a gravitational background, at least in the conformal coordinates. 

\begin{figure}
    \centering
      \begin{subfigure}[t]{0.51\textwidth} 

    \includegraphics[scale=0.5]{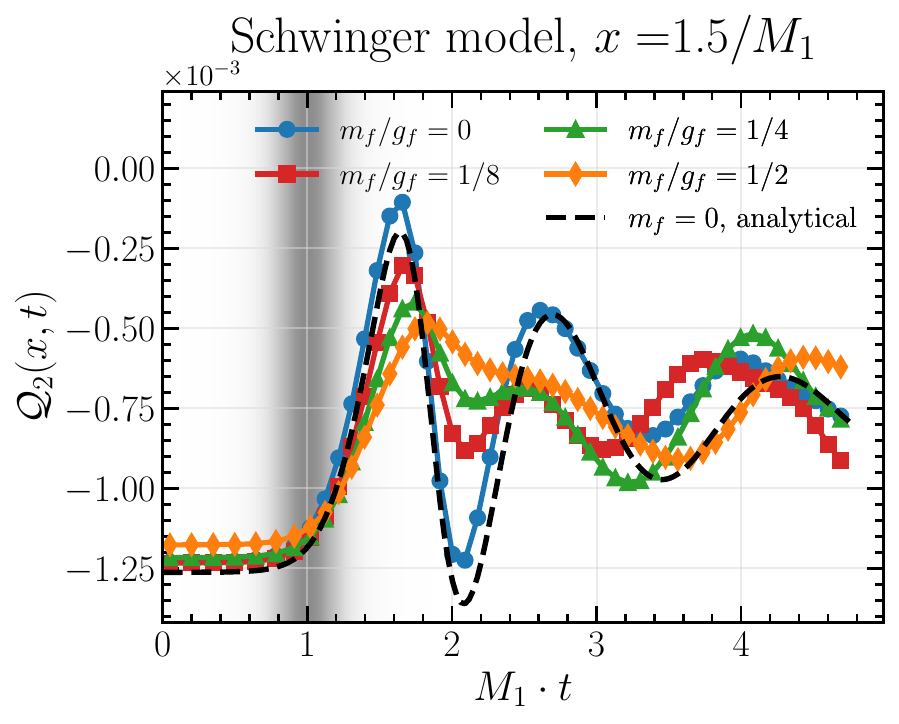}
      \end{subfigure}
      \begin{subfigure}[t]{0.48\textwidth}
        \includegraphics[scale=0.9]{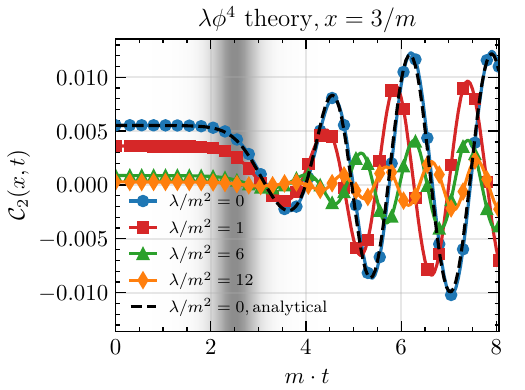}
      \end{subfigure}
\caption{Left: two-point charge density correlator ${\cal Q}_2^{\rm lat.}(n,t)$, given by Eq.~(\ref{eq:Q2_lat}), in the Schwinger model, as a function of time  for a fixed value of spatial separation $x=2an=6$. Fermion mass-to-coupling ratio is varied, while the lightest meson mass is fixed, $M_1 = 0.25$. Lattice parameters are $N=100, a=0.3$. Shaded region indicates when the expansion quench happens, with the shading intensity proportional to $\partial_t\Omega^2$. Right: same for the two-point field correlator ${\cal C}_2^{\rm lat.}(n,t)$, given by Eq.~(\ref{eq:C2_lat}),  calculated in the $\lambda\phi^4$ theory with $m=1$; $x=an=3$ in this case. Lattice parameters are $N=100, a=0.25$. In both panels, black dashed lines show the analytical continuum results for a free scalar theory of the corresponding mass, computed as described in App. \ref{app:free_2pt}.}   \label{fig:2pt_vs_t}
\end{figure}

 It is instructive to compare the results at zero fermion mass with the ones in a non-interacting scalar theory, to test the bosonization correspondence in the presence of a gravitational background. In curved space, the bosonization dictionary reads 
\begin{equation}
    j^0 = \frac{1}{\sqrt{\pi}}\frac{1}{\sqrt{-g}}\partial_x\varphi \,,
\end{equation}
and the charge density is obtained by further multiplying by $\sqrt{-g}$ (as e.g. the total charge is given by $\int dx \sqrt{-g} j^0$)
\begin{equation}
    q(x,t)\equiv \sqrt{-g(x,t)}j^0(x,t) = \frac{1}{\sqrt{\pi}} \partial_x\varphi(x,t) \ .
  \end{equation}
We can thus directly compare ${\cal Q}_2^{\rm lat.}(n,t)$ to 
\begin{align}
{\cal Q}_2^{\rm cont.}(x,t) \equiv \frac{1}{\pi}\langle\partial_x \varphi(0,t) \partial_x\varphi(x,t)\rangle
\end{align}
in the free scalar theory. The details of the calculation of ${\cal Q}_2^{\rm cont.}$ are presented in App. \ref{app:free_2pt}. 

In the left panel of Fig.~\ref{fig:2pt_vs_x}, we show the spatial dependence of the ${\cal Q}_2^{\rm lat.}(n,t)$ in the massless Schwinger model for several moments in time, with the correspondence $x=2na$. Comparison between the results obtained at different values of lattice spacing $a$ suggests that in the continuum limit, $a\rightarrow0$, the system moves towards the analytical predictions, shown with solid (or dashed, depending on the sign of the correlator) black lines. Except for minor deviations at distances of order of the lattice spacing, the agreement with the analytical results is again remarkable. This result is yet more significant than the one in the $\lambda\phi^4$ theory. Indeed, bosonization is usually discussed in flat spacetime. This agreement provides solid evidence that the equivalence between the massless Schwinger model and a free massive scalar theory extends to scenarios where both theories are placed in a gravitational background.

As with the scalar theory, to study the effect of the interaction in the scalar field, sourced by the fermion mass in the Schwinger model, on the dynamics of correlation functions, we study the dynamics of  ${\cal Q}_2^{\rm lat.}(n,t)$ at a given spatial separation $x=6$. The results, displayed in the left panel of Fig. \ref{fig:2pt_vs_t}, are obtained with the lattice  parameters $a=0.3, N = 100$. The analytical benchmark corresponds to the free scalar boson with $m_s = 0.25$. The lightest meson mass of the Schwinger model is tuned to the same value $M_1=0.25$ for all $\mf/\coupling$ by the choice of $\coupling$ presented in Table \ref{tab:m_to_g_set}. We find a confirmation that for $\mf=0$, the Schwinger model agrees with the analytical expectation quite accurately until a certain time, signaling accumulation of lattice effects over time. 

Again, we observe that during dynamical evolution  significant differences develop in theories with different values of the coupling. 
As in the case of the $\lambda\phi^4$ theory, we find that with stronger interaction the variation of the correlation function becomes smaller in magnitude and slower in time.
This behavior provides a qualitative indication of suppressed particle production in the presence of interactions, as the dynamics of the correlator is driven by particle creation. We confirm this in the next section.

\section{Measures of particle production} \label{sec:prob}

\subsection{Survival probability}

\begin{figure*}
  \centering
  \begin{subfigure}[t]{0.45\textwidth}
    \centering
    \includegraphics[width=\textwidth]{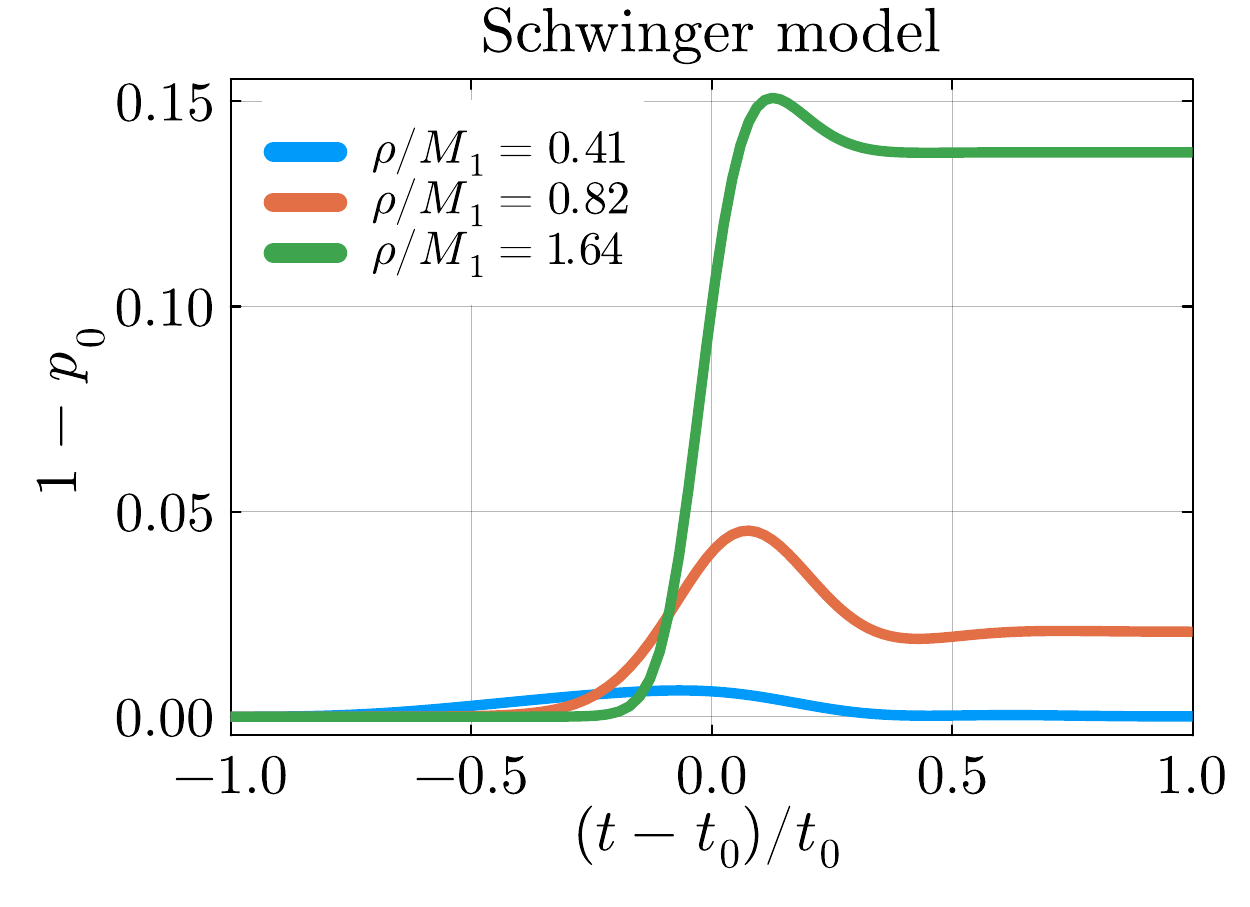}
  \end{subfigure}
  \hfill
  \begin{subfigure}[t]{0.45\textwidth}
    \centering
    \includegraphics[width=\textwidth]{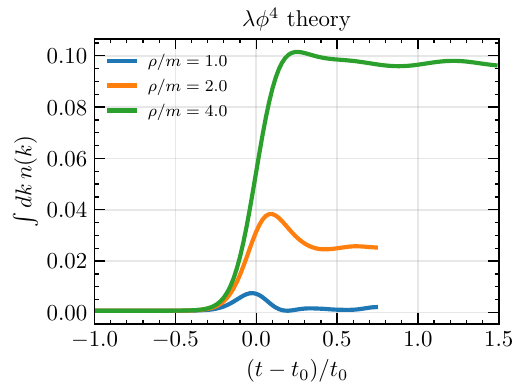}
  \end{subfigure}
  \caption{Comparison of the amount of excitations in a system for different values of the expansion rate $\rho$. Left: Excitation probability in the Schwinger model with parameters $\coupling =0.5, \mf=0.5$, corresponding to $M_1 = 1.22$; $N=40, a=1$. Right: integrated occupation number, defined in Sec. \ref{subsec:occupation}, in the $\lambda\phi^4$ theory with parameters $m=1$, $\lambda=1$, $N=100$, $a=1$. }
  \label{fig:p0_vs_rho}
\end{figure*}

Studying the two-point correlation functions in the previous section, we found first hints that self-interaction suppresses gravitational particle production. Is it possible to verify this hypothesis by examining the energy spectrum of produced states? For a system evolving under a time-dependent Hamiltonian, given either by Eq.~(\ref{eq:lphi4disc}) or Eq.~(\ref{eq:coord_Ham}), one can construct the instantaneous eigenbasis of the Hamiltonian, $\{|n_t\rangle\}$ where $n$ is the level index.  If the system evolved adiabatically, in the present context with the rate of spacetime expansion $\rho\rightarrow0$, it would remain in the instantaneous ground state, meaning no particle production. For a nonzero $\rho$, we find that the probability for the system to stay in the ground state, or the ``survival probability"
\begin{align}
    p_{0}(t) \equiv |\langle\Psi_t|0_t\rangle|^2 < 1,
\end{align}
which is a signal of particle production. In the left panel of Fig. \ref{fig:p0_vs_rho} we display the complement of the survival probability to unity, $1 - p_{0}(t)$, or ``excitation probability", as a function of time, for several values of the expansion rate parameter $\rho$, computed in the Schwinger model. In agreement with expectations, particle production is suppressed for smaller expansion rates. In the right panel of the same figure, we display the integrated value of the occupation number in the $\lambda\phi^4$ theory, defined below in Eq.~(\ref{eq:occupation_by_mode}). Again, we see that going towards the adiabatic limit, $\rho\rightarrow 0$, strongly suppresses particle production measured by this complementary probe.

\begin{figure*}
  \centering
  \begin{subfigure}[t]{0.45\textwidth}
    \centering
    \includegraphics[width=0.9\textwidth]{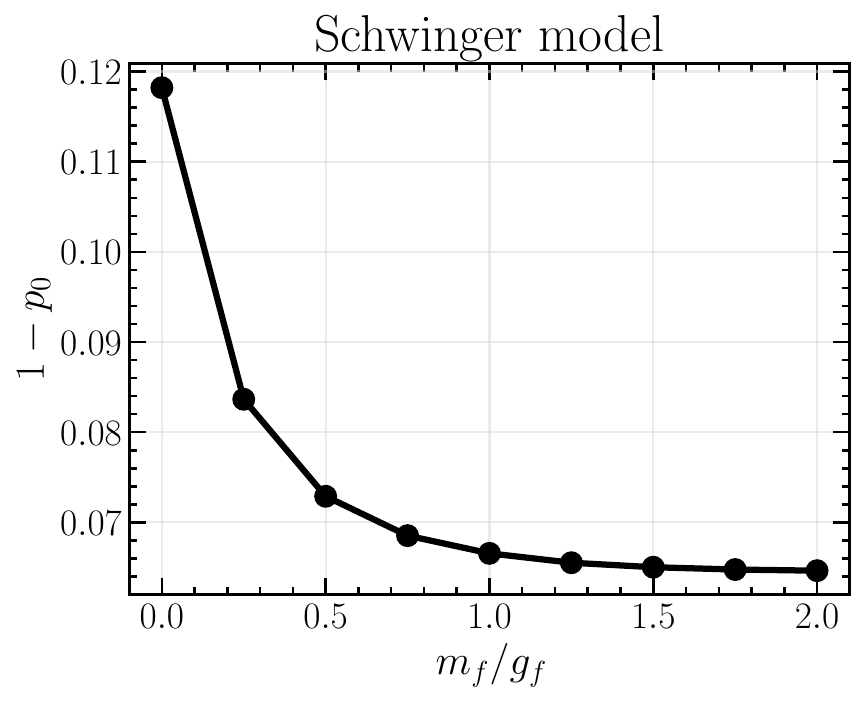}
  \end{subfigure}
  \hfill
  \begin{subfigure}[t]{0.45\textwidth}
    \centering
    \includegraphics[width=\textwidth]{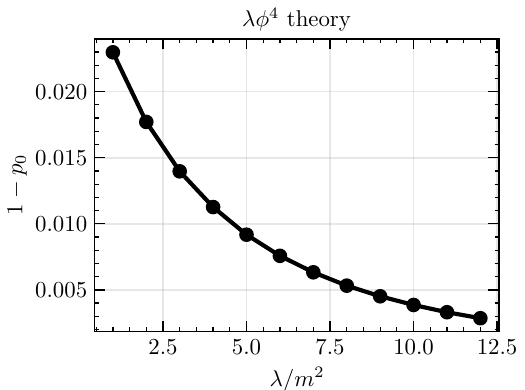}
  \end{subfigure}
  \caption{Excitation probability at late time as a function of the interaction strength. Left: Schwinger model with $N=40, a=1$; $\mf$ and $\coupling$ are adjusted so that the lightest meson mass $M_1=0.25$. The ratio $\mf/\coupling$ controls the strength of cosine interaction in the bosonized form. Right: $\lambda\phi^4$ theory with $N=100, a=0.25$, and  $m=1$.}
  \label{fig:p0_vs_interaction}
\end{figure*}

We further study the dependence of the survival probability on the interaction strength, controlled by $\lambda$ in the $\phi^4$ theory and by $\mf/\coupling$ in the Schwinger model. We characterize the particle production by the excitation probability at late times that we refer to as $p_{\rm exc} = 1-p_0({t\rightarrow\infty})$. Note that as the Hamiltonian and therefore its spectrum become time-independent at late times, the excitation probability has a well-defined late-time limit, unlike the correlation functions studied in the previous section. In Fig. \ref{fig:p0_vs_interaction} we show the excitation probability as a function of coupling strength both in the Schwinger model (left) and in the $\lambda\phi^4$ theory (right). In both theories, we find a confirmation that interaction suppresses particle production. In the Schwinger model we note that for $\mf/\coupling\geq1$ the excitation probability is nearly constant. At large $\mf/\coupling$, the Schwinger model in finite volume reduces to a free fermion theory, and we conclude that it approaches that regime with $\mf/\coupling\approx1$. As our primary interest is in exploring the dynamics of the dual boson, we restrict ourselves to $\mf/\coupling \leq 1/2$, where the theory is genuinely bosonic, in the remainder of this study.

Let us note that the survival probability loses its meaning in the thermodynamic limit, where it approaches zero as $e^{-N}$, with $N$ the number of degrees of freedom. This behavior is explained in Appendix \ref{app:survival}, using free theory as an example. Below, we explore other measures quantifying the amount of excitations in the system, that remain meaningful in the thermodynamic limit. One possibility is to examine the structure of excited state generation, which is explored later in Section \ref{subsec:excited} in the Schwinger model. The other option is to consider occupation number defined using momentum-space two-point functions, that we study in $\lambda\phi^4$ theory in Section \ref{subsec:occupation}.

\subsection{Excited-state tomography} \label{subsec:excited}

To characterize particle production in more detail than contained in the survival probability, we further study the internal structure of excitations. This study requires finding a substantial, and growing with the system size, number of excited states of a time-dependent Hamiltonian. While this can be achieved with DMRG, a systematic study of this kind requires resources beyond the scope of this exploratory work. Therefore, we use the exact diagonalization method on small Hilbert spaces (with dimension $\lesssim 10^6$) that still enables us to find the excited states of the Schwinger model on lattices of up to 20 sites. In practice, finding around 30 lowest-lying eigenstates is sufficient for our purposes with such lattice size. Each excited state is characterized by the expectation values of the Hamiltonian, square of the quasimomentum (\ref{eq:P_def}) and the approximate $\cal C$-parity (\ref{eq:S_R_def}), as discussed in Section \ref{sec:prelim_lattice}. Together, they allow us to characterize the particle content of each excited state, or at least what it should map to in the continuum and infinite volume limit. With the instantaneous eigenbasis, we compute
\begin{align}
    p_n^\pm(t)\equiv |\langle\Psi_t|n^\pm_t\rangle|^2\,, \label{eq:p_n_def}
\end{align}
where we have changed the notation slightly to denote the states with their spectral number in both positive and negative parity sectors independently. In particular, $|0^+\rangle$ is the vacuum, $|0^-\rangle$ is the first pseudoscalar mesonic excitation and $|1^+\rangle$ is the first scalar mesonic excitation.

\begin{figure}
    \centering
    \includegraphics[width=1.0\linewidth]{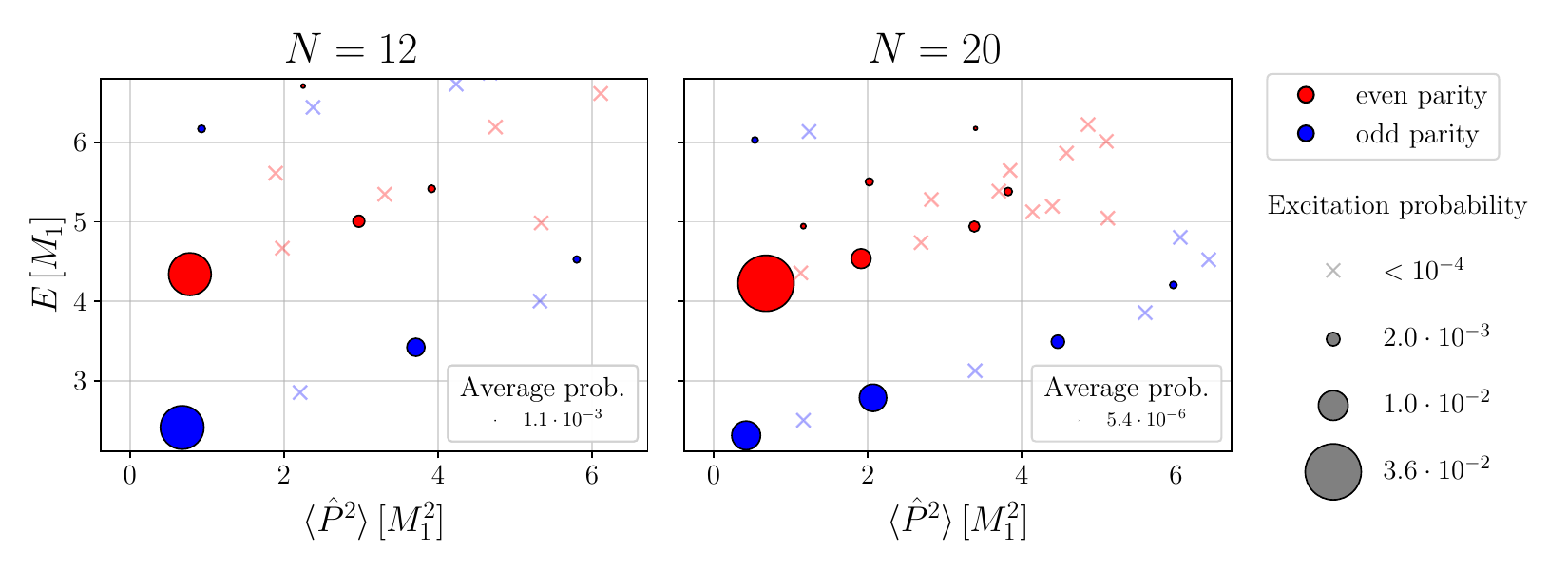}
    \caption{Structure of excitations at late times in the Schwinger model with $\mf=0, \coupling = 0.443, a = 1$ and $N=12$ (left) or $N=20$ (right). Energy $E$, with the vacuum value subtracted, is shown against (quasi)momentum squared, $\langle \hat P^2\rangle$ for the first excited states. Marker size indicates the late-time probability for this state to be generated by expansion from a vacuum state, given by Eq. (\ref{eq:p_n_def}) at $t=t_f$. States with the probability less than $10^{-4}$ are shown with a cross. Color refers to approximate charge parity defined in Eq.(\ref{eq:S_R_def}). For reference, on each panel we show the average probability to select a given state randomly in the fixed-charge $Q$ Hilbert subspace ${\cal H}_Q$, computed as $[{\rm dim}({\cal H}_Q)]^{-1}$.}
    \label{fig:exc_prob}
\end{figure}

Fig. \ref{fig:exc_prob} displays the dispersion relation for the low-lying excited states that are generated during the expansion in the massless Schwinger model. The size of a marker depends on the probability for this state to be found in the final state, given by Eq. (\ref{eq:p_n_def}) at $t=t_f$. Color encodes the approximate charge parity, with red for positive and blue for negative. The general structure of the dispersion relation is consistent with the one expected in the lattice Schwinger model: the lowest branch corresponds to pseudoscalar single-particle states of increasing momentum, while the upper branch describes two-particle states with positive parity. In the massless Schwinger model the latter correspond to free particle states, while at $\mf\neq 0$ Schwinger bosons are interacting, and these two-particle states correspond to bound or scattering states. Let us note that we observe that about half of the states on either branch have a probability numerically consistent with zero, possibly indicating some symmetry structure that was not apparent to us. Comparison between systems of different volume, $Na=12$ and $Na=20$, in Fig. \ref{fig:exc_prob} reveals that the role of negative-parity excitations diminishes at larger volume. It is consistent with the expectation that in the infinite volume parity becomes an exact quantum number irrespective of the boundary conditions, and negative-parity states could not be excited from the positive-parity vacuum in that limit. 

For a free massive scalar field in the continuum, subject to the same kind of spacetime expansion, the probability of finding any given particle content in the final state  is known analytically \cite{Bernard:1977pq, Birrell:1982ix} (see the discussion in Appendix \ref{app:free:analytic}). Because of momentum conservation, particles are produced in pairs of opposite momenta. When dealing with pseudoscalar particles, like in the Schwinger model, this additionally ensures that the final state has the same parity as the initial, vacuum state. As the positive-parity branch in Fig. \ref{fig:exc_prob} corresponds to the two-particle states, we can compare the probabilities on that branch with the analytical result for two-particle states with vanishing total momentum in the outgoing basis, $|1_k^{\rm out} 1_{-k}^{\rm out}\rangle$, where we adopted the notation for states in multiparticle Fock space:
\begin{equation}
|n^1_{k_1}\dots n^{I}_{k_I}\rangle = \bigotimes_{i=1}^I |n^i_{k_i}\rangle \,,
\end{equation}
where $|n^i_{k_i}\rangle$ is the state with occupation number $n^i$ in the single-particle Fock space of momentum mode $k_i$ and $I$ is the total number of excited modes in the state under consideration. To establish the correspondence with the Schwinger model, we identify $\langle n^+_{t\rightarrow\infty}|\hat P^2|n^+_{t\rightarrow\infty}\rangle - \langle 0^+_{t\rightarrow\infty}|\hat P^2|0^+_{t\rightarrow\infty}\rangle$, where the second term represents the vacuum-value subtraction, with $k^2$ of the state $|1_k^{\rm out} 1_{-k}^{\rm out}\rangle$. 

\begin{figure}
    \centering
    \begin{subfigure}{0.475\textwidth}
        \includegraphics[width=\textwidth]{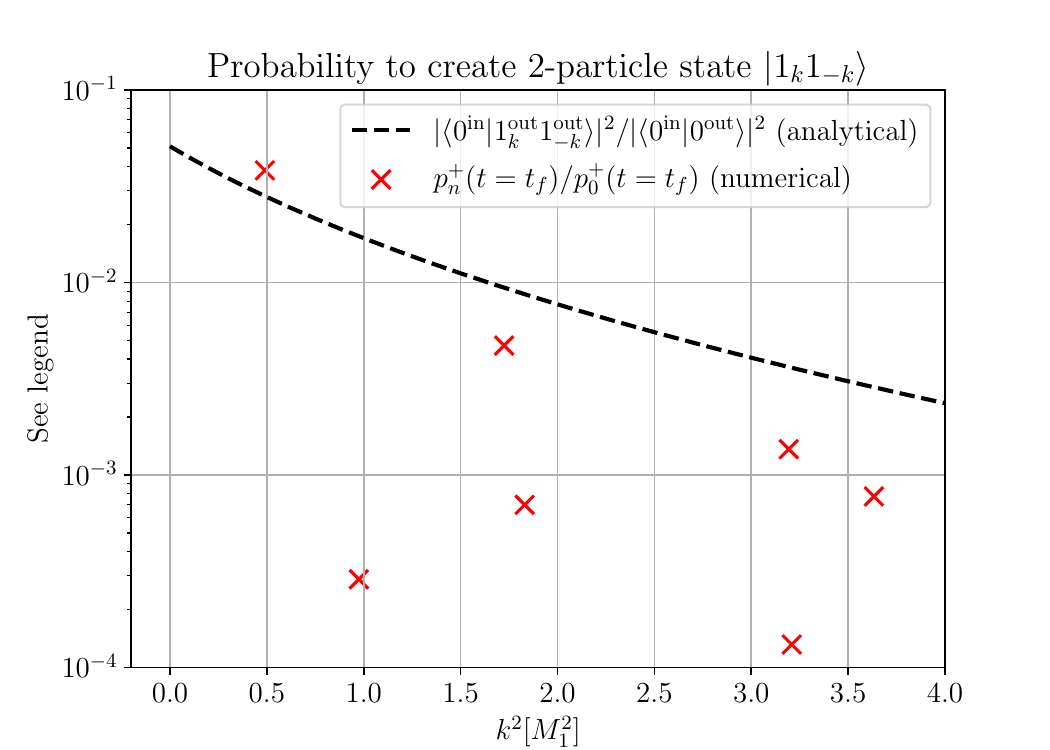}
    \end{subfigure}
    \hfill
    \begin{subfigure}{0.475\textwidth}
        \includegraphics[width=\textwidth]{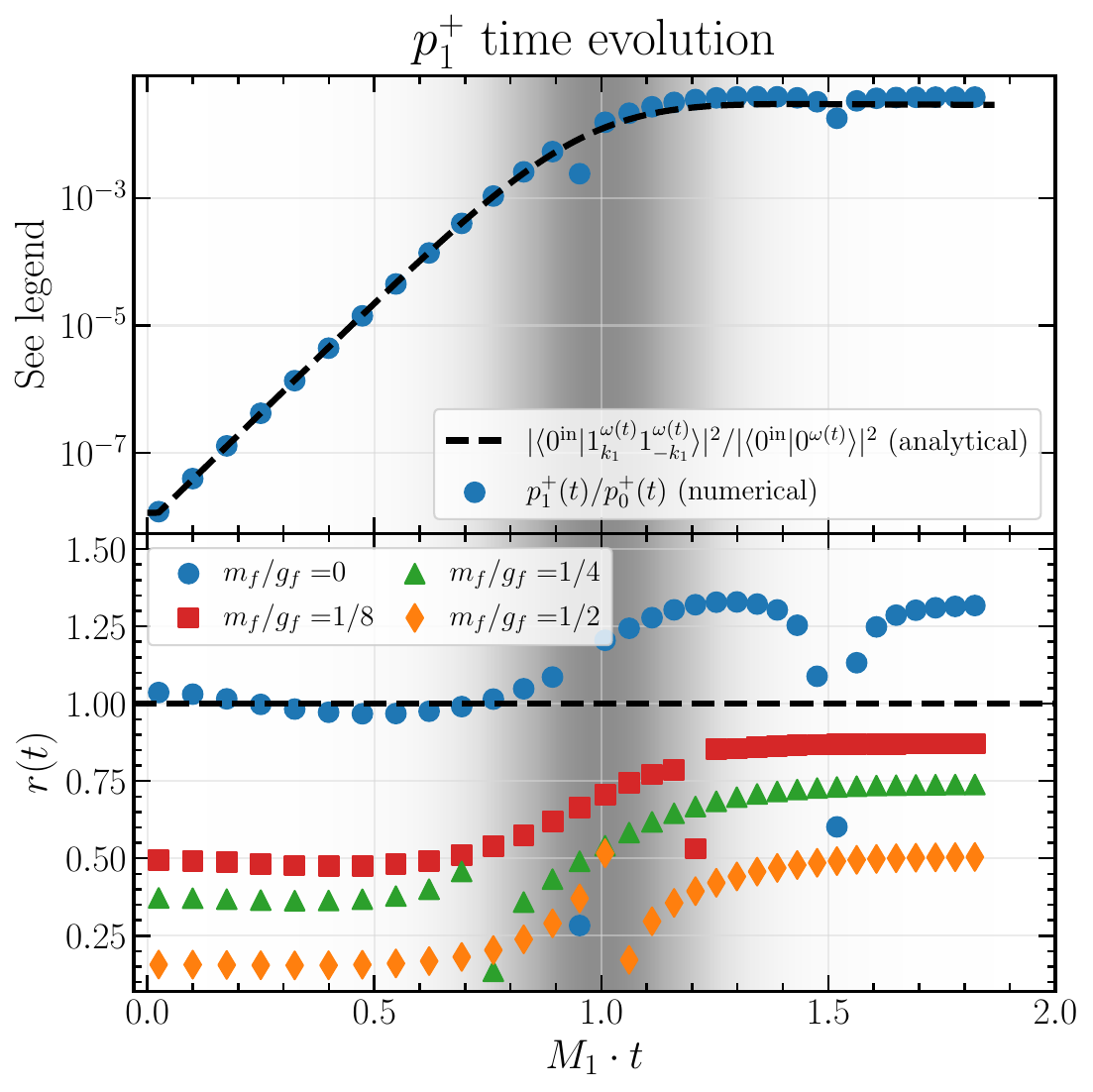}
    \end{subfigure}  
    \caption{Left: probability of creating a 2-particle state with momenta $\pm k$ in a continuum free massive boson theory (black dashed line) and in the massless lattice Schwinger model (red markers). $k^2$ for excited states in Schwinger model is computed as $\langle n^+_{t\rightarrow\infty}|\hat P^2|n^+_{t\rightarrow\infty}\rangle - \langle 0^+_{t\rightarrow\infty}|\hat P^2|0^+_{t\rightarrow\infty}\rangle$. Top right: evolution of the probability of the first parity-even excited state in massless Schwinger model compared to the analytical prediction for the same quantity, computed using quasimomentum extracted from the final state. Bottom right: the ratio of measured probability to the one computed in free theory for the corresponding quasimomentum as a function of time. In the massless case one expects this ratio to be close to 1, which is satisfied very well at early times. Nonzero interaction of the scalar field, introduced through the fermion mass, makes this ratio go down, signaling suppression of particle production by the interaction. Short sudden changes in the probability ratio are due to state mixing at proximity of level crossing and do not represent genuine effects. In both panels we study the system with $N=20, a=1$ and the fermion mass and coupling constant tuned so that the lightest meson mass is $M_1 = 0.25$. Shaded region indicates when the expansion quench happens, with the shading intensity proportional to $\partial_t\Omega^2$.}
    \label{fig:prob_vs_analytics}
\end{figure}

The probability of finding the system in the state $|1_k^{\rm out} 1_{-k}^{\rm out}\rangle$ is derived in Appendix~\ref{app:free:analytic} and is given by Eq.~(\ref{eq:one_pair_prob}). To define a measure of excitation which has a finite value in the thermodynamic limit, we divide it by the survival probability $p_0$, and do the same in the Schwinger model calculation, namely we consider $p_n^+(t=t_f)/p_0^+(t=t_f)$. The comparison is presented in the left panel of Fig. \ref{fig:prob_vs_analytics}. We find that the analytical prediction in the free theory works well for the lowest-momentum state. For larger momenta the qualitative behavior of the numerical results still matches the analytical prediction, but there is no close agreement. The reason may be that the difference between quasimomentum and momentum becomes more pronounced when either of them is large. It can also signal that the interpretation of the excited state as a two-particle state is only valid at small momentum. Performing the same analysis with periodic boundary conditions, with a well-defined notion of momentum, would help gain further understanding of this matter, and we leave it to future work.

Focusing on the state where we find the best agreement for the final probability, namely $|1^+\rangle$, we further track how the probability to find the system in such (instantaneous) state evolves in time. We again find the corresponding value of $k$ from the vacuum-subtracted expectation value of the square of quasimomentum operator in the final $|1^+\rangle$ state:
\begin{align}
    k_1^2 = \langle 1^+_{t\rightarrow\infty}|\hat P^2|1^+_{t\rightarrow\infty}\rangle - \langle 0^+_{t\rightarrow\infty}|\hat P^2|0^+_{t\rightarrow\infty}\rangle
\end{align}
We then study the following observable
\begin{align}
    r(t) \equiv \frac{p_1^+(t)/p_0^+(t)}{|\langle 0^{\rm in}|1_{k_1}^{\omega(t)}1_{-k_1}^{\omega(t)}\rangle|^2/|\langle0^{\rm in}|0^{\omega(t)}\rangle|^2}    \,,
\end{align}
where the numerator is measured in the massless Schwinger model simulations and the denominator is computed analytically from Eq.~(\ref{eq:analytic_prob_vs_t}) derived in Appendix~\ref{app:Schroedinger_Heisenberg}. It formally allows one to examine excitations in the instantaneous eigenbasis of the free Hamiltonian $H(t)$, corresponding to Eq. (\ref{eq:phi4_cont_Ham}) at $\lambda=0$, with the dispersion relation $\omega(t)^2 = k^2  + \Omega(t)^2 m^2$. In general, these eigenstates have no special meaning as they are not the asymptotic states of the model, so they cannot be assigned a particle interpretation. However, such states become physically meaningful in the adiabatic expansion limit, $\rho\rightarrow 0$, and are often considered within this adiabatic context. Here we also consider them as in the Hamiltonian simulation they can be accessed directly, without requiring any notion of a particle detector.

For the massless Schwinger model, we find $r(t)$ to be very close to 1 at early times, in excellent agreement with the analytical predictions, see Fig. \ref{fig:prob_vs_analytics}, right top panel. At later times we see a slight overproduction compared to the analytical prediction, which could be the artifact of approximations we are making. Note that sudden short changes in the value of probability are not genuine physical effects but signal state mixing in the proximity of level crossing. Moving on to study the nonzero values of $\mf/\coupling$, we observe that they all possess such late-time relative enhancement, but $r(t)$ at $\mf\neq0$ is systematically below the one at $\mf=0$ and decreases at larger $\mf/\coupling$, see the bottom right panel of Fig. \ref{fig:prob_vs_analytics}. As the fermion mass controls the interaction strength in the bosonic theory, these results provide another confirmation that particle production is suppressed by the interaction, as the probabilities found in the interacting theory are consistently lower than the ones in free theory. 

\subsection{Occupation number} \label{subsec:occupation}

 \begin{figure}
  \centering
\includegraphics[width=\textwidth]{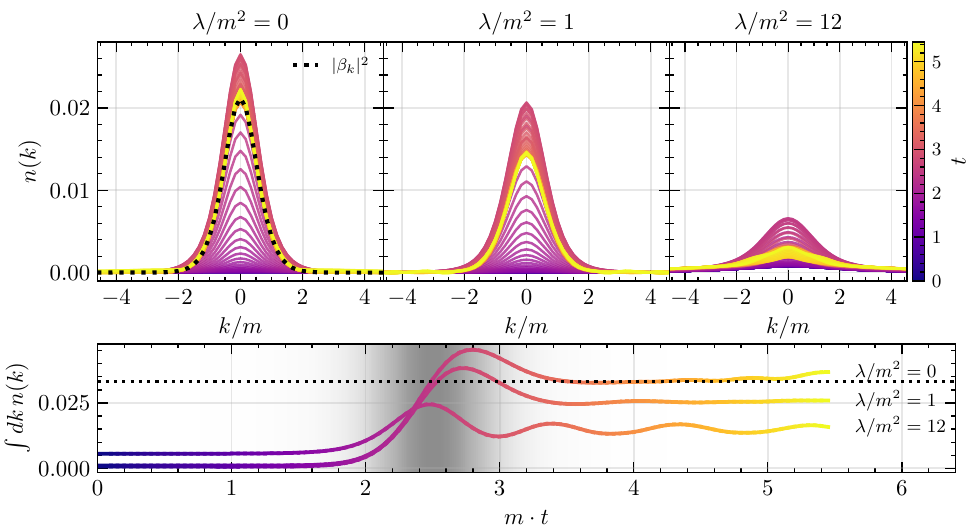}
  \caption{Top panel: evolution of the occupation number per mode, given by Eq.~(\ref{eq:occupation_by_mode}), for three values of the coupling constant. Lattice momenta are converted to physical momenta by means of Eq. (\ref{eq:lattice_dispersion}). In the $\lambda=0$ plot the dashed line corresponds to the analytical expectation in a free theory at late times. Bottom panel: evolution of the integrated occupation number. Dashed horizontal line corresponds to the analytical expectation in a free theory. The free theory converges to this value before deviating from it at $t\approx 5$, where lattice artifacts due to the propagation from the boundary start affecting the convergence. Simulations parameters are $N=100, a=0.25, m=1$. }
  \label{fig:spectrum_evolution_combined}
\end{figure}

In the $\lambda \phi^4$ theory we have direct access to the two-point correlation functions of the field and its canonical momentum, and we can compute the ``occupation number", as described in Section \ref{sec:prelim_lattice_lambda}.  Because the simulation uses open boundary conditions, we first compute the two-point correlator in the bulk region of the lattice $x\in[x_{\rm cut},L-x_{\rm cut}]$, excluding sites within a boundary zone to avoid boundary effects. We find that taking $x_{\text{cut}}=3.0/m$ ensures  that the results do not depend on this cutoff. Fixing a reference point $n_0$ at the left edge of this bulk region, we measure the connected correlators 
\begin{align}
C_{\cal O}(r) = \langle {\cal O}_{n_0}{\cal O}_{n_0+r}\rangle - \langle{\cal O}_{n_0}\rangle\langle{\cal O}_{n_0+r}\rangle \,,
\end{align}
where ${\cal O} = \{\Phi, \Pi\}$, as functions of the separation in lattice spacing units, $r$. The momentum-space correlators are obtained via the cosine transform
 \begin{align} 
{\tilde C}_{\cal O}(k) = a\bigg[C_{\cal O}(0) +
  2\sum_{r=1}^{r_{\max}} C_{\cal O}(r)\cos(kra)\bigg]\,,
  \end{align}
evaluated on the grid of lattice momenta $k_n^{\rm lat} = n\pi/L_{\text{corr}}$ with $n\in[-r_{\rm max},r_{\rm max}]$, where $L_{\text{corr}} = L-2x_{\rm cut} = r_{\max} \cdot a$ is the extent of the bulk correlator domain. The occupation number for each mode is then computed as \begin{align}
n(k) = \frac{1}{2}\bigg[\omega_k \tilde{C}_\Phi(k) + \frac{\tilde{C}_\Pi(k)}{\omega_k}\bigg] - \frac{1}{2}\,, \label{eq:occupation_by_mode}
\end{align}
where $\omega_k = \sqrt{k_{\text{phys}}^2 + m_{\text{eff}}^2(t)}$ with the time-dependent effective mass $m_{\text{eff}}^2(t) = \Omega(t)^2(m^2 + \frac{\lambda}{2}\langle\Phi^2\rangle)$. 

To ensure the correct comparison between the lattice and continuum theories, we convert the set of lattice momenta $k_n^{\rm lat}$ into physical momenta, using lattice dispersion relation 
\begin{align}
k_n^{\text{phys}} = \frac{2}{a}\bigg|\sin\bigg(\frac{k^{\rm lat}_n a}{2}\bigg)\bigg|\,. \label{eq:lattice_dispersion}
\end{align}
In Fig. \ref{fig:spectrum_evolution_combined} we display the time evolution of the occupation number per mode for several values of the coupling constant $\lambda$. In the free theory in the continuum the occupation number per mode can be computed analytically at late times \cite{Birrell:1982ix} (see Appendix \ref{app:free:analytic} for notation)
\begin{align}
n_c^{\lambda=0}(k) = \langle 0^{\rm in}|\hat N_k^{\rm out}|0^{\rm in}\rangle = |\beta_k|^2 = \frac{\sinh^2(\pi\omega_-/\rho)}{\sinh(\pi\omega_{\rm in}/\rho)\sinh(\pi\omega_{\rm out}/\rho)}\,, 
\end{align}
and it is shown with the dashed line on the $\lambda=0$ panel of Fig. \ref{fig:spectrum_evolution_combined}. Lattice results at late times are found to be in perfect agreement with $n_c^{\lambda=0}(k)$. Upon increasing the coupling constant, we find the occupation number in each mode decreasing, signaling decreased particle production in agreement with our findings of increasing survival probability and decreasing excitation probability at stronger coupling, reported earlier in this Section. In the bottom panel of Fig. \ref{fig:spectrum_evolution_combined} we display the integrated value of the occupation number, computed by numerically integrating the curves in the top panel. The corresponding late-time analytical value, shown with a dashed line, is
\begin{align}
    \int dk\, n_c^{\lambda=0}(k) = \int_{-\frac{\pi}{a}}^{\frac{\pi}{a}} dk \, |\beta_k|^2 \,.
\end{align}
We again find perfect agreement between the lattice and continuum results at $\lambda=0$, which continues until $t\approx 5$ where deviations appear. These deviations originate from the discretization, in particular from the presence of the boundary, and do not represent genuine physical effects. We find that with the correlators evaluated much closer to the boundary there is a faster buildup of lattice artifacts, compared to the two-point function shown in Fig. \ref{fig:2pt_vs_t}, where no comparable artifacts are seen up to $t\approx 8$.

\begin{figure}
  \centering
\includegraphics{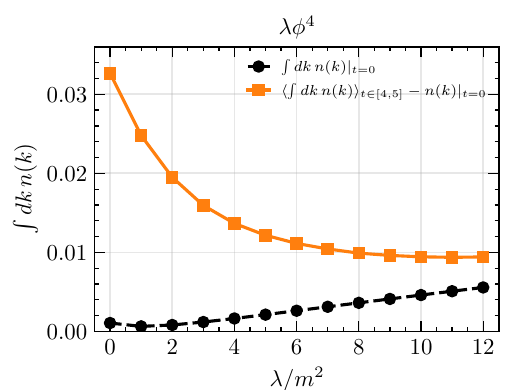}
  \caption{Total occupation number in $\lambda\phi^4$ model at late times, shown as a function of $\lambda$. To mitigate fluctuations and numerical artifacts, a time average within $t\in[4,5]$ is performed. The occupation number at $t=0$, shown with the black dashed line, is subtracted in order to characterize the particle production clearly. Simulations parameters are $N=100, a=0.25, m=1$.}
  \label{fig:occupation_vs_lambda}
\end{figure}

The late-time integrated occupation number is shown as a function of $\lambda$ in Fig. \ref{fig:occupation_vs_lambda}. To avoid lattice artifacts and to soften the effect of oscillations, seen e.g. in Fig. \ref{fig:spectrum_evolution_combined}  at $\lambda=12$, we take the average value of $\int dk\,n(k)$, evaluated over the interval $t\in[4,5]$. Furthermore, to quantify the production of particles, we subtract the occupation number present in the initial state, which is of the same order of magnitude for large $\lambda$; it is shown for comparison in the same figure. Once again, we find strong numerical evidence that the particle production is suppressed by the interaction.

\section{Dynamics of entanglement} \label{sec:entanglement}

We conclude this work by studying entanglement production. Quantum entanglement in quantum field theories and quantum many-body systems has attracted considerable attention in recent years; see, for example, the reviews~ \cite{Amico:2007ag, Calabrese:2009qy, Calabrese:2005zw, Eisert:2008ur, Hollands:2017dov, Witten:2018zxz}. In free QFTs~\cite{Casini:2009sr} and in conformal field theories (CFTs)~\cite{Calabrese:2009qy}, the entanglement entropy can be computed analytically in different settings, including both ground-state properties and post-quench dynamics~\cite{Calabrese:2005in}. For interacting gapped QFTs, ground-state entanglement measures can also be obtained numerically using Euclidean lattice methods. However, Euclidean approaches cannot directly access real-time evolution, making Hamiltonian formulations a natural framework for studying nonequilibrium entanglement dynamics in interacting QFTs.

Entanglement plays an increasingly important role in cosmology; see~\cite{Belfiglio:2025cst} for a recent review. In particular, entanglement may provide insights into the transition of quantum fluctuations in the early Universe into classical inhomogeneities, which seed large-scale structure formation at later stages of the cosmological evolution. In addition, entanglement is closely connected to black hole entropy~\cite{Bekenstein:1973ur, Hawking:1975vcx, Solodukhin:2011gn}. These considerations motivate the study of entanglement dynamics in models of gravitational particle production. Entanglement production in an expanding spacetime has previously been considered in free bosonic and fermionic theories in momentum space~\cite{Ball:2005xa, Fuentes:2010dt}. Other measures linking quantum state evolution to universal quantum computation, such as quantum magic, have also been considered in the cosmological context~\cite{Haque:2025pav}. However, to our knowledge, position-space entanglement, which is a natural notion in local QFT and generalizes straightforwardly to interacting theories, has not been studied in the literature. Here, we address this gap using Hamiltonian simulation methods which provide direct access to position-space entanglement.

\begin{figure}
    \centering
    \includegraphics[width=0.4\linewidth]{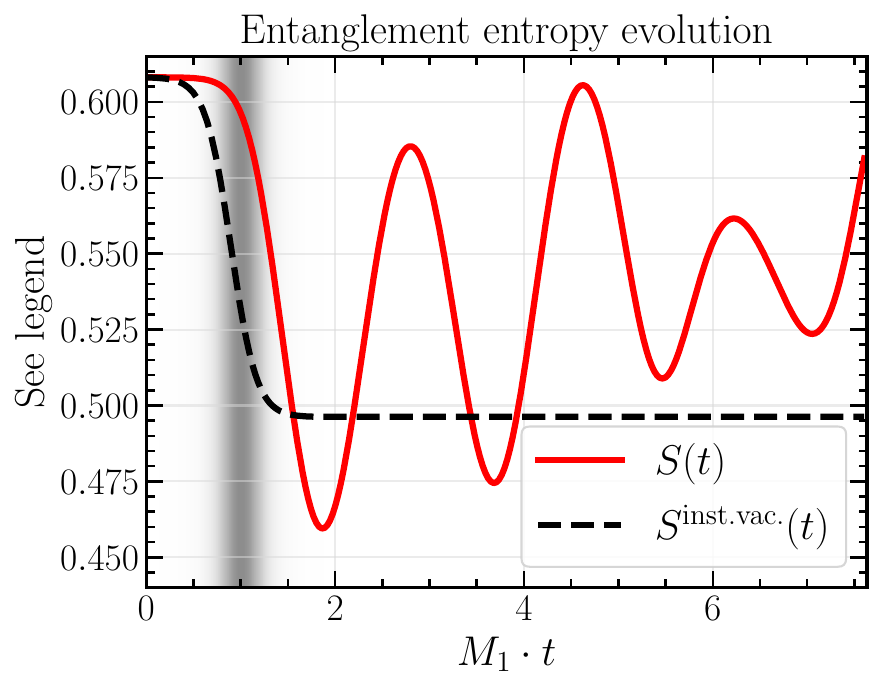}
    \caption{Evolution of the entanglement entropy (solid red line) and the corresponding instantaneous ground-state entanglement entropy (dashed black line), obtained in the Schwinger model with $\mf/\coupling = 1/8$, $M_1 = 0.25$, $N=30$, $a=1$. Shaded region indicates when the expansion quench happens, with the shading intensity proportional to $\partial_t\Omega^2$. An initial decrease in dynamical entanglement entropy mirrors the behavior of the instantaneous-ground-state entropy, followed by oscillations generated by particle production.}
    \label{fig:EE_evol_vs_inst}
\end{figure}

Let us partition the system into two halves with an  interface at the center. The entanglement entropy between the left (L) and right (R) subsystems is
\begin{align}
    S = -\tr\rho_L \log \rho_L \,, \label{eq:EE_def}
\end{align}
where $\rho_L$ is the reduced density matrix of the left half: 
\begin{equation}
\rho_L(t) = \tr_R \rho(t)  = \tr_R |\Psi_t\rangle\langle\Psi_t|
\end{equation}
and $\tr_R$ denotes taking the trace over the degrees of freedom residing in the right (R) half. This definition applies uniformly to both the $\lambda \phi^4$ theory and the Schwinger model, as the latter is formulated entirely in terms of fermionic degrees of freedom without gauge links, avoiding potential ambiguities in assigning degrees of freedom on the middle link to either of the two halves. While other bipartitions -- in particular an interval of finite length -- may also be of interest, here we focus exclusively on the half-system bipartition. 

Out of equilibrium, quantum systems typically exhibit an increase in entanglement as they evolve. This behavior has been extensively documented in QFT and lattice models, see e.g.~\cite{Calabrese:2005in,DeChiara:2005wb, Calabrese:2016xau, Rigobello:2021fxw, Florio:2023dke, Florio:2024aix,Florio:2025hoc}. In the context of cosmological particle production, one has to take extra steps to observe a similar entanglement growth. First, we note that if the system evolved adiabatically, it would remain in the ground state of the instantaneous Hamiltonian, whose entanglement entropy decreases due to the shrinking correlation length in lattice units. This can be seen as a manifestation of the UV-divergent nature of entanglement entropy. At finite expansion rate $\rho$, remnants of this adiabatic decrease remain visible, as shown in Fig.~\ref{fig:EE_evol_vs_inst}.

\begin{figure*}
  \centering
  \begin{subfigure}[t]{0.49\textwidth}
    \centering
    \includegraphics[width=\textwidth]{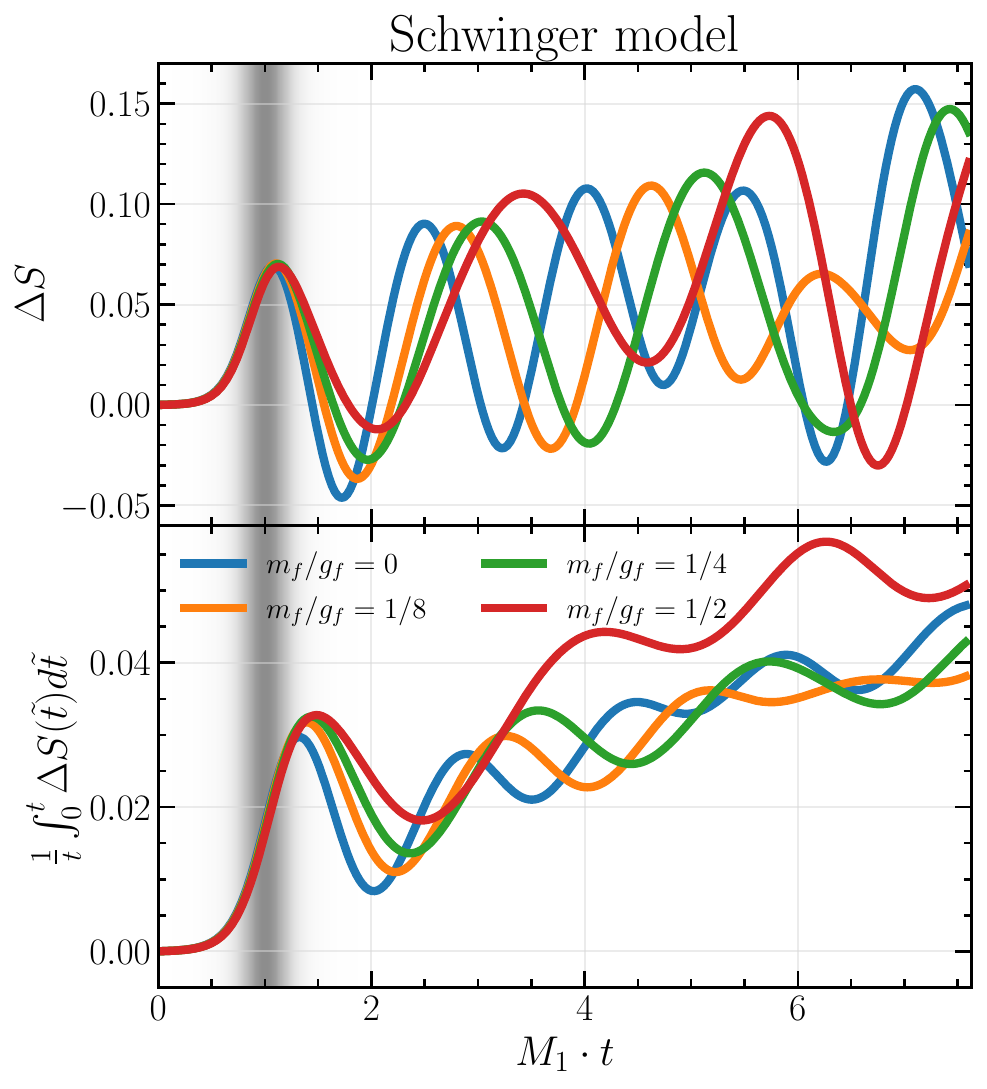}
  \end{subfigure}
  \hfill
  \begin{subfigure}[t]{0.49\textwidth}
    \centering
    \includegraphics[width=\textwidth]{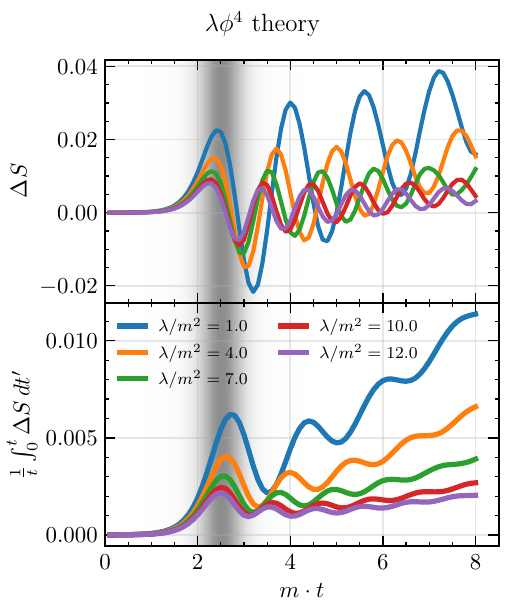}
  \end{subfigure}
  \caption{Left, top: dynamics of the entanglement entropy in the Schwinger model for several values of the fermion mass-to-coupling ratio, $\mf/\coupling$. Entanglement entropy of the instantaneous vacuum state is subtracted. Bottom: time-averaged vacuum-subtracted entanglement entropy, defined to reduce the oscillatory behavior and highlight long-time trends. Simulation parameters in the Schwinger model are $N=30, a=1, M_1=0.25$; the values of $\coupling$ corresponding to each $\mf/\coupling$ are presented in Table \ref{tab:m_to_g_set}. Shaded region indicates when the expansion quench happens, with the shading intensity proportional to $\partial_t\Omega^2$. Right: same quantities in the $\lambda\phi^4$ theory with simulation parameters $N=100, a=0.25, m=1$.}
  \label{fig:EE}
\end{figure*}

To isolate the entanglement generated by dynamical particle production rather than by a kinematical change in the Hamiltonian, we subtract the baseline, given by the left-right entanglement entropy of the instantaneous vacuum:
\begin{equation}
\Delta S(t) \equiv S(t) - S^{\rm inst. vac.}(t) \,,
\end{equation}
where
\begin{align}
    S^{\rm inst. vac.}(t) = -\tr_L \bigg[&\rho_L^{\rm inst. vac.}(t) \log \rho_L^{\rm inst. vac.}(t)\bigg] \,,\ \\ \rho_L^{\rm inst. vac.}(t) &\equiv \tr_R|0_t\rangle\langle0_t| \,.
\end{align}
The vacuum-subtracted entanglement entropy is shown in the upper panels of Fig. \ref{fig:EE}. The left panel displays the results in the Schwinger model for various values of $\mf/\coupling$. The right panel shows results in  $\lambda\phi^4$ model. After an initial period of growth,  the subtracted entropy exhibits oscillations, with periods matching those observed in two-point functions (Sec.~\ref{sec:2-point}). These oscillations can be interpreted as originating from produced particles that propagate through the system and reflect from its boundaries. With this interpretation, entanglement entropy peaks when a particle crosses the bipartition interface. 

For comparisons across different coupling strengths, the oscillatory behavior presents a challenge. We therefore define a time-averaged quantity,
\begin{equation}
    \overline{ \Delta S}(t) \equiv\frac{1}{t}\int_0^t \Delta S(t')\, dt' \,,
\end{equation}
shown in the lower panels of Fig. \ref{fig:EE}. The averaged entropy displays a clear long-time increase, as expected for post-quench dynamics of a quantum system.

In the $\lambda\phi^4$ theory, stronger coupling leads to reduced entanglement generation, consistent with the interaction-induced suppression of particle production observed in previous sections. In the Schwinger model, however, the dependence is more subtle. For $\mf/\coupling = \{0,1/8,1/4\}$, the long-time entropy-production rate is approximately constant, while for $\mf/\coupling=1/2$ it is significantly larger. Although stronger interactions suppress particle production, they also enhance the interactions among produced particles, which in turn increases entanglement. Our results indicate that the balance between these competing effects differs between $\lambda\phi^4$ and the $\cos\phi$ interaction of the bosonized Schwinger model: in $\lambda\phi^4$, suppression of particle production dominates across the couplings studied, whereas in the Schwinger model, interaction-induced enhancement of entanglement overtakes particle-production suppression once $\mf/\coupling$ becomes sufficiently large.

\section{Conclusion and outlook} \label{sec:conclusion}

In this paper, we performed the first fully nonperturbative study of gravitational particle production in interacting scalar quantum field theories, focusing on the $\lambda\phi^4$ theory and the massive Schwinger model in 1+1 dimensions. We developed a Hamiltonian lattice formulation of these models valid for arbitrary curved backgrounds, and we verified that both models reproduce known analytical results for two-point correlation functions in the free-field limit. For the Schwinger model, this agreement has profound implications: it demonstrates that the bosonization correspondence between the massless fermionic $U(1)$ gauge theory and a massive free scalar theory persists in time-dependent curved spacetime without additional modification. This result provides a solid justification for interpreting the Schwinger model as an interacting scalar theory in our gravitational setting.

Our central physical finding is the suppression of gravitational particle production by the self-interaction. In both models, stronger interactions --- parametrized by $\lambda$ in the $\lambda\phi^4$ theory and by the fermion mass in the Schwinger model --- lead to reduced fluctuations in the two-point functions and to a particle spectrum more strongly biased toward the vacuum configuration. By comparing occupation numbers and final-state probabilities with exact free-theory predictions, we obtain clear numerical evidence of this suppression. The behavior of real-space entanglement entropy is consistent with this picture: interactions suppress entanglement growth in the $\lambda\phi^4$ theory, while in the Schwinger model we observe a nontrivial interplay between reduced particle production and stronger correlations among the produced excitations. We note that some observables, particularly momentum-space occupation numbers, exhibit finite-volume effects at intermediate and late times. Since the systems studied here are homogeneous, infinite-MPS methods such as iMPS, iDMRG, and infinite-time evolution algorithms offer a natural path forward, working directly in the thermodynamic limit~\cite{Vidal:2006ofj, Orus:2008zsh}. Applying these techniques would eliminate finite-volume artifacts altogether and enable more robust access to late-time dynamics.

In the present study we focused on FLRW-type expansions with asymptotically static metrics, ensuring a well-defined notion of particle number. Extending our framework to genuinely expanding geometries such as de Sitter space is highly desirable for cosmological applications. While, as in the classical case, an exponentially growing background poses the challenge of numerically resolving exponentially separated scales, it promises valuable insights into interacting quantum field dynamics in de Sitter spacetime.

The long-term goal is the genuine extension to higher dimensions. Gravitational theories in (2+1) dimensions provide access to cosmic microwave background (CMB)-like angular observables, yet do not possess dynamical gravity and, in particular, lack gravitational waves~\cite{Gott:1982qg,Deser:1983tn,Jackiw:1984je,Witten:1988hc}. Going to (3+1) dimensions will allow to directly match to phenomenology and compute true CMB observables as well as gravitational-wave signatures associated with gravitational particle production. 
From the computational standpoint, even though tensor network methods can still be applied in higher dimensions, see e.g.~\cite{Catterall:2025ofz} for a recent example, we note that their utility beyond 1+1D is limited.
This provides a compelling example where quantum simulators will eventually be able to outperform classical approaches. 

In the nearer-term, and having  future phenomenological applications in mind, we  note that the FLRW metric under consideration offers a unique opportunity to study theories in  higher spacetime dimensions, already with classical methods. Indeed, owing to isotropy, the leading effects associated with expansion can be reduced to a $(1+1)$-dimensional theory of radial modes. Since our Hamiltonian formulation treats homogeneous and inhomogeneous metrics on equal footing, these effective theories are accessible. Such first-principles computation describing the transition from the quantum-dominated regime to the semi-classical regime will lead to new insights into the emergence of classicality.

Along similar lines, dimensionally reduced theories also open up the possibility of considering the semiclassical backreaction of matter on the FLRW metric. We envision this being achieved by solving an effective Friedmann equation, sourced by the appropriate quantum expectation values of the energy-momentum tensor (EMT) of the matter content under consideration.  Beyond its role in backreaction, the EMT is a fundamental local operator that encodes information about excitations even in situations where the notion of particles is ambiguous, such as in non-asymptotically static spacetimes or at strong coupling. This makes the EMT a particularly natural observable to study in curved-spacetime quantum field theory. At the same time, it is well known that renormalizing the energy-momentum tensor in curved backgrounds introduces significant conceptual and technical subtleties \cite{Birrell:1982ix}. A systematic treatment of these issues within our real-time lattice framework is another important direction for future work.

\section*{Acknowledgements}

We thank João Barata for collaboration in early stages of this work and many valuable discussions. We are grateful to Taku Izubuchi and Rob Pisarski for insightful comments and to Jorinde van de Vis for an interesting discussion.

This material is based upon work supported by the U.S. Department of Energy, Office of Science, Office of Nuclear Physics under contract number DE-SC0012704, and Laboratory Directed Research and Development funds (project No. 25-033B) from Brookhaven Science Associates. The work of E.B. was supported by the U.S. Department of Energy, Office of Science, Office of Workforce Development for Teachers and Scientists (WDTS) under the Science Undergraduate Laboratory Internships Program (SULI). A.F. is supported by the Deutsche Forschungsgemeinschaft (DFG, German Research Foundation) through the Emmy Noether Programme Project No. 545261797 and the  CRC-TR 211 "Strong-interaction matter under extreme conditions," Project No. 315477589-TRR 211. The authors gratefully acknowledge the computing time made available to them on the high-performance computer Noctua 2 at the NHR Center Paderborn Center for Parallel Computing (PC2). This center is jointly supported by the Federal Ministry of Research, Technology and Space and the state governments participating in the National High-Performance Computing (NHR) joint funding program (www.nhr-verein.de/en/our-partners).

\clearpage
\appendix 

\section{Hamiltonian Schwinger model in gravitational background} \label{app:Hamiltonian}

In Section \ref{sec:prelim_lattice} we argue, mostly from dimensional analysis, that the lattice Hamiltonian describing Schwinger model in FLRW gravitational background is given by Eq. (\ref{eq:coord_Ham}). The goal of this Appendix is to derive it, starting from the standard action of Schwinger model in arbitrary gravitational field in the continuum.

\subsection{Continuum formulation}

We begin with the action for the Schwinger model in a curved background described by a metric $g_{\mu\nu}$ \cite{Blaschke:2014ioa} 
\begin{equation}
    S = \int d^2 x\sqrt{-g} \left[\bar\psi(i \gamma^a \invweibein{\mu}{a} \overleftrightarrow\nabla_\mu - \mf)\psi - \frac{1}{4\coupling^2}F_{\mu\nu}F^{\mu\nu}\right], \label{eq:cont_action}
\end{equation}
where the fermion covariant derivative includes the gravitational and gauge connections: $\nabla_\mu \psi = (\partial_\mu + \omega_\mu +i  A_\mu)\psi$; $\overleftrightarrow\nabla \equiv \dfrac{1}{2}(\overrightarrow \nabla_\mu - \overleftarrow\nabla_\mu)$; the gauge field strength tensor is $F_{\mu\nu} \equiv D_\mu A_\nu - D_\nu A_\mu = \partial_\mu A_\nu - \partial_\nu A_\mu$. The spin connection $\omega$ is defined as
\begin{align}
    \omega_\mu &= \frac{1}{8}[\gamma^a, \gamma^b] \omega_{ab\mu} \,,  \\ \omega_{ab\mu} &= \invweibein{\nu}{[a}\eta_{b]c}\nabla_\mu \zweibein{c}{~\nu}\,,
\end{align}
where $\zweibein{a}{~\mu}$ is the local zweibein and $\invweibein{\mu}{a}$ is its inverse
\begin{align}
    \zweibein{a}{~\mu}\zweibein{b}{~\nu}\eta_{ab} = g_{\mu\nu} \,, \\
    \invweibein{\mu}{a} = \eta_{ab}g^{\mu\nu} \zweibein{b}{~\nu} \,,
\end{align}
and the (1+1)-dimensional Minkowski metric is $\eta_{ab} = {\rm diag}(1,-1)$. In (1+1) dimensions one can always choose conformally flat coordinates with the general form of the metric:
\begin{equation}
    g_{\mu\nu}(x,t) = \Omega^2(x,t)\eta_{\mu\nu}.
\end{equation}
Then the zweibein fields are
\begin{align}
    \zweibein{a}{~\mu} &= {\rm diag}(\Omega, \Omega)\,, \\
    \invweibein{\mu}{a} &= {\rm diag}(\Omega^{-1}, \Omega^{-1}) \,,
\end{align}
and the nonzero Christoffel symbols are
\begin{align}
    \Gamma^{0}_{00} &= \Gamma^1_{01} = \Gamma^0_{11} = \frac{\Omega_t}{\Omega} \,, \\
    \Gamma^{1}_{00} &= \Gamma^0_{01} = \Gamma^1_{11} = \frac{\Omega_x}{\Omega},
\end{align}
with the shorthand notation $\Omega_\mu\equiv \partial_\mu\Omega$. The spin connection is
\begin{align}
    \omega_{010} &= \frac{\Omega_x}{\Omega} \,, \\
    \omega_{011} &= \frac{\Omega_t}{\Omega}.
\end{align}
We find that in the fermionic part of the action the only effect of the conformally flat metric is multiplication by an overall factor $1/\Omega$ \cite{Lewis:2019oyx}:
\begin{equation}
  \bar\psi i \gamma^a \invweibein{\mu}{a}\overleftrightarrow\nabla\psi = \frac{1}{\Omega} \delta^\mu_a \,\bar\psi(i\gamma^a\overleftrightarrow\partial_\mu - \gamma^a A_\mu)\psi.
\end{equation}

Now we reformulate the theory in Hamiltonian terms. Later we will work in the temporal gauge, $A_0 = 0$, but we do not enforce it yet. The canonical momentum for the $A_1$ field is  $\Pi_E$:
\begin{equation}
  \Pi_E\equiv \frac{\partial \cal L}{\partial(\partial_0 A_1)} = \frac{\sqrt{-g}}{\coupling^2}(\partial^0A^1 - \partial^1 A^0) =  \frac{1}{\Omega^2\coupling^2}(\partial_0 A_1 - \partial_1 A_0),
\end{equation}
while the canonical momenta for the fermion fields are given by the left derivatives with respect to the corresponding time derivatives of the fields:
\begin{align}
    \pi_\psi &= \frac{\overrightarrow\partial^L}{\partial(\partial_0\psi)}{\cal L} =  -\frac{i\Omega}{2}\psi^\dagger \,, \nonumber \\
    \pi_{\psi^\dagger} &= \frac{\overrightarrow\partial^L}{\partial(\partial_0\psi^\dagger)}{\cal L} = -\frac{i\Omega}{2}\psi. \label{eq:psi_momentum}
\end{align}
The Hamiltonian density reads
\begin{align}
    {\cal H} &= \Pi_E \partial_0 A_1 + (\partial_0\psi)  \pi_\psi + (\partial_0\psi^\dagger) \pi_{\psi^\dagger} - {\cal L} \\
    &= \frac{1}{2} \coupling^2\Omega^2 \Pi_E^2 + \Pi_E \partial_1A_0 + \Omega^2 \mf \bar\psi\psi \nonumber + \Omega\,\bar\psi(-i\gamma^1\overleftrightarrow\partial_1 + \gamma^0A_0 + \gamma^1A_1)\psi.
\end{align}
Enforcing the gauge $A_0 = 0$, we find the curved spacetime version of the Gauss's law from the factor multiplying $A_0$ in the above Hamiltonian
\begin{equation}
    \partial_x \Pi_E - \Omega \bar\psi\gamma^0\psi = 0, \label{eq:cont_Gauss_app}
\end{equation}
and establish the final version of the Schwinger model Hamiltonian in the continuum:
\begin{align}
  H &= \int \, dx\, \bigg(\frac{1}{2} \coupling^2\Omega(x)^2 \Pi_E(x)^2 + \Omega(x)\bar\psi(x)[-i\gamma^1\overleftrightarrow\partial_1 + \gamma^1A_1(x)]\psi(x) + \Omega(x)^2 \mf \bar\psi(x)\psi(x) \bigg)  \label{eq:cont_Ham}\\
  &= \int \, dx\, \bigg(\frac{1}{2} \coupling^2\Omega(x)^2 \Pi_E(x)^2 + \bar{\tilde\psi}(x)[-i\gamma^1\partial_1 + \gamma^1A_1(x)]\tilde\psi(x) + \Omega(x) \mf \bar{\tilde\psi}(x)\tilde\psi(x) \bigg) 
\end{align}
where in the second line we introduced a rescaled fermionic field $\tilde\psi(x) = \sqrt{\Omega(x)}\psi(x)$ and then integrated by parts the fermionic gradient term. This form of the Hamiltonian makes it readily apparent that the varying metric manifests as a rescaling of $\coupling$ and $\mf$ by $\Omega(x)$.

\subsection{Canonical anticommutators: an exercise in constrained Hamiltonian systems}

Before we move on to placing the system described by Eq. (\ref{eq:cont_Ham}) on the lattice, some care is required in dealing with the canonical structure of the theory, as it is a constrained system. Indeed, naively imposing the canonical anticommutator between $\psi$ and $\pi_\psi$ and using the definition of the canonical fermionic momenta in Eq. (\ref{eq:psi_momentum}), one would find the anticommutator between $\psi, \psi^\dagger$ with an unusual factor of 1/2. The correct procedure is to perform the Dirac-Bergmann \cite{Dirac:1950pj, Anderson:1951ta, Dirac:1958sq} analysis of constraints in the system, leading to the standard result with no extra $1/2$ factor. 
As this point is not commonly discussed in the literature, for completeness, we demonstrate this procedure using a simple example.

We begin with a classical theory of a single complex Grassmann variable $\eta$, formulated with the simplest model Lagrangian, capturing the Hamiltonian structure of Eq. (\ref{eq:cont_action}):
\begin{equation}
    \mathcal L = \frac{i}{2}(\eta^\dagger\partial_t\eta - \eta^\dagger\overleftarrow{\partial_t}\eta)
\end{equation}
In all conventions we will follow \cite{Henneaux:1994lbw}. Canonical momenta $\pi_\eta$, $\pi_{\eta^\dagger}$ are defined as the left derivatives of the Lagrangian with respect to $\partial_t\eta, \partial_t\eta^\dagger$, respectively, yielding two primary constraints:
\begin{align}
    \phi_1 &= \pi_\eta + \frac{i}{2}\eta^\dagger \approx 0 \,, \\
    \phi_2 &= \pi_{\eta^\dagger} + \frac{i}{2}\eta \approx 0 \,.
\end{align}
No secondary constraints arise in this simple theory. Using the canonical Poisson bracket (PB) between two Grassmann-odd functions, $\{\eta, \pi_\eta\}_{\rm PB} = -1$, we find that the constraints $\phi_1, \phi_2$ are second-class, namely their Poisson bracket does not vanish even weakly:
\begin{equation}
    \{\phi_1, \phi_2\}_{\rm PB} =  \{\phi_2, \phi_1\}_{\rm PB} = -i
\end{equation}
In other words, the matrix of Poisson brackets of the second-class constraints is 
\begin{equation}
    M_{ab} = \{\phi_a,\phi_b\}_{\rm PB} = \begin{pmatrix}
        0 & -i \\
        -i & 0
    \end{pmatrix}
\end{equation}
The Dirac bracket (DB) is given by
\begin{equation}
    \{\eta, \eta^\dagger\}_{\rm DB} = \{\eta, \eta^\dagger\}_{\rm PB} - \{\eta, \phi_a\}_{\rm PB} M_{ab}^{-1}\{\phi_2, \eta^\dagger\}_{\rm PB} = -i.
\end{equation}
Upon quantizing the theory, the anticommutator is obtained by multiplying the Dirac bracket by $i\hbar$, leading to $\{\hat\eta, \hat\eta^\dagger\} = 1$, the canonical anticommutator of fermionic operators. For the theory described  by canonical momenta in Eq. (\ref{eq:psi_momentum}), the only difference is in the factor of $\Omega$, which is not a function of $\psi$ or $\pi_\psi$, so it does not affect the canonical structure other than by an overall factor.

\subsection{Lattice formulation}

To place the system on the lattice, we employ the Kogut-Susskind staggered fermion formulation  \cite{Susskind:1976jm} to avoid the doubling problem. In accordance with the discussion of the Hamiltonian formalism for constrained systems in the previous subsection, the anticommutation relation in the continuum reads
\begin{equation}
    \{\psi^i(x),\psi_j^\dagger(y)\} = \frac{1}{\Omega(y)}\delta(x-y)\delta_j^i =  \frac{1}{\sqrt{\Omega(x)\Omega(y)}}\delta(x-y)\delta_j^i\, ,
\end{equation}
where the second representation is equivalent to the first one because of the delta-function. It is satisfied for the following discretization \cite{Lewis:2019oyx}:
\begin{align}
    \psi^1(x_n) &= \frac{1}{\sqrt{a\Omega(t,x_n)}}\chi_n~,~ \text{even } n \nonumber \,, \nonumber \\
    \psi^2(x_n) &= \frac{1}{\sqrt{a\Omega(t,x_n)}}\chi_n ~,~ \text{odd } n, \label{eq:staggering} 
\end{align}
where the spinor components are defined by $\psi = \begin{pmatrix}
    \psi^1 \\ \psi^2
\end{pmatrix}$, in the basis where the gamma-matrices are $\gamma^0=\sigma_z, \gamma^1 = i\sigma_y$, with the canonical anticommutation relations for $\chi$: $$\{\chi^\dagger_n, \chi_m\} = \delta_{nm}.$$
Apart from the factors of $\Omega$ in the Hamiltonian (\ref{eq:cont_Ham}) and in the discretization in Eq. (\ref{eq:staggering}), the Hamiltonian is that of the standard Schwinger model, and the mapping to the lattice follows a standard route \cite{Casher:1974vf,Susskind:1976jm} (for a recent review, see e.g. \cite{Davoudi:2025kxb}) and one only needs to keep track of the powers of $\Omega$ at every step. Furthermore, in the lattice Schwinger model with open boundary conditions, the gauge field does not have dynamical degrees of freedom and it can be integrated out using Gauss's law, while the gauge link degrees of freedom can be gauged away.
The resulting Hamiltonian is
\begin{align}
    H = \frac{a\coupling^2}{2}\sum_{n=1}^{N-1}\Omega_n^2(t) E_n^2 - \frac{i}{2a} \sum_{n=1}^{N-1} (\chi^\dagger_n\chi_{n+1} - \chi^\dagger_{n+1}\chi_n) + \mf \sum_{n=1}^N(-1)^n\Omega_n(t)\chi_n^\dagger\chi_n, \label{eq:lat_Ham}
\end{align}
where $\Omega_n(t) = \Omega(x_n, t)$. In the above Hamiltonian, the discretized electric field is defined  as $E_n = \Pi_E(x_n)$, and from Eqs. (\ref{eq:cont_Gauss_app}) and (\ref{eq:staggering}) it follows that Gauss' law looks exactly like in Minkowski spacetime:
\begin{equation}
    E_{n}-E_{n-1} = Q_n \equiv \chi^\dagger_n\chi_n - \frac{1-(-1)^n}{2}\,, \label{eq:lat_Gauss_app}
\end{equation}
allowing to represent $E_n = \sum_{i=1}^n Q_n$ for a system with open boundary conditions. For a spatially uniform metric, $\Omega_n(t) = \Omega(t)$, the Hamiltonian (\ref{eq:lat_Ham}) together with the Gauss's law (\ref{eq:lat_Gauss_app}) reduces to Eq. (\ref{eq:coord_Ham}), used in the main text.

In the main text we also use the expression for the electric charge density operator that is derived below. The continuum expression for the vector current in gravitational background is
\begin{equation}
    j^\mu = \bar \psi \invweibein{\mu}{a} \gamma^a\psi = \frac{1}{\Omega(t)}\delta^\mu_a\bar\psi\gamma^a\psi \,.
\end{equation}
When discretizing this expression, the prescription (\ref{eq:staggering}) introduces an additional factor of $1/\Omega$, and we find for the temporal component of the current
\begin{equation}
    j^0_n = \frac{1}{a\,\Omega^2(t)}\bigg(\chi_n^\dagger\chi_n - \frac{1-(-1)^n}{2} \bigg)\,.
\end{equation}
To find the charge density (that computes the total charge when integrated over $x^1$) we introduce a factor of $\sqrt{-g} = \Omega^2(t)$:
\begin{equation}
    q_n \equiv \sqrt{-g}j^0_n = \frac{1}{a}\bigg(\chi_n^\dagger\chi_n - \frac{1-(-1)^n}{2} \bigg)\,,
\end{equation}
which is the same expression as the one without a gravitational background.

\subsection{A different coordinate choice}

We are interested in the Schwinger model placed in the background of a spatially uniform expanding ``Universe", described by the metric
\begin{equation}
    g_{\mu\nu} = \Omega^2(t)\eta_{\mu\nu} \,, \label{eq:metric}
\end{equation}
which is the Friedmann–Lemaitre–Robertson–Walker (FLRW) Universe, in the conformal time coordinate. In terms of physical time $\tau$, given by the relation $d\tau = \Omega(t) dt$, the metric takes the more familiar FLRW form:
\begin{equation}
    ds^2 = d\tau^2 - \tilde\Omega(\tau)^2dx^2 \,,
\end{equation}
with $\tilde\Omega(\tau) = \Omega[t(\tau)]$, where $t(\tau)$ is the function inverse to $\tau(t) = \int_0^t dt'\Omega(t')$.

Note that one could work in a different coordinate system: the physical frame, in which the metric is the Minkowski one. In order to describe the expanding system, one would then introduce a time-dependent lattice discretization:
\begin{equation}
    a'(\tau) = a\,\tilde\Omega(\tau)
\end{equation}
where we use $\tau$ as the physical time variable. With this replacement, the Hamiltonian 
Eq.~(\ref{eq:coord_Ham})  becomes:
\begin{align}
    H'(\tau) = \frac{a \,\tilde\Omega(\tau) \coupling^2}{2}H_E & - \frac{1}{2a \,  \tilde\Omega(\tau)}H_k + \mf H_m\,. \label{eq:phys_Ham}
\end{align}
Note that
\begin{equation}
    H'(\tau) = H[t(\tau)]/\tilde\Omega(\tau) \label{eq:Ham_relation}
\end{equation}
There are two different Hamiltonians describing the same system because they correspond to different time (as well as  space) coordinates. Namely, Eq. (\ref{eq:coord_Ham}) describes evolution in the conformal time $t$. On the other hand, Eq. (\ref{eq:phys_Ham}) leads to the evolution of the system in the physical time, with $d\tau = \Omega(t) dt$. Importantly, Eq.~(\ref{eq:Ham_relation}) ensures that the time evolution operator is identical in both coordinate systems:
\begin{equation}
    {\cal T}e^{\int i H(t) dt} = {\cal T}e^{\int i H'(\tau) d\tau} \,.
\end{equation}

\section{Free massive scalar in the continuum} \label{app:free:analytic}

The case of a free massive scalar field in an expanding Universe is a textbook example of cosmological particle production \cite{Bernard:1977pq,Birrell:1982ix}; see also~\cite{Kolb:2023ydq} for a recent review. For a self-contained discussion, here we briefly review it and highlight the points where a comparison is made to the results reported in the main text.

Consider a free massive scalar field, minimally coupled to the gravitational background described by the metric $g_{\mu\nu}$:
\begin{equation}
    S = \int d^2x\sqrt{-g}\left(\frac{1}{2}g^{\mu\nu}\partial_\mu \phi\partial_\nu\phi  - \frac{1}{2}m_s^2 \phi^2\right)\,,
\end{equation}
where the notation for the mass, $m_s$, is chosen to distinguish the mass of the free scalar field from the other mass parameters mentioned elsewhere. The quantum field operator $\hat\phi$ can be expanded in different bases of creation-annihilation operators. If the metric is a function of time only, $g_{\mu\nu}(t)$, momentum is a good quantum number and the field can be expanded in the modes with well-defined momenta. Yet, there is still freedom in choosing the modes, namely the mass in the dispersion relation. In free QFT in Minkowski space, this freedom is eliminated by stating that the Hamiltonian should be diagonal in the particle number basis, leading to the usual notion of the Fock space of non-interacting particles such that the energy in a state with several particles is the sum of  their individual energies. 

Let us briefly remind the argument. For now we will be in Minkowski spacetime. The Hamiltonian operator of a free massive scalar is given by
\begin{equation}
    \hat H = \frac{1}{2}\int dx \left[\hat\pi(x)^2 + (\partial_x\hat\phi(x))^2 + m_s^2\hat\phi(x)^2\right]
\end{equation}
The mode decomposition in terms of creation-annihilation operators is fixed by matching the commutators $[\hat\phi(x),\hat\pi(x')] = i\delta(x-x')$ and $[a_k, a_{k'}^\dagger] = 2\pi\delta(k-k')$ and takes form
\begin{align}
    \hat\phi(x) &= \int \frac{dk}{2\pi \sqrt{2\omega_k}}(a_ke^{ikx} + a_k^\dagger e^{-ikx}) \,, \\
    \hat\pi(x) &= \int \frac{dk}{2\pi} \sqrt{\frac{\omega_k}{2}}(-i)(a_ke^{ikx} - a_k^\dagger e^{-ikx}) \, ,
\end{align}
where $\omega_k$ is \textit{a priori} an arbitrary function of momentum (more precisely, the absolute value of momentum if we use rotational invariance). Evaluating the Hamiltonian in terms of creation and annihilation operators, we obtain
\begin{align}
    \hat H = \int \frac{dk}{8\pi\,\omega_k}[&(a_k^\dagger a_k +a_k a_k^\dagger)(\omega_k^2+k^2+m_s^2)
    +(a_k a_{-k} +a_k^\dagger a_{-k}^\dagger)(\omega_k^2 - k^2 - m_s^2)] \,.
\end{align}
The second term is non-diagonal in the particle number basis, and to cancel it we choose the usual relativistic dispersion relation, $\omega_k^2 = k^2 + m_s^2$. The remaining Hamiltonian then takes the familiar form $\hat H = \int\frac{dk}{2\pi}(a_k^\dagger a_k +\frac{1}{2})\omega_k$.

Now we return to the case of the expanding Universe, focusing on the FLRW case in the conformal coordinates, $g_{\mu\nu} = \Omega^2(t)\eta_{\mu\nu}$, where the Lagrangian density reads 
\begin{equation}
    {\cal L} = \frac{1}{2}\partial_\mu\phi\partial_\nu\phi \eta^{\mu\nu} - \frac{1}{2}\Omega(t)^2m_s^2\phi^2,
\end{equation}
so the mass parameter, determining the dispersion relation, becomes time-dependent. Therefore, at each moment in time there is an instantaneous particle basis diagonalizing the instantaneous Hamiltonian, defined by the dispersion relation $\omega_k(t)^2  = k^2 + \Omega(t)^2m_s^2$. 

While the field operator can be decomposed in any of those bases, of particular importance are the ones with a straightforward particle interpretation. Particles are defined as asymptotic states in QFT, so it only makes sense to discuss particles if $g_{\mu\nu}\rightarrow\rm{const}$ as $t\rightarrow\pm\infty$. Then the field operator can be decomposed in the ``in" and ``out" bases, namely
\begin{align}
    \hat\phi(x,t) = \int\frac{dk}{2\pi}(a_k^{\rm in} u_k^{\rm in}(x,t) + {a_k^{\rm in}}^\dagger u_k^{\rm in }(x,t)^*)= \int\frac{dk}{2\pi}(a_k^{\rm out} u_k^{\rm out}(x,t) + {a_k^{\rm out}}^\dagger u_k^{\rm out}(x,t)^*) \,, \label{eq:field_op}
\end{align}
where the mode functions $u_k^{\rm in}(x,t), u_k^{\rm out}(x,t)$ in addition to the standard $x$-dependence($\sim e^{ikx}$) contain the time dependence, determined by the equation of motion for $\hat \phi$ in Heisenberg representation. For the mode functions it translates into the following Klein-Gordon equation with time-dependent mass:
\begin{equation}
    \partial_t^2 u_k(t,x) + (k^2 + \Omega(t)^2m_s^2)u_k(x,t) = 0\,. \label{eq:time-dep-KG}
\end{equation}
Here we will focus on the metric profile discussed in the main text that has the virtue of admitting analytical solutions to the above equation in terms of hypergeometric functions \cite{Bernard:1977pq,Birrell:1982ix}. Namely, we consider
\begin{equation}
    \Omega(t)^2 = A + B\, \tanh(\rho\, t)\,.
\end{equation}
Note that the metric goes to Minkowski metric asymptotically, so it makes sense to define particle states in the asymptotic past and the asymptotic future. The solutions to the mode equation (\ref{eq:time-dep-KG}) that approach the free particle (plane wave) solutions at $t\rightarrow\pm\infty$ read, respectively
\begin{align}
    u_k^{\rm in}(t,x) = \sqrt{\frac{1}{4\pi\omega_{\rm in}}}\exp\left[ikx-i\omega_+ t - \frac{i\omega_-}{\rho}\log(2\cosh\rho t)\right] 
    {}_2F_1\left(1+\frac{i\omega_-}{\rho},\frac{i\omega_-}{\rho};1-\frac{i\omega_{\rm in}}{\rho};\frac{1+\tanh\rho t}{2}\right) \, \label{eq:u_k_in},\\
     u_k^{\rm out}(t,x) = \sqrt{\frac{1}{4\pi\omega_{\rm out}}}\exp\left[ikx-i\omega_+ t - \frac{i\omega_-}{\rho}\log(2\cosh\rho t)\right] {}_2F_1\left(1+\frac{i\omega_-}{\rho},\frac{i\omega_-}{\rho};1+\frac{i\omega_{\rm out}}{\rho};\frac{1-\tanh\rho t}{2}\right)\,,
\end{align}
where ($k$-dependence in $\omega$ is implied, but suppressed in the notation for brevity)
\begin{align}
    \omega_{\rm in}^2 &= k^2 + (A-B)\,m_s^2 \,,  \label{eq:omega_in} \\ 
    \omega_{\rm out}^2 &= k^2 + (A+B)\,m_s^2 \,, \label{eq:omega_out} \\ 
    \omega_\pm &= \frac{1}{2}(\omega_{\rm out} \pm \omega_{\rm in}) \,. \label{eq:omega_pm}
\end{align}
The mode functions can be expressed via each other as follows:
\begin{equation}
    u_k^{\rm in}(t,x) = \alpha_k u_k^{\rm out}(t,x) + \beta_ku_{-k}^{\rm out}(t,x)^* \,,
\end{equation}
with the coefficients 
\begin{align}
    \alpha_k = \sqrt{\frac{\omega_{\rm out}}{\omega_{\rm in}}}\frac{\Gamma(1-i\omega_{\rm in}/\rho)\Gamma(-i\omega_{\rm out}/\rho)}{\Gamma(-i\omega_+/\rho) \Gamma(1-i\omega_+/\rho)} \, , \label{eq:alpha_k} \\ 
    \beta_k = \sqrt{\frac{\omega_{\rm out}}{\omega_{\rm in}}}\frac{\Gamma(1-i\omega_{\rm in}/\rho)\Gamma(i\omega_{\rm out}/\rho)}{\Gamma(i\omega_-/\rho) \Gamma(1+i\omega_-/\rho)}\,. \label{eq:beta_k}
\end{align}
This implies the following Bogoliubov transformation between the creation and annihilation operators in the incoming and outgoing bases:
\begin{align}
    a_k^{\rm out} = \alpha_k a_k^{\rm in} + \beta_k^*{a_{-k}^{\rm in}}^\dagger \, \\
    {a_k^{\rm out}}^\dagger = \alpha_k^*{a_k^{\rm in}}^\dagger + \beta_k a_{-k}^{\rm in} \,. \label{eq:Bogol_op}
\end{align}
The commutation relations between $a^{\rm out}$ and ${a^{\rm out}}^\dagger$ are canonical due to the property
\begin{equation}
    |\alpha_k|^2 - |\beta_k|^2 = 1 \,, \label{eq:alpha_beta_relation}
\end{equation}
obeyed by (\ref{eq:alpha_k}) and (\ref{eq:beta_k}).

Note that the field operator in Eq. (\ref{eq:field_op}) is explicitly time-dependent as we work in the Heisenberg representation that is the conventional choice in the canonical quantization of free QFT. Therefore, the quantum state is time-independent in this formalism. If we start with the vacuum state in the asymptotic past, $|0^{\rm in}\rangle$, we remain in the same state in the asymptotic future when the notion of particles makes sense again. Note, however, that the notion of particles is different now and the state is not a vacuum state with respect to that notion. From (\ref{eq:Bogol_op}), this state in the \textit{out} basis is given by \cite{Bernard:1977pq} 
\begin{equation}
    |0^{\rm in}\rangle = \langle 0^{\rm out}|0^{\rm in}\rangle \prod_k \exp\left[\frac{1}{2} \frac{\beta^*(k)}{\alpha^*(k)}{a_k^{\rm out}}^\dagger {a_{-k}^{\rm out}}^\dagger\right]  |0^{\rm out}\rangle\,. \label{eq:squeezed_state}
\end{equation}
Owing to momentum conservation, particles are always produced in pairs with exactly opposite momenta. The prefactor, expressed as the amplitude to stay in vacuum, accounts for normalization. 

From the Bogoliubov transformation (\ref{eq:Bogol_op}) it is straightforward to find the occupation number of the outgoing quanta at late times:
\begin{align}
    \langle N_k^{\rm out}\rangle &= \langle 0_{\rm in}|a_k^{\rm out}{^\dagger a_k^{\rm out}}|0_{\rm in}\rangle = |\beta_k|^2 = \frac{\sinh^2(\pi\omega_-/\rho)}{\sinh(\pi\omega_{\rm in}/\rho)\sinh(\pi\omega_{\rm out}/\rho)} \,. \label{eq:asymptotic_particle_number}
\end{align}
If the particle production rate is small, $N_k$ is mostly saturated with single-pair production and serves as a very good approximation of the probability of producing exactly one pair. More precisely, from Eq. (\ref{eq:squeezed_state}) we obtain
\begin{equation}
    p(k, t\rightarrow\infty) = |\langle 1_k^{\rm out}1_{-k}^{\rm out}|0^{\rm in}\rangle|^2 = p_0\frac{|\beta_k|^2}{|\alpha_k|^2}
    \,, \label{eq:one_pair_prob}
\end{equation}
where the survival probability is $p_0=|\langle 0^{\rm out}|0^{\rm in}\rangle|^2$ .

\section{Analytical calculation of two-point correlators}\label{app:free_2pt}

Here we provide some details on the calculation of the correlators ${\cal C}_2^{\rm cont.}(x,t)$ and ${\cal Q}_2^{\rm cont.}(x,t)$ in a free massive scalar theory based on the exact solution of the field given by the mode expansion of App. \ref{app:free:analytic}. In the Heisenberg representation with the time-independent state fixed as $|0_{\rm in}\rangle$, the ``in" basis represents the most convenient choice for the mode expansion of the field operator:
\begin{equation}
    \hat\phi(x,t) = \int\frac{dk}{2\pi}(a_k^{\rm in} u_k^{\rm in}(x,t) + {a_k^{\rm in}}^\dagger u_k^{\rm in }(x,t)^*) \,,
\end{equation}
with the expression for $u_k^{\rm in}(t,x)$ given in Eq. (\ref{eq:u_k_in}). We find
\begin{align}
    &{\cal C}_2^{\rm cont.}(x,t) = \langle \hat \phi(0,t) \hat\phi(x,t)\rangle = \int_{-\infty}^\infty dk \,u_{k}^{\rm in}(0,t) u_{k}^{\rm in}(x,t)^* \nonumber \\
    &=\int_{-\infty}^\infty \frac{dk}{4\pi\omega_k^{\rm in}}e^{-ikx}\bigg|{}_2F_1\left[1+\frac{i\omega_k^-}{\rho},\frac{i\omega_k^-}{\rho};1-\frac{i\omega_k^{\rm in}}{\rho}; \frac{1+\tanh\rho t}{2}\right]\bigg|^2\,. \label{eq:int_2pt_c} \\
    &{\cal Q}_2^{\rm cont.}(x,t) = \frac{1}{\pi}\langle\partial_x \hat\phi(0,t) \partial_x \hat\phi(x,t)\rangle = \frac{1}{\pi} \int_{-\infty}^\infty dk \partial_x u_{k}^{\rm in}(0,t) \partial_x u_{k}^{\rm in}(x,t)^* \nonumber \\
    &=\int_{-\infty}^\infty \frac{dk}{4\pi^2\omega_k^{\rm in}}k^2e^{-ikx}\bigg|{}_2F_1\left[1+\frac{i\omega_k^-}{\rho},\frac{i\omega_k^-}{\rho};1-\frac{i\omega_k^{\rm in}}{\rho}; \frac{1+\tanh\rho t}{2}\right]\bigg|^2\,. \label{eq:int_2pt_q}
\end{align}
At large momentum, $k\rightarrow\infty$, $\omega_k^{\rm in}\sim\omega_k^{\rm out}\propto k$, $\omega_k^-\propto 1/k$ and $|_2F_1[...]|^2\sim 1+O(1/k^2)$. The integral in Eq.~(\ref{eq:int_2pt_c}) is convergent and is straightforwardly evaluated numerically. The integral in Eq. (\ref{eq:int_2pt_q}) has a divergent part, which we evaluate separately with an $\epsilon$-regularization:
\begin{align}
    \lim_{\epsilon\rightarrow0}\int_{-\infty}^\infty \frac{dk}{\sqrt{k^2 + m_{\rm in}^2}}k^2 e^{-ikx-\epsilon k} =
    -\frac{\partial^2}{\partial x^2}2\lim_{\epsilon\rightarrow0}\int_0^\infty \frac{dk}{\sqrt{k^2 + m_{\rm in}^2}}\cos kx = -2m_{\rm in}^2 K_0''(m_s^{\rm in} \,x) \,,
\end{align}
where $m_{\rm in} = m_s\sqrt{A-B}$ and $K_0$ is the modified Bessel function of second kind. So, for the full correlator we find
\begin{align}
    &{\cal Q}_2^{\rm cont.}(x,t) = -\frac{m_s^2}{2\pi^2}(A-B)K_0''(m_s\sqrt{A-B}\,x) \nonumber \\ &+ \int_0^\infty\frac{dk}{2\pi^2\omega_k^{\rm in}} k^2\left[\bigg|{}_2F_1\left(1+\frac{i\omega_k^-}{\rho},\frac{i\omega_k^-}{\rho};1-\frac{i\omega_k^{\rm in}}{\rho}; \frac{1+\tanh\rho t}{2}\right)\bigg|^2 - 1\right] \cos kx
\end{align}
where the integral in the second line is convergent and is readily evaluated numerically. The result of this calculation is used in Figs. \ref{fig:2pt_vs_x} and \ref{fig:2pt_vs_t} in the main text to display the analytical results.

\section{Equivalence of Schr\"odinger and Heisenberg time evolution} \label{app:Schroedinger_Heisenberg}

In quantum field theory, we usually represent time evolution in Heisenberg picture, while in the Hamiltonian lattice simulation the more natural representation is the Schr\"odinger one. The equivalence between these two representations becomes nontrivial in quantum field theory in curved spacetime, even in the case of asymptotically static metric \cite{Agullo:2015qqa}. Without addressing the UV issues that spoil the conventional correspondence, in this Appendix we will demonstrate the equivalence in a lattice setting, which corresponds to a finite number of time-dependent harmonic oscillators. The motivation for this consideration is that even in that setting, it seems not entirely obvious that Eq.~(\ref{eq:asymptotic_particle_number}) for the expectation value of the outgoing number of quanta,
\begin{equation}
    \langle N_{\rm out}\rangle_{t\rightarrow\infty} = \langle 0_{\rm in}|\hat N_{\rm out}|0_{\rm in}\rangle \,, \label{eq:initially_confusing}
\end{equation}
is truly formulated in the Heisenberg picture. Naively, one can ask: where are the time-evolution operators in $\hat N_{\rm out}$, given that it is defined (see Eq. (\ref{eq:asymptotic_particle_number})) at late times in a time-independent manner? Is this consistent with the state $|0_{\rm in}\rangle$ being defined at a different time, $t\rightarrow-\infty$?

The purpose of this Appendix is to establish the equivalence between Schr\"odinger and Heisenberg representations and to show that Eq.~(\ref{eq:initially_confusing}) is formulated correctly. We further derive a generalization of Eq.~(\ref{eq:initially_confusing}) valid for arbitrary times and general quanta. We begin with a simplified setup containing all essential elements, namely a one-dimensional quantum harmonic oscillator with a time-dependent frequency. We assume asymptotic behavior $\omega(t\rightarrow-\infty) = \omega_{\rm in}$, $\omega(t\rightarrow\infty) = \omega_{\rm out}$, similarly to the asymptotically static spacetime expansion. This setup allows us to sidestep questions about which quanta are physically observable: in quantum mechanics, one can in principle define quanta of a harmonic oscillator with any frequency $\omega$ and imagine idealized measurement devices for all such modes. We briefly comment on the relevance of this construction in the cosmological and Hamiltonian simulation contexts  at the end of this Appendix.

Let us begin with Schr\"odinger (S) representation. The Hamiltonian is
\begin{equation}
    \hat H_S = \frac{\hat p_S^2}{2m} + \frac{m\,\omega(t)^2 \hat x_S^2}{2} \,,
\end{equation}
where $\hat x_S, \hat p_S$ are the canonical position and momentum operators in Schr\"odinger representation. Defining an initial state $|\psi_0\rangle_S$ at some fixed time $t_0$, we obtain the unitary evolution operator to find the state at later times
\begin{equation}
    |\psi(t)\rangle_S = U(t,t_0)|\psi_0\rangle_S = {\cal T}e^{-i\int_{t_0}^t dt' H_S(t')}|\psi_0\rangle_S.
\end{equation}
The expectation value of any Hermitian operator $\cal {\hat O}_S$ at time $t$ is given by 
\begin{equation}
    {\cal O}(t) = \langle \psi(t)|_S{\cal \hat O}_S|\psi(t)\rangle_S \,.
\end{equation}
Alternatively, one may work in the Heisenberg (H) representation, where the state is fixed as $|\psi\rangle_H = |\psi_0\rangle_S$, and all time dependence is encoded in the operators
\begin{equation}
    \hat{\cal O}_H(t) = U(t,t_0)^\dagger \hat{\cal O}_S U(t,t_0) \,,
\end{equation}
with the same reference time $t_0$. In Heisenberg representation the operators satisfy the Heisenberg equation of motion, 
\begin{equation}
    \frac{d \hat{\cal O}_H}{dt} = i[\hat H_H, {\cal\hat O}_H] \,,
\end{equation}
where we assume that the operator has no explicit time dependence. For the position and momentum operators of the harmonic oscillator it reduces to
\begin{align}
    \frac{d \hat x_H}{dt} &= \frac{p_H(t)}{m} \,, \label{eq:x_H_EOM} \\ 
    \frac{d \hat p_H}{dt} &= -m\omega(t)^2 x_H(t) \,. \label{eq:p_H_EOM}
\end{align}
We expand the Heisenberg-picture operators in the basis of the Schr\"odinger-picture operators $\hat x_S$ and $\hat p_S$ as 
\begin{align}
    \hat x_H(t) &= f_{xx}\hat x_S + f_{xp}\hat p_S \,,\\
    \hat p_H(t) &= f_{px}\hat x_S + f_{pp}\hat p_S\,.
\end{align}
The equations of motion (\ref{eq:x_H_EOM}), (\ref{eq:p_H_EOM}) imply relations among the expansion coefficients which ultimately reduce to
\begin{equation}
    f_{xi}''+\omega(t)^2 f_{xi} = 0\,, \label{eq:TDHO}
\end{equation}
where $ i\in\{x,p\}$ and a prime denotes differentiation with respect to time. While the general theory of the classical harmonic oscillator equation with time-dependent frequency, including its invariants, has been developed, see e.g.~\cite{Coelho:2024zuf} and references therein, we limit ourselves to an analytically solvable case, motivated by our cosmological setup,
\begin{equation}
    \omega(t)^2 = A+ B\tanh(\rho t) \,.
\end{equation}
This profile describes a smooth interpolation between the initial frequency $\omega_{\rm in} = \sqrt{A-B}$ and the final frequency $\omega_{\rm out} = \sqrt{A+B}$. We choose the initialization time $t_0^{\rm in}$, in the asymptotic past region, and define $\hat x_H(t_0^{\rm in}) = \hat x_S$, $\hat p_H(t_0^{\rm in}) = \hat p_S$. The particular choice of $t_0^{\rm in}$ affects only the overall phase, since the Hamiltonian is asymptotically time independent. To eliminate this phase, we choose $t_0^{\rm in} = -\frac{2\pi N}{\omega_{\rm in}}$ with integer $N$. The Heisenberg-picture position and momentum operators take the form
\begin{align}
    \hat x_H(t) &= {\rm Re} f_{\rm in}(t)\,\hat x_S -  \frac{{\rm Im}f_{\rm in}(t)}{m \,\omega_{\rm in}} \hat p_S \,, \\
    \hat p_H(t) &= m\,{\rm Re} f_{\rm in}'(t)\, \hat x_S - \frac{{\rm Im}f_{\rm in}'(t)}{ \omega_{\rm in}} \hat p_S \,.
\end{align}
Here $f_{\rm in}$ denotes the solution to Eq.~(\ref{eq:TDHO}) with asymptotic behavior $f_{\rm in}(t)\xrightarrow{t\rightarrow-\infty} e^{-i\omega_{\rm in}t}$, known in closed form:
\begin{align}
    f_{\rm in}(t) = \exp\left[-i\omega_+t-\frac{i\omega_-}{\rho}\log(2\cosh\rho t)\right]
    {}_2F_1\left(1+\frac{i\omega_-}{\rho}, \frac{i\omega_-}{\rho};1-\frac{i\omega_{\rm in}}{\rho};\frac{1+\tanh\rho t}{2}\right)\,,
\end{align}
where ${}_2F_1$ denotes the Gaussian hypergeometric function, and we have defined
\begin{align}
    \omega_{\pm} &= \frac{1}{2}(\omega_{\rm out} \pm\omega_{\rm in}) \,.
\end{align}
Now we can compute the time evolution of the creation and annihilation operators in arbitrary basis. The Schr\"odinger-picture operator describing annihilation of quanta with arbitrary frequency $\omega$ is related to the position and momentum operators as
\begin{equation}
    \hat a_{\omega}^S = \frac{1}{\sqrt{2}} \left(\sqrt{m\,\omega}\,\hat x_S + \frac{i}{\sqrt{m\,\omega}}\,\hat p_S\right) \,.
\end{equation}
In the Heisenberg representation, this operator reads
\begin{align}
    \hat a_\omega^H(t) &= \frac{1}{\sqrt{2}} \left(\sqrt{m\,\omega}\,\hat x_H(t) + \frac{i}{\sqrt{m\,\omega}}\,\hat p_H(t)\right) \nonumber \\
    &= \sqrt{\frac{m\,\omega}{2}}\left({\rm Re} f_{\rm in}(t) + \frac{i \,{\rm Re}f_{\rm in}'(t)}{\omega}\right) \hat x_S 
    - \sqrt{\frac{\omega}{2\,m\,\omega_{\rm in}^2}}\left({\rm Im} f_{\rm in}(t) + \frac{i\, {\rm Im}f_{\rm in}'(t)}{\omega}\right) \hat p_S \,.
\end{align}
In principle, a harmonic oscillator can be prepared in an arbitrary initial state, for instance with occupation number $n$ in a basis specified by $\omega$. To connect directly with the cosmological context, we specialize to a particular initial state: the vacuum with respect to the initial quanta $|0_{\rm in}\rangle$. It is defined as the state annihilated by $\hat a^S_{\omega_{\rm in}}$. To find the occupation number in arbitrary basis at later times, we will need the following expectation values of operators quadratic in position and momentum
\begin{align}
    &\langle 0_{\rm in}|\hat x_S^2|0_{\rm in}\rangle = \frac{1}{2\,m\,\omega_{\rm in}} \,,\\
    &\langle 0_{\rm in}|\hat p_S^2|0_{\rm in}\rangle = \frac{m\,\omega_{\rm in}}{2} \,,\\
     &\langle 0_{\rm in}|\hat x_S \hat p_S|0_{\rm in}\rangle = \frac{i}{2}
\end{align}
Constructing a number operator for quanta with frequency $\omega$, $\hat N_{\omega}^H \equiv \hat a_\omega^H{}^\dagger \hat a_\omega^H$, we find the following time dependence of its expectation value in the Heisenberg representation
\begin{equation}
    \langle 0_{\rm in}|\hat N_\omega^H(t)|0_{\rm in}\rangle = \frac{\omega}{4\omega_{\rm in}}\bigg|f_{\rm in}(t) + \frac{f_{\rm in}'(t)}{i\omega}\bigg|^2 \,. \label{eq:<N^H_omega>}
\end{equation}
We can now verify that this formalism reproduces Eq.~(\ref{eq:initially_confusing}) by computing the occupation number in the outgoing basis at asymptotically late times, $\langle\hat N_{\rm out}(t\rightarrow\infty)\rangle$. At late times, the ``in" solution has the generic asymptotic form
\begin{equation}
    f_{\rm in}(t\rightarrow\infty) \rightarrow \alpha e^{-i\omega_{\rm out}t} + \beta e^{i\omega_{\rm out}t} \,. \label{eq:in_asymptotics}
\end{equation}
To determine the Bogoliubov coefficients $\alpha$ and $\beta$, we first construct the outgoing modes of the oscillator
\begin{align}
    &f_{\rm out}(t) = \exp\left[-i\omega_+t-\frac{i\omega_-}{\rho}\log(2\cosh\rho t)\right] {}_2F_1\left(1+\frac{i\omega_-}{\rho}, \frac{i\omega_-}{\rho};1+\frac{i\omega_{\rm out}}{\rho};\frac{1-\tanh\rho t}{2}\right) \,,
\end{align}
such that $f_{\rm out}(t)\xrightarrow{t\rightarrow\infty} e^{-i\omega_{\rm out}t}$. We then express the incoming mode in terms of the outgoing one:
\begin{equation}
    f_{\rm in}(t) = \alpha f_{\rm out}(t) + \beta f_{\rm out}^*(t)
\end{equation}
with the same Bogoliubov coefficients as in Eq.~(\ref{eq:in_asymptotics}). From the properties of hypergeometric functions, we obtain \cite{Birrell:1982ix}
\begin{align}
    \alpha= \frac{\Gamma(1-i\omega_{\rm in}/\rho)\Gamma(-i\omega_{\rm out}/\rho)}{\Gamma(-i\omega_+/\rho) \Gamma(1-i\omega_+/\rho)} \, , \\ 
    \beta = \frac{\Gamma(1-i\omega_{\rm in}/\rho)\Gamma(i\omega_{\rm out}/\rho)}{\Gamma(i\omega_-/\rho) \Gamma(1+i\omega_-/\rho)}\,.
\end{align}
From Eqs.~(\ref{eq:<N^H_omega>}) and (\ref{eq:in_asymptotics}) we find 
\begin{align}
    \langle\hat N_{\rm out}(t\rightarrow\infty)\rangle &= \frac{\omega_{\rm out}}{4\omega_{\rm in}}~|2\beta|^2 = \frac{\sinh^2(\pi\omega_-/\rho)}{\sinh(\pi\omega_{\rm in}/\rho)\sinh(\pi\omega_{\rm out}/\rho)} \,,
\end{align}
in full agreement with Eq. (\ref{eq:asymptotic_particle_number}) (note that the definition of $\beta$ differs from the one in Appendix \ref{app:free:analytic} by a constant factor due to a different normalization of the mode functions). Thus we reproduce Eq.~(\ref{eq:initially_confusing}) within the Heisenberg picture, explicitly demonstrating its equivalence to the Schr\"odinger formulation.

This discussion is straightforwardly generalized to the case of a finite set of uncoupled harmonic oscillators, for which the full time-evolution operator factorizes into a finite product of single-oscillator unitaries. The cosmological particle production problem reduces to a finite set of oscillators for a finite set of momenta, which can be realized by placing the system in finite volume and on a lattice. Since our numerical method is based precisely on such a lattice formulation, we conclude that our Schr\"odinger-picture numerical method must be equivalent to the conventional Heisenberg-picture QFT description. In the main text, we report strong numerical evidence supporting this claim.

Finally, we combine Eq.~(\ref{eq:<N^H_omega>}) with the analysis relating the occupation number to the one-pair creation probability at the end of Appendix \ref{app:free:analytic} to derive the analytical prediction for the time-dependent one-pair creation probability:
\begin{equation}
    p_{1 -\rm pair}(k, t) = |\langle 1_k^{\omega_k(t)} 1_{-k}^{\omega_{k}(t)}|0^{\rm in}\rangle|^2 = p_0(t) \frac{\langle N_k^{\omega(t)}\rangle}{1+\langle N_k^{\omega(t)}\rangle} \,.\label{eq:analytic_prob_vs_t}
\end{equation}
Here the superscript $\omega_k(t) = \sqrt{k^2 + m_s^2\Omega(t)^2}$ denotes the basis that diagonalizes the instantaneous Hamiltonian in the Fock space. This expression is used to obtain the analytical prediction shown in Fig. \ref{fig:prob_vs_analytics} of the main text.

\section{Scaling of the survival probability} \label{app:survival}

\begin{figure}
    \centering
    \includegraphics[width=0.5\linewidth]{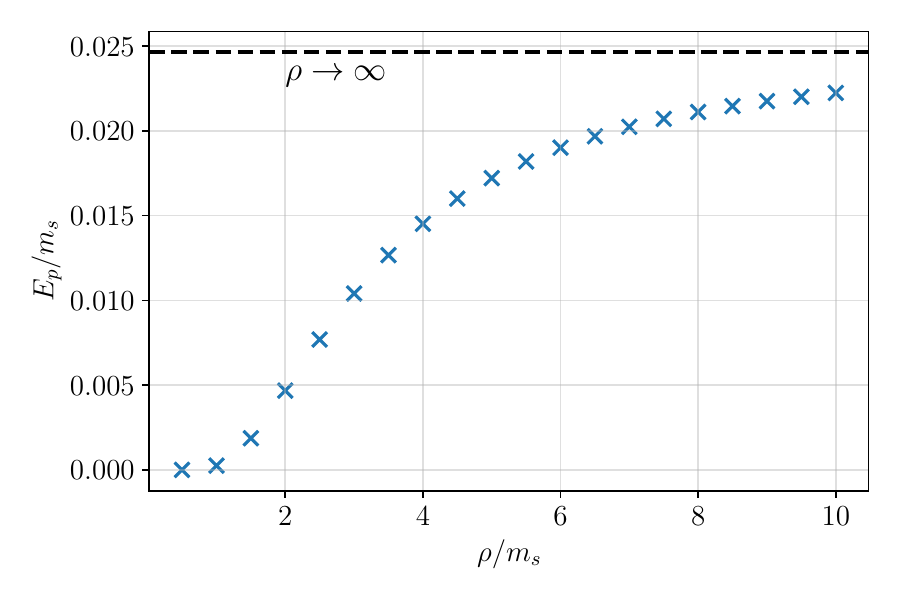}
    \caption{Characteristic energy scale for spatial volume dependence of the survival probability, $E_p$, as a function of the expansion speed $\rho$, both measured in the units of the free boson mass, $m_s$. Dashed horizontal line shows the limit of the instantaneous expansion $\rho\rightarrow\infty$. Parameters of the expanding metric are $A=2, B=1$.}
    \label{fig:E_p_vs_rho}
\end{figure}

In Sec. \ref{sec:prob} we discuss various measures of particle production, including the survival probability - the probability for the system to remain in the (instantaneous) ground state, $p_0=|\langle0^{\rm in}|0^{\rm out}\rangle|^2$, and its complement to unity --- excitation probability, $1-p_0$. Here we derive the predictions for these quantities in the free, analytically solvable case. First, let us find  the survival probability. It is fixed by the requirement that the sum of probabilities of all the possible final states is unity.  From the expression (\ref{eq:squeezed_state}) one finds the probability of creating the state with the set of occupation numbers $\{n_k\}$ corresponding to a set of momenta $\{k\}$ (which we assume to be finite) is
\begin{equation}
    p(\{k\},\{n_k\}) = p_0\prod_k \frac{|\beta_k|^{2n_k}}{|\alpha_k|^{2n_k}}\,. \label{eq:general_state_prob}
\end{equation}
Summing over all the possible occupation number sets is straightforward because it factorizes into a product of sums for each mode. Using the relation in Eq.(\ref{eq:alpha_beta_relation}) we find
\begin{equation}
    p_0 = \prod_k \frac{1}{|\alpha_k|^2}\,.
\end{equation}
To move forward, let us make assumptions about the set of momenta. If the system is placed in a finite box of spatial size $L$ and discretized with the lattice step $a$, the set of momenta is 
\begin{align}
    \{k\} = \{-\frac{\pi}{a} + \frac{2\pi}{L}, -\frac{\pi}{a} + \frac{4\pi}{L},\ldots, \frac{\pi}{a}\} \,.
\end{align}
Then
\begin{align}
    p_0 = \exp\bigg[-\sum_k \log|\alpha_k|^2\bigg] \overset{L\rightarrow\infty}{\overset{a\rightarrow 0}{=}} \exp\bigg[-\frac{L}{2\pi}\int_{-\infty}^\infty dk \,\log|\alpha_k|^2\bigg]\,,
\end{align}
where the infinite volume limit in practice means $L\gg1/m_s$. From this expression it is evident that the survival probability in the thermodynamic limit decreases exponentially with the system size, $p_0 = \exp(-L E_p)$ where $E_p$ is the characteristic ``probability" energy scale, given by
\begin{equation}
    E_p = \frac{\int_{-\infty}^\infty dk \,\log\bigg(1+\dfrac{\sinh^2(\pi\omega_-/\rho)}{\sinh(\pi\omega_{\rm in}/\rho)\sinh(\pi\omega_{\rm out}/\rho)}\bigg)}{2\pi} \,,
\end{equation}
where we used the explicit expression for $|\alpha_k|^2$. From the definitions of mode frequencies $\omega_{\rm in}, \omega_{\rm out}, \omega_-$, given in Eqs.(\ref{eq:omega_in})--(\ref{eq:omega_pm}), it is clear that $E_p$ is a function of $m_s, \rho, A$ and $B$. Curiously, in the parameter regime explored in our work, this energy scale turns out to be very different from the only natural scale in the problem, $m_s$; namely, $E_p\ll m_s$. As shown in Fig. \ref{fig:E_p_vs_rho}, the characteristic energy scale of survival probability is about two orders of magnitude below the natural scale $m_s$, as a consequence of numerical factors. It changes from near 0 at $\rho\lesssim m_s$ to a fixed value in the $\rho\rightarrow\infty$ limit that corresponds to an instantaneous expansion, or in other words to a sudden quench in the Hamiltonian.

\begin{figure}[h!]
    \centering    \includegraphics[width=0.5\linewidth]{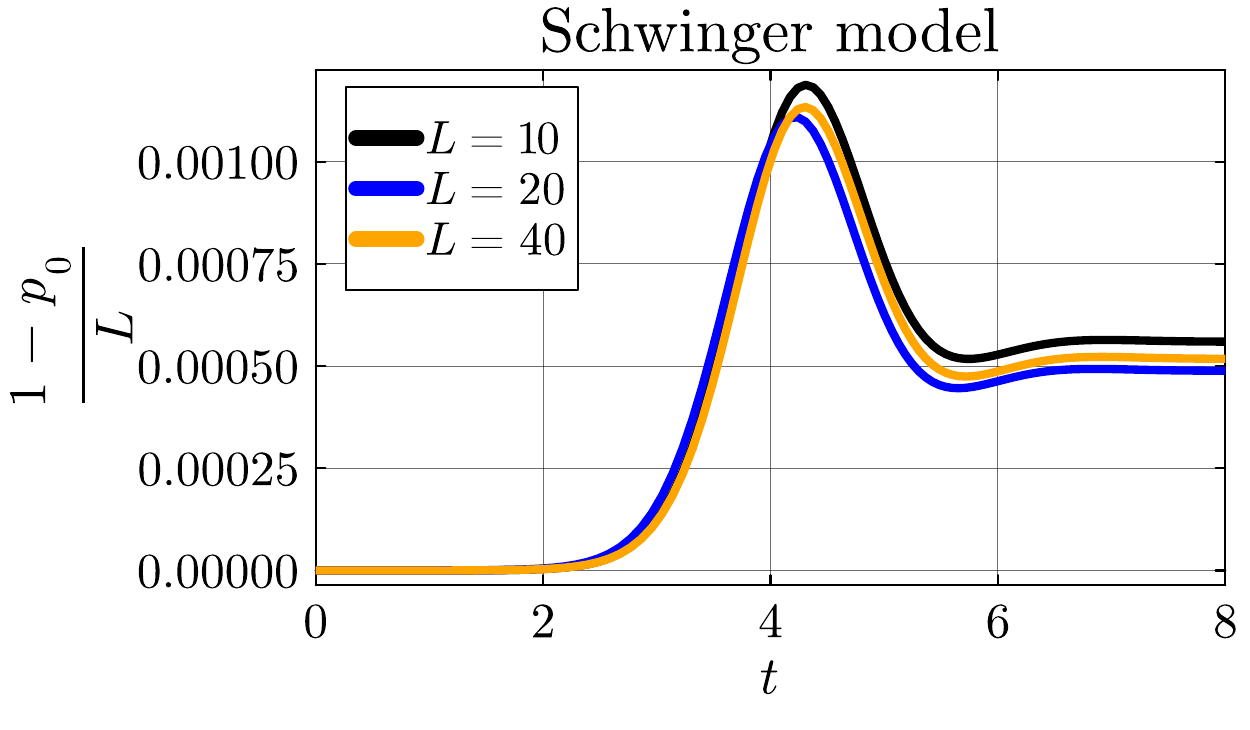}
    \caption{Excitation probability rescaled by the system volume $L$ as a function of time in the Schwinger model. Approximate scaling (\ref{eq:exc_prob_volume_scaling}) is satisfied for several values of the system size $L$. The system parameters are $\coupling a=0.5$, $\mf/\coupling = 1$, $a=1$.}
    \label{fig:p0_N_scaling}
\end{figure}

Now let us briefly discuss the scaling of the excitation probability. In small volumes, $L\ll1/E_p$, expanding the exponential yields linear scaling of the excitation probability with volume
\begin{equation}
    1-p_0\approx E_p \,L \,. \label{eq:exc_prob_volume_scaling}
\end{equation}
Needless to say, this behavior is only approximate, and for larger volumes should be replaced by the full expression, $1-\exp(-E_p L)\leq1$. In Fig. \ref{fig:p0_N_scaling} we demonstrate that this approximate linear scaling at small volume persists to good accuracy in an interacting system and as a function of time.

\bibliographystyle{utphys}
\bibliography{main}

\providecommand{\href}[2]{#2}\begingroup\raggedright\begin{thebibliography}{100}

\bibitem{Figueroa:2020rrl}
D.~G. Figueroa, A.~Florio, F.~Torrenti, and W.~Valkenburg, ``{The art of simulating the early Universe -- Part I},'' \href{http://dx.doi.org/10.1088/1475-7516/2021/04/035}{{\em JCAP} {\bfseries 04} (2021) 035}, \href{http://arxiv.org/abs/2006.15122}{{\ttfamily arXiv:2006.15122 [astro-ph.CO]}}.

\bibitem{Baeza-Ballesteros:2025tme}
J.~Baeza-Ballesteros, D.~G. Figueroa, A.~Florio, J.~Lizarraga, N.~Loayza, K.~Marschall, T.~Opferkuch, B.~A. Stefanek, F.~Torrent{\'\i}, and A.~Urio, ``{The art of simulating the early Universe. Part II},'' \href{http://arxiv.org/abs/2512.15627}{{\ttfamily arXiv:2512.15627 [astro-ph.CO]}}.

\bibitem{Bekenstein:1973ur}
J.~D. Bekenstein, ``{Black holes and entropy},'' \href{http://dx.doi.org/10.1103/PhysRevD.7.2333}{{\em Phys. Rev. D} {\bfseries 7} (1973) 2333--2346}.

\bibitem{Hawking:1975vcx}
S.~W. Hawking, ``{Particle Creation by Black Holes},'' \href{http://dx.doi.org/10.1007/BF02345020}{{\em Commun. Math. Phys.} {\bfseries 43} (1975) 199--220}. [Erratum: Commun.Math.Phys. 46, 206 (1976)].

\bibitem{Heinz:2000bk}
U.~W. Heinz and M.~Jacob, ``{Evidence for a new state of matter: An Assessment of the results from the CERN lead beam program},'' \href{http://arxiv.org/abs/nucl-th/0002042}{{\ttfamily arXiv:nucl-th/0002042}}.

\bibitem{Unruh:1980cg}
W.~G. Unruh, ``{Experimental black hole evaporation},'' \href{http://dx.doi.org/10.1103/PhysRevLett.46.1351}{{\em Phys. Rev. Lett.} {\bfseries 46} (1981) 1351--1353}.

\bibitem{Garay:1999sk}
L.~J. Garay, J.~R. Anglin, J.~I. Cirac, and P.~Zoller, ``{Black holes in Bose-Einstein condensates},'' \href{http://dx.doi.org/10.1103/PhysRevLett.85.4643}{{\em Phys. Rev. Lett.} {\bfseries 85} (2000) 4643--4647}, \href{http://arxiv.org/abs/gr-qc/0002015}{{\ttfamily arXiv:gr-qc/0002015}}.

\bibitem{Visser:2001fe}
M.~Visser, C.~Barcelo, and S.~Liberati, ``{Analog models of and for gravity},'' \href{http://dx.doi.org/10.1023/A:1020180409214}{{\em Gen. Rel. Grav.} {\bfseries 34} (2002) 1719--1734}, \href{http://arxiv.org/abs/gr-qc/0111111}{{\ttfamily arXiv:gr-qc/0111111}}.

\bibitem{Novello:2002qg}
M.~Novello, M.~Visser, and G.~Volovik, eds., {\em {Artificial black holes}}.
\newblock 2002.

\bibitem{Barcelo:2003wu}
C.~Barcelo, S.~Liberati, and M.~Visser, ``{Probing semiclassical analog gravity in Bose-Einstein condensates with widely tunable interactions},'' \href{http://dx.doi.org/10.1103/PhysRevA.68.053613}{{\em Phys. Rev. A} {\bfseries 68} (2003) 053613}, \href{http://arxiv.org/abs/cond-mat/0307491}{{\ttfamily arXiv:cond-mat/0307491}}.

\bibitem{Fedichev:2003bv}
P.~O. Fedichev and U.~R. Fischer, ``{'Cosmological' quasiparticle production in harmonically trapped superfluid gases},'' \href{http://dx.doi.org/10.1103/PhysRevA.69.033602}{{\em Phys. Rev. A} {\bfseries 69} (2004) 033602}, \href{http://arxiv.org/abs/cond-mat/0303063}{{\ttfamily arXiv:cond-mat/0303063}}.

\bibitem{Fedichev:2003dj}
P.~O. Fedichev and U.~R. Fischer, ``{Observer dependence for the phonon content of the sound field living on the effective curved space-time background of a Bose-Einstein condensate},'' \href{http://dx.doi.org/10.1103/PhysRevD.69.064021}{{\em Phys. Rev. D} {\bfseries 69} (2004) 064021}, \href{http://arxiv.org/abs/cond-mat/0307200}{{\ttfamily arXiv:cond-mat/0307200}}.

\bibitem{Fedichev:2003id}
P.~O. Fedichev and U.~R. Fischer, ``{Gibbons-Hawking effect in the sonic de Sitter space-time of an expanding Bose-Einstein-condensed gas},'' \href{http://dx.doi.org/10.1103/PhysRevLett.91.240407}{{\em Phys. Rev. Lett.} {\bfseries 91} (2003) 240407}, \href{http://arxiv.org/abs/cond-mat/0304342}{{\ttfamily arXiv:cond-mat/0304342}}.

\bibitem{Fischer:2004bf}
U.~R. Fischer and R.~Schutzhold, ``{Quantum simulation of cosmic inflation in two-component Bose-Einstein condensates},'' \href{http://dx.doi.org/10.1103/PhysRevA.70.063615}{{\em Phys. Rev. A} {\bfseries 70} (2004) 063615}, \href{http://arxiv.org/abs/cond-mat/0406470}{{\ttfamily arXiv:cond-mat/0406470}}.

\bibitem{Jain:2007gg}
P.~Jain, S.~Weinfurtner, M.~Visser, and C.~W. Gardiner, ``{Analogue model of a FRW universe in Bose-Einstein condensates: Application of the classical field method},'' \href{http://dx.doi.org/10.1103/PhysRevA.76.033616}{{\em Phys. Rev. A} {\bfseries 76} (2007) 033616}, \href{http://arxiv.org/abs/0705.2077}{{\ttfamily arXiv:0705.2077 [cond-mat.other]}}.

\bibitem{Viermann:2022wgw}
C.~Viermann {\em et~al.}, ``{Quantum field simulator for dynamics in curved spacetime},'' \href{http://dx.doi.org/10.1038/s41586-022-05313-9}{{\em Nature} {\bfseries 611} no.~7935, (2022) 260--264}, \href{http://arxiv.org/abs/2202.10399}{{\ttfamily arXiv:2202.10399 [cond-mat.quant-gas]}}.

\bibitem{Lv:2021dgf}
C.~Lv, R.~Zhang, Z.~Zhai, and Q.~Zhou, ``{Curving the space by non-Hermiticity},'' \href{http://dx.doi.org/10.1038/s41467-022-29774-8}{{\em Nature Commun.} {\bfseries 13} no.~1, (2022) 2184}, \href{http://arxiv.org/abs/2106.02477}{{\ttfamily arXiv:2106.02477 [cond-mat.quant-gas]}}.

\bibitem{Birrell:1982ix}
N.~D. Birrell and P.~C.~W. Davies, \href{http://dx.doi.org/10.1017/CBO9780511622632}{{\em {Quantum Fields in Curved Space}}}.
\newblock Cambridge Monographs on Mathematical Physics. Cambridge University Press, Cambridge, UK, 1982.

\bibitem{Cabass:2022avo}
G.~Cabass, M.~M. Ivanov, M.~Lewandowski, M.~Mirbabayi, and M.~Simonovi{\'c}, ``{Snowmass white paper: Effective field theories in cosmology},'' \href{http://dx.doi.org/10.1016/j.dark.2023.101193}{{\em Phys. Dark Univ.} {\bfseries 40} (2023) 101193}, \href{http://arxiv.org/abs/2203.08232}{{\ttfamily arXiv:2203.08232 [astro-ph.CO]}}.

\bibitem{Kogut:1982ds}
J.~B. Kogut, ``{A Review of the Lattice Gauge Theory Approach to Quantum Chromodynamics},'' \href{http://dx.doi.org/10.1103/RevModPhys.55.775}{{\em Rev. Mod. Phys.} {\bfseries 55} (1983) 775}.

\bibitem{Brambilla:2014jmp}
N.~Brambilla {\em et~al.}, ``{QCD and Strongly Coupled Gauge Theories: Challenges and Perspectives},'' \href{http://dx.doi.org/10.1140/epjc/s10052-014-2981-5}{{\em Eur. Phys. J. C} {\bfseries 74} no.~10, (2014) 2981}, \href{http://arxiv.org/abs/1404.3723}{{\ttfamily arXiv:1404.3723 [hep-ph]}}.

\bibitem{Banuls:2019bmf}
M.~C. Ba{\~n}uls {\em et~al.}, ``{Simulating Lattice Gauge Theories within Quantum Technologies},'' \href{http://dx.doi.org/10.1140/epjd/e2020-100571-8}{{\em Eur. Phys. J. D} {\bfseries 74} no.~8, (2020) 165}, \href{http://arxiv.org/abs/1911.00003}{{\ttfamily arXiv:1911.00003 [quant-ph]}}.

\bibitem{Bauer:2022hpo}
C.~W. Bauer {\em et~al.}, ``{Quantum Simulation for High-Energy Physics},'' \href{http://dx.doi.org/10.1103/PRXQuantum.4.027001}{{\em PRX Quantum} {\bfseries 4} no.~2, (2023) 027001}, \href{http://arxiv.org/abs/2204.03381}{{\ttfamily arXiv:2204.03381 [quant-ph]}}.

\bibitem{DiMeglio:2023nsa}
A.~Di~Meglio {\em et~al.}, ``{Quantum Computing for High-Energy Physics: State of the Art and Challenges},'' \href{http://dx.doi.org/10.1103/PRXQuantum.5.037001}{{\em PRX Quantum} {\bfseries 5} no.~3, (2024) 037001}, \href{http://arxiv.org/abs/2307.03236}{{\ttfamily arXiv:2307.03236 [quant-ph]}}.

\bibitem{Klco:2018kyo}
N.~Klco, E.~F. Dumitrescu, A.~J. McCaskey, T.~D. Morris, R.~C. Pooser, M.~Sanz, E.~Solano, P.~Lougovski, and M.~J. Savage, ``{Quantum-classical computation of Schwinger model dynamics using quantum computers},'' \href{http://dx.doi.org/10.1103/PhysRevA.98.032331}{{\em Phys. Rev. A} {\bfseries 98} no.~3, (2018) 032331}, \href{http://arxiv.org/abs/1803.03326}{{\ttfamily arXiv:1803.03326 [quant-ph]}}.

\bibitem{Kharzeev:2020kgc}
D.~E. Kharzeev and Y.~Kikuchi, ``{Real-time chiral dynamics from a digital quantum simulation},'' \href{http://dx.doi.org/10.1103/PhysRevResearch.2.023342}{{\em Phys. Rev. Res.} {\bfseries 2} no.~2, (2020) 023342}, \href{http://arxiv.org/abs/2001.00698}{{\ttfamily arXiv:2001.00698 [hep-ph]}}.

\bibitem{Ciavarella:2021nmj}
A.~Ciavarella, N.~Klco, and M.~J. Savage, ``{Trailhead for quantum simulation of SU(3) Yang-Mills lattice gauge theory in the local multiplet basis},'' \href{http://dx.doi.org/10.1103/PhysRevD.103.094501}{{\em Phys. Rev. D} {\bfseries 103} no.~9, (2021) 094501}, \href{http://arxiv.org/abs/2101.10227}{{\ttfamily arXiv:2101.10227 [quant-ph]}}.

\bibitem{Rigobello:2021fxw}
M.~Rigobello, S.~Notarnicola, G.~Magnifico, and S.~Montangero, ``{Entanglement generation in (1+1)D QED scattering processes},'' \href{http://dx.doi.org/10.1103/PhysRevD.104.114501}{{\em Phys. Rev. D} {\bfseries 104} no.~11, (2021) 114501}, \href{http://arxiv.org/abs/2105.03445}{{\ttfamily arXiv:2105.03445 [hep-lat]}}.

\bibitem{deJong:2021wsd}
W.~A. de~Jong, K.~Lee, J.~Mulligan, M.~P{\l}osko{\'n}, F.~Ringer, and X.~Yao, ``{Quantum simulation of nonequilibrium dynamics and thermalization in the Schwinger model},'' \href{http://dx.doi.org/10.1103/PhysRevD.106.054508}{{\em Phys. Rev. D} {\bfseries 106} no.~5, (2022) 054508}, \href{http://arxiv.org/abs/2106.08394}{{\ttfamily arXiv:2106.08394 [quant-ph]}}.

\bibitem{Zhou:2021kdl}
Z.-Y. Zhou, G.-X. Su, J.~C. Halimeh, R.~Ott, H.~Sun, P.~Hauke, B.~Yang, Z.-S. Yuan, J.~Berges, and J.-W. Pan, ``{Thermalization dynamics of a gauge theory on a quantum simulator},'' \href{http://dx.doi.org/10.1126/science.abl6277}{{\em Science} {\bfseries 377} no.~6603, (2022) abl6277}, \href{http://arxiv.org/abs/2107.13563}{{\ttfamily arXiv:2107.13563 [cond-mat.quant-gas]}}.

\bibitem{Czajka:2021yll}
A.~M. Czajka, Z.-B. Kang, H.~Ma, and F.~Zhao, ``{Quantum simulation of chiral phase transitions},'' \href{http://dx.doi.org/10.1007/JHEP08(2022)209}{{\em JHEP} {\bfseries 08} (2022) 209}, \href{http://arxiv.org/abs/2112.03944}{{\ttfamily arXiv:2112.03944 [hep-ph]}}.

\bibitem{Farrell:2022wyt}
R.~C. Farrell, I.~A. Chernyshev, S.~J.~M. Powell, N.~A. Zemlevskiy, M.~Illa, and M.~J. Savage, ``{Preparations for quantum simulations of quantum chromodynamics in 1+1 dimensions. I. Axial gauge},'' \href{http://dx.doi.org/10.1103/PhysRevD.107.054512}{{\em Phys. Rev. D} {\bfseries 107} no.~5, (2023) 054512}, \href{http://arxiv.org/abs/2207.01731}{{\ttfamily arXiv:2207.01731 [quant-ph]}}.

\bibitem{Farrell:2022vyh}
R.~C. Farrell, I.~A. Chernyshev, S.~J.~M. Powell, N.~A. Zemlevskiy, M.~Illa, and M.~J. Savage, ``{Preparations for quantum simulations of quantum chromodynamics in 1+1 dimensions. II. Single-baryon {\ensuremath{\beta}}-decay in real time},'' \href{http://dx.doi.org/10.1103/PhysRevD.107.054513}{{\em Phys. Rev. D} {\bfseries 107} no.~5, (2023) 054513}, \href{http://arxiv.org/abs/2209.10781}{{\ttfamily arXiv:2209.10781 [quant-ph]}}.

\bibitem{Davoudi:2022xmb}
Z.~Davoudi, A.~F. Shaw, and J.~R. Stryker, ``{General quantum algorithms for Hamiltonian simulation with applications to a non-Abelian lattice gauge theory},'' \href{http://dx.doi.org/10.22331/q-2023-12-20-1213}{{\em Quantum} {\bfseries 7} (2023) 1213}, \href{http://arxiv.org/abs/2212.14030}{{\ttfamily arXiv:2212.14030 [hep-lat]}}.

\bibitem{Florio:2023dke}
A.~Florio, D.~Frenklakh, K.~Ikeda, D.~Kharzeev, V.~Korepin, S.~Shi, and K.~Yu, ``{Real-Time Nonperturbative Dynamics of Jet Production in Schwinger Model: Quantum Entanglement and Vacuum Modification},'' \href{http://dx.doi.org/10.1103/PhysRevLett.131.021902}{{\em Phys. Rev. Lett.} {\bfseries 131} no.~2, (2023) 021902}, \href{http://arxiv.org/abs/2301.11991}{{\ttfamily arXiv:2301.11991 [hep-ph]}}.

\bibitem{Farrell:2023fgd}
R.~C. Farrell, M.~Illa, A.~N. Ciavarella, and M.~J. Savage, ``{Scalable Circuits for Preparing Ground States on Digital Quantum Computers: The Schwinger Model Vacuum on 100 Qubits},'' \href{http://arxiv.org/abs/2308.04481}{{\ttfamily arXiv:2308.04481 [quant-ph]}}.

\bibitem{Belyansky:2023rgh}
R.~Belyansky, S.~Whitsitt, N.~Mueller, A.~Fahimniya, E.~R. Bennewitz, Z.~Davoudi, and A.~V. Gorshkov, ``{High-Energy Collision of Quarks and Mesons in the Schwinger Model: From Tensor Networks to Circuit QED},'' \href{http://dx.doi.org/10.1103/PhysRevLett.132.091903}{{\em Phys. Rev. Lett.} {\bfseries 132} no.~9, (2024) 091903}, \href{http://arxiv.org/abs/2307.02522}{{\ttfamily arXiv:2307.02522 [quant-ph]}}.

\bibitem{Barata:2023jgd}
J.~a. Barata, W.~Gong, and R.~Venugopalan, ``{Realtime dynamics of hyperon spin correlations from string fragmentation in a deformed four-flavor Schwinger model},'' \href{http://arxiv.org/abs/2308.13596}{{\ttfamily arXiv:2308.13596 [hep-ph]}}.

\bibitem{Farrell:2024fit}
R.~C. Farrell, M.~Illa, A.~N. Ciavarella, and M.~J. Savage, ``{Quantum Simulations of Hadron Dynamics in the Schwinger Model using 112 Qubits},'' \href{http://arxiv.org/abs/2401.08044}{{\ttfamily arXiv:2401.08044 [quant-ph]}}.

\bibitem{Davoudi:2024wyv}
Z.~Davoudi, C.-C. Hsieh, and S.~V. Kadam, ``{Scattering wave packets of hadrons in gauge theories: Preparation on a quantum computer},'' \href{http://dx.doi.org/10.22331/q-2024-11-11-1520}{{\em Quantum} {\bfseries 8} (2024) 1520}, \href{http://arxiv.org/abs/2402.00840}{{\ttfamily arXiv:2402.00840 [quant-ph]}}.

\bibitem{Papaefstathiou:2024zsu}
I.~Papaefstathiou, J.~Knolle, and M.~C. Ba{\~n}uls, ``{Real-time scattering in the lattice Schwinger model},'' \href{http://dx.doi.org/10.1103/PhysRevD.111.014504}{{\em Phys. Rev. D} {\bfseries 111} no.~1, (2025) 014504}, \href{http://arxiv.org/abs/2402.18429}{{\ttfamily arXiv:2402.18429 [hep-lat]}}.

\bibitem{Florio:2024aix}
A.~Florio, D.~Frenklakh, K.~Ikeda, D.~E. Kharzeev, V.~Korepin, S.~Shi, and K.~Yu, ``{Quantum real-time evolution of entanglement and hadronization in jet production: Lessons from the massive Schwinger model},'' \href{http://dx.doi.org/10.1103/PhysRevD.110.094029}{{\em Phys. Rev. D} {\bfseries 110} no.~9, (2024) 094029}, \href{http://arxiv.org/abs/2404.00087}{{\ttfamily arXiv:2404.00087 [hep-ph]}}.

\bibitem{Grieninger:2024axp}
S.~Grieninger and I.~Zahed, ``{Quasifragmentation functions in the massive Schwinger model},'' \href{http://dx.doi.org/10.1103/PhysRevD.110.116009}{{\em Phys. Rev. D} {\bfseries 110} no.~11, (2024) 116009}, \href{http://arxiv.org/abs/2406.01891}{{\ttfamily arXiv:2406.01891 [hep-ph]}}.

\bibitem{Grieninger:2024cdl}
S.~Grieninger, K.~Ikeda, and I.~Zahed, ``{Quasiparton distributions in massive QED2: Toward quantum computation},'' \href{http://dx.doi.org/10.1103/PhysRevD.110.076008}{{\em Phys. Rev. D} {\bfseries 110} no.~7, (2024) 076008}, \href{http://arxiv.org/abs/2404.05112}{{\ttfamily arXiv:2404.05112 [hep-ph]}}.

\bibitem{Araz:2024dcy}
J.~Y. Araz, M.~Grau, J.~Montgomery, and F.~Ringer, ``{Hybrid quantum simulations with qubits and qumodes on trapped-ion platforms},'' \href{http://dx.doi.org/10.1103/kbv4-jj51}{{\em Phys. Rev. A} {\bfseries 112} no.~1, (2025) 012620}, \href{http://arxiv.org/abs/2410.07346}{{\ttfamily arXiv:2410.07346 [quant-ph]}}.

\bibitem{Barata:2024apg}
J.~Barata and S.~Mukherjee, ``{Probing celestial energy and charge correlations through real-time quantum simulations: Insights from the Schwinger model},'' \href{http://dx.doi.org/10.1103/PhysRevD.111.L031901}{{\em Phys. Rev. D} {\bfseries 111} no.~3, (2025) L031901}, \href{http://arxiv.org/abs/2409.13816}{{\ttfamily arXiv:2409.13816 [hep-ph]}}.

\bibitem{Farrell:2025nkx}
R.~C. Farrell, N.~A. Zemlevskiy, M.~Illa, and J.~Preskill, ``{Digital quantum simulations of scattering in quantum field theories using W states},'' \href{http://arxiv.org/abs/2505.03111}{{\ttfamily arXiv:2505.03111 [quant-ph]}}.

\bibitem{Florio:2025hoc}
A.~Florio, D.~Frenklakh, S.~Grieninger, D.~E. Kharzeev, A.~Palermo, and S.~Shi, ``{Thermalization from quantum entanglement: jet simulations in the massive Schwinger model},'' \href{http://arxiv.org/abs/2506.14983}{{\ttfamily arXiv:2506.14983 [hep-ph]}}.

\bibitem{Barata:2025jhd}
J.~Barata, D.~Frenklakh, and S.~Mukherjee, ``{Thermal modifications of mesons and energy-energy correlators from real-time simulations of a $U(1)$ lattice gauge theory},'' \href{http://arxiv.org/abs/2507.16890}{{\ttfamily arXiv:2507.16890 [hep-ph]}}.

\bibitem{Barata:2025rjb}
J.~Barata, J.~Hormaza, Z.-B. Kang, and W.~Qian, ``{Hadronic scattering in (1+1)D SU(2) lattice gauge theory from tensor networks},'' \href{http://arxiv.org/abs/2511.00154}{{\ttfamily arXiv:2511.00154 [hep-lat]}}.

\bibitem{Grieninger:2025mbm}
S.~Grieninger, J.~Montgomery, F.~Ringer, and I.~Zahed, ``{Tensor network simulations of quasi-GPDs in the massive Schwinger model},'' \href{http://arxiv.org/abs/2511.17752}{{\ttfamily arXiv:2511.17752 [hep-lat]}}.

\bibitem{Kang:2025xpz}
Z.-B. Kang, N.~Moran, P.~Nguyen, and W.~Qian, ``{Partonic distribution functions and amplitudes using tensor network methods},'' \href{http://dx.doi.org/10.1007/JHEP09(2025)176}{{\em JHEP} {\bfseries 09} (2025) 176}, \href{http://arxiv.org/abs/2501.09738}{{\ttfamily arXiv:2501.09738 [hep-ph]}}.

\bibitem{Steinhauer:2021fhb}
J.~Steinhauer, M.~Abuzarli, T.~Aladjidi, T.~Bienaim{\'e}, C.~Piekarski, W.~Liu, E.~Giacobino, A.~Bramati, and Q.~Glorieux, ``{Analogue cosmological particle creation in an ultracold quantum fluid of light},'' \href{http://dx.doi.org/10.1038/s41467-022-30603-1}{{\em Nature Commun.} {\bfseries 13} (2022) 2890}, \href{http://arxiv.org/abs/2102.08279}{{\ttfamily arXiv:2102.08279 [cond-mat.quant-gas]}}.

\bibitem{Fulgado-Claudio:2024xvk}
C.~Fulgado-Claudio, P.~Sala, D.~Gonz{\'a}lez-Cuadra, and A.~Bermudez, ``{Interacting Dirac fields in an expanding universe: dynamical condensates and particle production},'' \href{http://arxiv.org/abs/2408.06405}{{\ttfamily arXiv:2408.06405 [gr-qc]}}.

\bibitem{Schmidt:2024zpg}
C.~F. Schmidt, {\'A}.~Parra-L{\'o}pez, M.~Tolosa-Sime{\'o}n, M.~Sparn, E.~Kath, N.~Liebster, J.~Duchene, H.~Strobel, M.~K. Oberthaler, and S.~Floerchinger, ``{Cosmological particle production in a quantum field simulator as a quantum mechanical scattering problem},'' \href{http://dx.doi.org/10.1103/PhysRevD.110.123523}{{\em Phys. Rev. D} {\bfseries 110} no.~12, (2024) 123523}, \href{http://arxiv.org/abs/2406.08094}{{\ttfamily arXiv:2406.08094 [gr-qc]}}.

\bibitem{Kinoshita:2024ahu}
S.~Kinoshita, K.~Murata, D.~Yamamoto, and R.~Yoshii, ``{Spin systems as quantum simulators of quantum field theories in curved spacetimes},'' \href{http://dx.doi.org/10.1103/PhysRevResearch.7.023197}{{\em Phys. Rev. Res.} {\bfseries 7} no.~2, (2025) 023197}, \href{http://arxiv.org/abs/2410.07587}{{\ttfamily arXiv:2410.07587 [hep-th]}}.

\bibitem{Gong:2025lmj}
J.-Q. Gong and J.-C. Yang, ``{Digit quantum simulation of a fermion field in an expanding universe},'' \href{http://arxiv.org/abs/2502.14021}{{\ttfamily arXiv:2502.14021 [quant-ph]}}.

\bibitem{Kharel:2025lek}
P.~K. Kharel, M.~Ghimire, A.~Khanal, S.~Pudasaini, N.~Khatri, S.~Bhandari, D.~Rai, K.~Adhikari, and R.~Singh, ``{Entanglement and particle production from cosmological perturbations: a quantum optical simulation approach},'' \href{http://arxiv.org/abs/2508.04249}{{\ttfamily arXiv:2508.04249 [gr-qc]}}.

\bibitem{Chandran:2023ogt}
S.~M. Chandran, K.~Rajeev, and S.~Shankaranarayanan, ``{Real-space quantum-to-classical transition of time dependent background fluctuations},'' \href{http://dx.doi.org/10.1103/PhysRevD.109.023503}{{\em Phys. Rev. D} {\bfseries 109} no.~2, (2024) 023503}, \href{http://arxiv.org/abs/2307.13611}{{\ttfamily arXiv:2307.13611 [gr-qc]}}.

\bibitem{Steinhauer:2015saa}
J.~Steinhauer, ``{Observation of quantum Hawking radiation and its entanglement in an analogue black hole},'' \href{http://dx.doi.org/10.1038/nphys3863}{{\em Nature Phys.} {\bfseries 12} (2016) 959}, \href{http://arxiv.org/abs/1510.00621}{{\ttfamily arXiv:1510.00621 [gr-qc]}}.

\bibitem{Rodriguez-Laguna:2016kri}
J.~Rodriguez-Laguna, L.~Tarruell, M.~Lewenstein, and A.~Celi, ``{Synthetic Unruh effect in cold atoms},'' \href{http://dx.doi.org/10.1103/PhysRevA.95.013627}{{\em Phys. Rev. A} {\bfseries 95} (2017) 013627}, \href{http://arxiv.org/abs/1606.09505}{{\ttfamily arXiv:1606.09505 [cond-mat.quant-gas]}}.

\bibitem{Hu:2018psq}
J.~Hu, L.~Feng, Z.~Zhang, and C.~Chin, ``{Quantum simulation of Unruh radiation},'' \href{http://dx.doi.org/10.1038/s41567-019-0537-1}{{\em Nature Phys.} {\bfseries 15} no.~8, (2019) 785--789}, \href{http://arxiv.org/abs/1807.07504}{{\ttfamily arXiv:1807.07504 [physics.atom-ph]}}.

\bibitem{Lewis:2019oyx}
A.~G.~M. Lewis and G.~Vidal, ``{Classical Simulations of Quantum Field Theory in Curved Spacetime I: Fermionic Hawking-Hartle Vacua from a Staggered Lattice Scheme},'' \href{http://dx.doi.org/10.22331/q-2020-10-28-351}{{\em Quantum} {\bfseries 4} (2020) 351}, \href{http://arxiv.org/abs/1911.12978}{{\ttfamily arXiv:1911.12978 [gr-qc]}}.

\bibitem{Ikeda:2025yqq}
K.~Ikeda and Y.~Oz, ``{Quantum Simulation of Fermions in $AdS_2$ Black Hole: Chirality, Entanglement, and Spectral Crossovers},'' \href{http://arxiv.org/abs/2509.21410}{{\ttfamily arXiv:2509.21410 [quant-ph]}}.

\bibitem{Ikeda:2025lig}
K.~Ikeda and Y.~Oz, ``{Geometry Induced Chiral Transport and Entanglement in $AdS_2$ Background},'' \href{http://arxiv.org/abs/2511.09714}{{\ttfamily arXiv:2511.09714 [hep-th]}}.

\bibitem{Catterall:2025ofz}
S.~Catterall, A.~F. Kemper, Y.~Meurice, A.~Samlodia, and G.~C. Toga, ``{Quantum Ising Model on $(2+1)-$Dimensional Anti$-$de Sitter Space using Tensor Networks},'' \href{http://arxiv.org/abs/2512.20838}{{\ttfamily arXiv:2512.20838 [hep-lat]}}.

\bibitem{Barcelos-Neto:1985yfn}
J.~Barcelos-Neto and A.~K. Das, ``{Path Integrals and the Solution of Schwinger Model in Curved Space-time},'' \href{http://dx.doi.org/10.1103/PhysRevD.33.2262}{{\em Phys. Rev. D} {\bfseries 33} (1986) 2262}.

\bibitem{Eboli:1987mu}
O.~J.~P. Eboli, ``{Abelian Bosonization in Curved Space},'' \href{http://dx.doi.org/10.1103/PhysRevD.36.2408}{{\em Phys. Rev. D} {\bfseries 36} (1987) 2408}.

\bibitem{Orus:2013kga}
R.~Orus, ``{A Practical Introduction to Tensor Networks: Matrix Product States and Projected Entangled Pair States},'' \href{http://dx.doi.org/10.1016/j.aop.2014.06.013}{{\em Annals Phys.} {\bfseries 349} (2014) 117--158}, \href{http://arxiv.org/abs/1306.2164}{{\ttfamily arXiv:1306.2164 [cond-mat.str-el]}}.

\bibitem{Akhmedov:2022whm}
E.~T. Akhmedov and P.~A. Anempodistov, ``{Loop corrections to cosmological particle creation},'' \href{http://dx.doi.org/10.1103/PhysRevD.105.105019}{{\em Phys. Rev. D} {\bfseries 105} no.~10, (2022) 105019}, \href{http://arxiv.org/abs/2204.01388}{{\ttfamily arXiv:2204.01388 [hep-th]}}.

\bibitem{Schwinger:1962tp}
J.~S. Schwinger, ``{Gauge Invariance and Mass. 2.},'' \href{http://dx.doi.org/10.1103/PhysRev.128.2425}{{\em Phys. Rev.} {\bfseries 128} (1962) 2425--2429}.

\bibitem{Casher:1974vf}
A.~Casher, J.~B. Kogut, and L.~Susskind, ``{Vacuum polarization and the absence of free quarks},'' \href{http://dx.doi.org/10.1103/PhysRevD.10.732}{{\em Phys. Rev. D} {\bfseries 10} (1974) 732--745}.

\bibitem{Coleman:1975pw}
S.~R. Coleman, R.~Jackiw, and L.~Susskind, ``{Charge Shielding and Quark Confinement in the Massive Schwinger Model},'' \href{http://dx.doi.org/10.1016/0003-4916(75)90212-2}{{\em Annals Phys.} {\bfseries 93} (1975) 267}.

\bibitem{Coleman:1976uz}
S.~R. Coleman, ``{More About the Massive Schwinger Model},'' \href{http://dx.doi.org/10.1016/0003-4916(76)90280-3}{{\em Annals Phys.} {\bfseries 101} (1976) 239}.

\bibitem{Byrnes:2002nv}
T.~Byrnes, P.~Sriganesh, R.~J. Bursill, and C.~J. Hamer, ``{Density matrix renormalization group approach to the massive Schwinger model},'' \href{http://dx.doi.org/10.1103/PhysRevD.66.013002}{{\em Phys. Rev. D} {\bfseries 66} (2002) 013002}, \href{http://arxiv.org/abs/hep-lat/0202014}{{\ttfamily arXiv:hep-lat/0202014}}.

\bibitem{Banuls:2013jaa}
M.~C. Ba\~nuls, K.~Cichy, K.~Jansen, and J.~I. Cirac, ``{The mass spectrum of the Schwinger model with Matrix Product States},'' \href{http://dx.doi.org/10.1007/JHEP11(2013)158}{{\em JHEP} {\bfseries 11} (2013) 158}, \href{http://arxiv.org/abs/1305.3765}{{\ttfamily arXiv:1305.3765 [hep-lat]}}.

\bibitem{Dempsey:2025wia}
R.~Dempsey, A.-M.~E. Gl{\"u}ck, S.~S. Pufu, and B.~T. S{\o}gaard, ``{Infinite matrix product states for $(1+1)$-dimensional gauge theories},'' \href{http://arxiv.org/abs/2508.16363}{{\ttfamily arXiv:2508.16363 [hep-th]}}.

\bibitem{Felder:2000hr}
G.~N. Felder and L.~Kofman, ``{The Development of equilibrium after preheating},'' \href{http://dx.doi.org/10.1103/PhysRevD.63.103503}{{\em Phys. Rev. D} {\bfseries 63} (2001) 103503}, \href{http://arxiv.org/abs/hep-ph/0011160}{{\ttfamily arXiv:hep-ph/0011160}}.

\bibitem{Itou:2024psm}
E.~Itou, A.~Matsumoto, and Y.~Tanizaki, ``{DMRG study of the theta-dependent mass spectrum in the 2-flavor Schwinger model},'' \href{http://dx.doi.org/10.1007/JHEP09(2024)155}{{\em JHEP} {\bfseries 09} (2024) 155}, \href{http://arxiv.org/abs/2407.11391}{{\ttfamily arXiv:2407.11391 [hep-lat]}}.

\bibitem{Jordan:1928wi}
P.~Jordan and E.~P. Wigner, ``{About the Pauli exclusion principle},'' \href{http://dx.doi.org/10.1007/BF01331938}{{\em Z. Phys.} {\bfseries 47} (1928) 631--651}.

\bibitem{Bernard:1977pq}
C.~W. Bernard and A.~Duncan, ``{Regularization and Renormalization of Quantum Field Theory in Curved Space-Time},'' \href{http://dx.doi.org/10.1016/0003-4916(77)90210-X}{{\em Annals Phys.} {\bfseries 107} (1977) 201}.

\bibitem{White:1993zza}
S.~R. White, ``{Density-matrix algorithms for quantum renormalization groups},'' \href{http://dx.doi.org/10.1103/PhysRevB.48.10345}{{\em Phys. Rev. B} {\bfseries 48} (1993) 10345--10356}.

\bibitem{Schollwock:2005zz}
U.~Schollwock, ``{The density-matrix renormalization group},'' \href{http://dx.doi.org/10.1103/RevModPhys.77.259}{{\em Rev. Mod. Phys.} {\bfseries 77} (2005) 259--315}, \href{http://arxiv.org/abs/cond-mat/0409292}{{\ttfamily arXiv:cond-mat/0409292}}.

\bibitem{Haegeman:2011zz}
J.~Haegeman, J.~I. Cirac, T.~J. Osborne, I.~Pizorn, H.~Verschelde, and F.~Verstraete, ``{Time-Dependent Variational Principle for Quantum Lattices},'' \href{http://dx.doi.org/10.1103/PhysRevLett.107.070601}{{\em Phys. Rev. Lett.} {\bfseries 107} (2011) 070601}, \href{http://arxiv.org/abs/1103.0936}{{\ttfamily arXiv:1103.0936 [cond-mat.str-el]}}.

\bibitem{Haegeman:2016gfj}
J.~Haegeman, C.~Lubich, I.~Oseledets, B.~Vandereycken, and F.~Verstraete, ``{Unifying time evolution and optimization with matrix product states},'' \href{http://dx.doi.org/10.1103/PhysRevB.94.165116}{{\em Phys. Rev. B} {\bfseries 94} no.~16, (2016) 165116}.

\bibitem{itensor}
M.~Fishman, S.~R. White, and E.~M. Stoudenmire, ``{The ITensor Software Library for Tensor Network Calculations},'' \href{http://dx.doi.org/10.21468/SciPostPhysCodeb.4}{{\em SciPost Phys. Codebases} (2022) 4}. \url{https://scipost.org/10.21468/SciPostPhysCodeb.4}.

\bibitem{itensor-r0.3}
M.~Fishman, S.~R. White, and E.~M. Stoudenmire, ``{Codebase release 0.3 for ITensor},'' \href{http://dx.doi.org/10.21468/SciPostPhysCodeb.4-r0.3}{{\em SciPost Phys. Codebases} (2022) 4--r0.3}. \url{https://scipost.org/10.21468/SciPostPhysCodeb.4-r0.3}.

\bibitem{Sugihara:2004qr}
T.~Sugihara, ``{Density matrix renormalization group in a two-dimensional lambda phi4 Hamiltonian lattice model},'' \href{http://dx.doi.org/10.1088/1126-6708/2004/05/007}{{\em JHEP} {\bfseries 05} (2004) 007}, \href{http://arxiv.org/abs/hep-lat/0403008}{{\ttfamily arXiv:hep-lat/0403008}}.

\bibitem{Amico:2007ag}
L.~Amico, R.~Fazio, A.~Osterloh, and V.~Vedral, ``{Entanglement in many-body systems},'' \href{http://dx.doi.org/10.1103/RevModPhys.80.517}{{\em Rev. Mod. Phys.} {\bfseries 80} (2008) 517--576}, \href{http://arxiv.org/abs/quant-ph/0703044}{{\ttfamily arXiv:quant-ph/0703044}}.

\bibitem{Calabrese:2009qy}
P.~Calabrese and J.~Cardy, ``{Entanglement entropy and conformal field theory},'' \href{http://dx.doi.org/10.1088/1751-8113/42/50/504005}{{\em J. Phys. A} {\bfseries 42} (2009) 504005}, \href{http://arxiv.org/abs/0905.4013}{{\ttfamily arXiv:0905.4013 [cond-mat.stat-mech]}}.

\bibitem{Calabrese:2005zw}
P.~Calabrese and J.~L. Cardy, ``{Entanglement Entropy and Quantum Field Theory: a Non-technical Introduction},'' \href{http://dx.doi.org/10.1142/s021974990600192x}{{\em Int. J. Quant. Inf.} {\bfseries 04} no.~03, (2006) 429--438}, \href{http://arxiv.org/abs/quant-ph/0505193}{{\ttfamily arXiv:quant-ph/0505193}}.

\bibitem{Eisert:2008ur}
J.~Eisert, M.~Cramer, and M.~B. Plenio, ``{Area laws for the entanglement entropy - a review},'' \href{http://dx.doi.org/10.1103/RevModPhys.82.277}{{\em Rev. Mod. Phys.} {\bfseries 82} (2010) 277--306}, \href{http://arxiv.org/abs/0808.3773}{{\ttfamily arXiv:0808.3773 [quant-ph]}}.

\bibitem{Hollands:2017dov}
S.~Hollands and K.~Sanders, ``{Entanglement measures and their properties in quantum field theory},'' \href{http://arxiv.org/abs/1702.04924}{{\ttfamily arXiv:1702.04924 [quant-ph]}}.

\bibitem{Witten:2018zxz}
E.~Witten, ``{APS Medal for Exceptional Achievement in Research: Invited article on entanglement properties of quantum field theory},'' \href{http://dx.doi.org/10.1103/RevModPhys.90.045003}{{\em Rev. Mod. Phys.} {\bfseries 90} no.~4, (2018) 045003}, \href{http://arxiv.org/abs/1803.04993}{{\ttfamily arXiv:1803.04993 [hep-th]}}.

\bibitem{Casini:2009sr}
H.~Casini and M.~Huerta, ``{Entanglement entropy in free quantum field theory},'' \href{http://dx.doi.org/10.1088/1751-8113/42/50/504007}{{\em J. Phys. A} {\bfseries 42} (2009) 504007}, \href{http://arxiv.org/abs/0905.2562}{{\ttfamily arXiv:0905.2562 [hep-th]}}.

\bibitem{Calabrese:2005in}
P.~Calabrese and J.~L. Cardy, ``{Evolution of entanglement entropy in one-dimensional systems},'' \href{http://dx.doi.org/10.1088/1742-5468/2005/04/P04010}{{\em J. Stat. Mech.} {\bfseries 0504} (2005) P04010}, \href{http://arxiv.org/abs/cond-mat/0503393}{{\ttfamily arXiv:cond-mat/0503393}}.

\bibitem{Belfiglio:2025cst}
A.~Belfiglio, O.~Luongo, and S.~Mancini, ``{Quantum entanglement in cosmology},'' \href{http://dx.doi.org/10.1016/j.physrep.2025.09.001}{{\em Phys. Rept.} {\bfseries 1146} (2025) 1--47}, \href{http://arxiv.org/abs/2506.03841}{{\ttfamily arXiv:2506.03841 [gr-qc]}}.

\bibitem{Solodukhin:2011gn}
S.~N. Solodukhin, ``{Entanglement entropy of black holes},'' \href{http://dx.doi.org/10.12942/lrr-2011-8}{{\em Living Rev. Rel.} {\bfseries 14} (2011) 8}, \href{http://arxiv.org/abs/1104.3712}{{\ttfamily arXiv:1104.3712 [hep-th]}}.

\bibitem{Ball:2005xa}
J.~L. Ball, I.~Fuentes-Schuller, and F.~P. Schuller, ``{Entanglement in an expanding spacetime},'' \href{http://dx.doi.org/10.1016/j.physleta.2006.07.028}{{\em Phys. Lett. A} {\bfseries 359} (2006) 550--554}, \href{http://arxiv.org/abs/quant-ph/0506113}{{\ttfamily arXiv:quant-ph/0506113}}.

\bibitem{Fuentes:2010dt}
I.~Fuentes, R.~B. Mann, E.~Martin-Martinez, and S.~Moradi, ``{Entanglement of Dirac fields in an expanding spacetime},'' \href{http://dx.doi.org/10.1103/PhysRevD.82.045030}{{\em Phys. Rev. D} {\bfseries 82} (2010) 045030}, \href{http://arxiv.org/abs/1007.1569}{{\ttfamily arXiv:1007.1569 [quant-ph]}}.

\bibitem{Haque:2025pav}
S.~S. Haque, G.~Jafari, and B.~Underwood, ``{Inflation is Not Magic},'' \href{http://arxiv.org/abs/2512.10126}{{\ttfamily arXiv:2512.10126 [hep-th]}}.

\bibitem{DeChiara:2005wb}
G.~De~Chiara, S.~Montangero, P.~Calabrese, and R.~Fazio, ``{Entanglement entropy dynamics in Heisenberg chains},'' \href{http://dx.doi.org/10.1088/1742-5468/2006/03/P03001}{{\em J. Stat. Mech.} {\bfseries 0603} (2006) P03001}, \href{http://arxiv.org/abs/cond-mat/0512586}{{\ttfamily arXiv:cond-mat/0512586}}.

\bibitem{Calabrese:2016xau}
P.~Calabrese and J.~Cardy, ``{Quantum quenches in 1 + 1 dimensional conformal field theories},'' \href{http://dx.doi.org/10.1088/1742-5468/2016/06/064003}{{\em J. Stat. Mech.} {\bfseries 1606} no.~6, (2016) 064003}, \href{http://arxiv.org/abs/1603.02889}{{\ttfamily arXiv:1603.02889 [cond-mat.stat-mech]}}.

\bibitem{Vidal:2006ofj}
G.~Vidal, ``{Classical simulation of infinite-size quantum lattice systems in one spatial dimension},'' \href{http://dx.doi.org/10.1103/PhysRevLett.98.070201}{{\em Phys. Rev. Lett.} {\bfseries 98} (2007) 070201}, \href{http://arxiv.org/abs/cond-mat/0605597}{{\ttfamily arXiv:cond-mat/0605597}}.

\bibitem{Orus:2008zsh}
R.~Or{\'u}s and G.~Vidal, ``{Infinite time-evolving block decimation algorithm beyond unitary evolution},'' \href{http://dx.doi.org/10.1103/PhysRevB.78.155117}{{\em Phys. Rev. B} {\bfseries 78} no.~15, (2008) 155117}.

\bibitem{Gott:1982qg}
J.~R. Gott and M.~Alpert, ``{General relativity in a (2+1)-dimensional space-time},'' \href{http://dx.doi.org/10.1007/BF00762539}{{\em Gen. Rel. Grav.} {\bfseries 16} (1984) 243--247}.

\bibitem{Deser:1983tn}
S.~Deser, R.~Jackiw, and G.~'t~Hooft, ``{Three-Dimensional Einstein Gravity: Dynamics of Flat Space},'' \href{http://dx.doi.org/10.1016/0003-4916(84)90085-X}{{\em Annals Phys.} {\bfseries 152} (1984) 220}.

\bibitem{Jackiw:1984je}
R.~Jackiw, ``{Lower Dimensional Gravity},'' \href{http://dx.doi.org/10.1016/0550-3213(85)90448-1}{{\em Nucl. Phys. B} {\bfseries 252} (1985) 343--356}.

\bibitem{Witten:1988hc}
E.~Witten, ``{(2+1)-Dimensional Gravity as an Exactly Soluble System},'' \href{http://dx.doi.org/10.1016/0550-3213(88)90143-5}{{\em Nucl. Phys. B} {\bfseries 311} (1988) 46}.

\bibitem{Blaschke:2014ioa}
D.~N. Blaschke, R.~Carballo-Rubio, and E.~Mottola, ``{Fermion Pairing and the Scalar Boson of the 2D Conformal Anomaly},'' \href{http://dx.doi.org/10.1007/JHEP12(2014)153}{{\em JHEP} {\bfseries 12} (2014) 153}, \href{http://arxiv.org/abs/1407.8523}{{\ttfamily arXiv:1407.8523 [hep-th]}}.

\bibitem{Dirac:1950pj}
P.~A.~M. Dirac, ``{Generalized Hamiltonian dynamics},'' \href{http://dx.doi.org/10.4153/CJM-1950-012-1}{{\em Can. J. Math.} {\bfseries 2} (1950) 129--148}.

\bibitem{Anderson:1951ta}
J.~L. Anderson and P.~G. Bergmann, ``{Constraints in covariant field theories},'' \href{http://dx.doi.org/10.1103/PhysRev.83.1018}{{\em Phys. Rev.} {\bfseries 83} (1951) 1018--1025}.

\bibitem{Dirac:1958sq}
P.~A.~M. Dirac, ``{Generalized Hamiltonian dynamics},'' \href{http://dx.doi.org/10.1098/rspa.1958.0141}{{\em Proc. Roy. Soc. Lond. A} {\bfseries 246} (1958) 326--332}.

\bibitem{Henneaux:1994lbw}
M.~Henneaux and C.~Teitelboim, {\em {Quantization of Gauge Systems}}.
\newblock Princeton University Press, 8, 1994.

\bibitem{Susskind:1976jm}
L.~Susskind, ``{Lattice Fermions},'' \href{http://dx.doi.org/10.1103/PhysRevD.16.3031}{{\em Phys. Rev. D} {\bfseries 16} (1977) 3031--3039}.

\bibitem{Davoudi:2025kxb}
Z.~Davoudi, ``{TASI/CERN/KITP Lecture Notes on ''Toward Quantum Computing Gauge Theories of Nature''},'' \href{http://arxiv.org/abs/2507.15840}{{\ttfamily arXiv:2507.15840 [hep-lat]}}.

\bibitem{Kolb:2023ydq}
E.~W. Kolb and A.~J. Long, ``{Cosmological gravitational particle production and its implications for cosmological relics},'' \href{http://dx.doi.org/10.1103/RevModPhys.96.045005}{{\em Rev. Mod. Phys.} {\bfseries 96} no.~4, (2024) 045005}, \href{http://arxiv.org/abs/2312.09042}{{\ttfamily arXiv:2312.09042 [astro-ph.CO]}}.

\bibitem{Agullo:2015qqa}
I.~Agullo and A.~Ashtekar, ``{Unitarity and ultraviolet regularity in cosmology},'' \href{http://dx.doi.org/10.1103/PhysRevD.91.124010}{{\em Phys. Rev. D} {\bfseries 91} no.~12, (2015) 124010}, \href{http://arxiv.org/abs/1503.03407}{{\ttfamily arXiv:1503.03407 [gr-qc]}}.

\bibitem{Coelho:2024zuf}
S.~S. Coelho, L.~Queiroz, and D.~T. Alves, ``{The time-dependent quantum harmonic oscillator: a pedagogical approach via the Lewis{\textendash}Riesenfeld dynamical invariant method},'' \href{http://dx.doi.org/10.1088/1361-6404/add972}{{\em Eur. J. Phys.} {\bfseries 46} no.~4, (2025) 045401}, \href{http://arxiv.org/abs/2411.12894}{{\ttfamily arXiv:2411.12894 [quant-ph]}}.

\end{thebibliography}\endgroup

\end{document}